\documentclass[twocolumn,10pt,letterpaper]{article}
\usepackage[top=2cm, bottom=2.2cm, left=1.8cm, right=1.8cm]{geometry}
\usepackage[ruled,algosection,noend,linesnumbered]{algorithm2e}
\usepackage{graphicx}
\usepackage{float}
\usepackage{paralist}
\usepackage{times}
\usepackage{booktabs}

\usepackage{flushend}
\usepackage{url}

\usepackage{amsmath}
\usepackage{amsfonts}
\usepackage[bf]{caption}
\usepackage[justification=justified,singlelinecheck=true]{caption}
\usepackage{multirow}
\usepackage[justification=centering]{subcaption}
\usepackage{xcolor}
\usepackage[toc,page]{appendix}
\usepackage{xpatch}

\makeatletter
\def\url@leostyle{%
  \@ifundefined{selectfont}{\def\UrlFont{}}%
  {\def\UrlFont{}}%
}
\makeatother	
\usepackage[hyphenbreaks]{breakurl}
\definecolor{darkgreen}{RGB}{47,109,79}
\definecolor{darkblue}{RGB}{57,79,99}
\usepackage[bookmarks=false, colorlinks=true, plainpages = false,  linkcolor = darkgreen,   citecolor = darkgreen, urlcolor = darkblue, filecolor = blue]{hyperref}

\newcommand{\E}{\ensuremath{\sf E}\xspace}
\newcommand{\jj}{\ensuremath{\sf j}\xspace}
\newcommand{\kk}{\ensuremath{\sf k}\xspace}
\newcommand{\ite}{\ensuremath{\sf i}\xspace}
\newcommand{\cy}{\ensuremath{\sf c}\xspace}
\newcommand{\locs}{\ensuremath{\sf M}\xspace}
\newcommand{\adverr}{\ensuremath{\sf AdvErr}\xspace}
\newcommand{\fone}{\ensuremath{\sf F1}\xspace}
\newcommand{\js}{\ensuremath{\sf JS}\xspace}
\newcommand{\Ss}{\ensuremath{\sf S}\xspace}
\newcommand{\s}{\ensuremath{\sf s}\xspace}
\newcommand{\Ssize}{\ensuremath{\sf |S|\xspace}}
\newcommand{\Tt}{\ensuremath{\sf T\xspace}}
\newcommand{\Ttsize}{\ensuremath{\sf |T|\xspace}}
\newcommand{\Tp}{{\ensuremath{\sf T'\xspace}}}
\newcommand{\ti}{\ensuremath{\sf t}\xspace}

\newcommand{\Tpsize}{\ensuremath{\sf{|T'|}}\xspace}
\newcommand{\Th}{\ensuremath{\sf{\tilde{\sf{T}}}\xspace}}
\newcommand{\us}{\ensuremath{\sf u}\xspace}
\newcommand{\Uu}{\ensuremath{\sf U}\xspace}
\newcommand{\Usize}{\ensuremath{\sf{|U|}}\xspace}
\newcommand{\agg}{\ensuremath{{\sf A}}\xspace}
\newcommand{\ag}{\ensuremath{\sf a}\xspace}
\newcommand{\pagg}{\ensuremath{{\sf A'}}\xspace}
\newcommand{\aggP}{\ensuremath{\sf{A}^{p}}\xspace}
\newcommand{\aP}{\ensuremath{\sf{a}^{p}}\xspace}
\newcommand{\gt}{\ensuremath{\sf{L}}\xspace}

\newcommand{\lb}{\ensuremath{\sf l}\xspace}
\newcommand{\lp}{\ensuremath{\sf l^{p}}\xspace}

\newcommand{\gtP}{\ensuremath{\sf{L}^{p}}\xspace}

\newcommand{\gs}{\ensuremath{\sf P}\xspace}
\newcommand{\gstar}{\ensuremath{\sf \hat{P}}\xspace}
\newcommand{\gsU}{\ensuremath{\sf P_{\sf{U}}}\xspace}
\newcommand{\loc}{\ensuremath{\sf LOC_{\sf{U}}}\xspace}
\newcommand{\adv}{{\ensuremath{\sf Adv}}\xspace}
\newcommand{\seas}{{\ensuremath{\sf SEAS}}\xspace}
\newcommand{\roiseas}{{\ensuremath{\sf ROI\_SEAS}}\xspace}
\newcommand{\timeseas}{{\ensuremath{\sf TIME\_SEAS}}\xspace}
\newcommand{\lastseas}{{\ensuremath{\sf LAST\_SEAS}}\xspace}
\newcommand{\freqroi}{\ensuremath{{\sf FREQ\_ROI}}\xspace}
\newcommand{\roiday}{\ensuremath{{\sf ROI\_DAY}}\xspace}
\newcommand{\roidayweek}{\ensuremath{{\sf ROI\_DAY\_WEEK}}\xspace}
\newcommand{\timeday}{\ensuremath{{\sf TIME\_DAY}}\xspace}
\newcommand{\timedayweek}{\ensuremath{{\sf TIME\_DAY\_WEEK}}\xspace}
\newcommand{\lastday}{\ensuremath{{\sf LAST\_DAY}}\xspace}
\newcommand{\lasthour}{\ensuremath{{\sf LAST\_HOUR}}\xspace}
\newcommand{\lastweek}{\ensuremath{{\sf LAST\_WEEK}}\xspace}
\newcommand{\bayes}{\ensuremath{{\sf BAYES}}\xspace}
\newcommand{\maxr}{\ensuremath{{\sf MAX\_ROI}}\xspace}
\newcommand{\maxu}{\ensuremath{{\sf MAX\_USER}}\xspace}
\newcommand{\all}{\ensuremath{{\sf ALL}}\xspace}
\newcommand{\pop}{\ensuremath{{\sf POP}}\xspace}
\newcommand{\sort}{\ensuremath{{\sf SORT}}\xspace}
\newcommand{\ind}{\ensuremath{{\sf INDEX}}\xspace}
\newcommand{\descr}[1]{\smallskip\noindent\textbf{#1}}
\newcommand{\descrit}[1]{\vspace{0.1cm}\noindent{\em #1}}
\newcommand{\reduxB}{\vspace{-0.075cm}}
\newcommand{\reduxA}{\vspace{-0.075cm}}
\newcommand{\beforesec}{\vspace{-0.0cm}}
\newcommand{\aftersec}{\vspace{-0.0cm}}

\makeatletter
\patchcmd{\ttlh@hang}{\parindent\z@}{\parindent\z@\leavevmode}{}{}
\patchcmd{\ttlh@hang}{\noindent}{}{}{}
\makeatother\renewcommand{\subparagraph}{}

\usepackage[compact]{titlesec}
\titlespacing*{\section}{0pt}{*2}{3.5pt} 
\titlespacing{\subsection}{0pt}{*2}{4pt}
\titlespacing{\subsubsection}{0pt}{*1.5}{3pt}

\usepackage{etoolbox}
\patchcmd{\thebibliography}
  {\settowidth}
  {\setlength{\itemsep}{0pt plus 0.1pt}\settowidth}
  {}{}

\usepackage{setspace}
\makeatletter
\def\@copyrightspace{\relax}
\makeatother
\begin{document} 

\sloppy

\title{\bf What Does The Crowd Say About You?\\Evaluating Aggregation-based  Location Privacy\thanks{A preliminary version of this paper appears at PETS 2017.}}
\author{Apostolos Pyrgelis$^{1}$, Carmela Troncoso$^{2}$, Emiliano De Cristofaro$^{1}$\\[0.25ex]
\normalsize $^{1}$University College London~~~~~$^{2}$IMDEA Software Institute}
\date{}

\maketitle

\begin{abstract}
Information about people's movements and the locations they visit enables
an increasing number of mobility analytics applications, e.g., in the context of urban and transportation planning, 
In this setting, rather than collecting or sharing raw data, entities often use aggregation 
as a privacy protection mechanism, aiming to hide {\em individual} users' location traces. 
Furthermore, to bound information leakage from the aggregates, they can
perturb the input of the aggregation or its output to ensure that these are differentially private.

In this paper, we set to evaluate the impact of releasing aggregate location 
time-series on the privacy of individuals contributing to the aggregation. We 
introduce a framework allowing us to reason about privacy against an adversary 
attempting to predict users' locations or recover their mobility patterns. We 
formalize these attacks as inference problems, and discuss a few strategies to 
model the adversary's prior knowledge based on the information she may have 
access to. We then use the framework to quantify the privacy loss stemming from 
aggregate location data, {\em with} and {\em without} the protection of differential privacy, using two real-world  
mobility datasets.
We find that aggregates do leak information about individuals' punctual 
locations and mobility profiles. The density of the observations, as 
well as timing, play important roles, e.g., regular patterns during peak 
hours are better protected than sporadic movements. Finally, our evaluation 
shows that both output and input perturbation offer little additional 
protection, unless they introduce large amounts of noise ultimately destroying 
the utility of the data.
\end{abstract}

\section{Introduction}\label{sec:introduction}
\aftersec

The availability of people's locations and movements supports progress in
``mobility analytics'' -- e.g., applications geared to improve urban 
planning~\cite{bocconi2015social}, study the effect of ``shocks'' on 
transport~\cite{silva2015predicting}, predict
events~\cite{horvitz2012prediction}, detect traffic 
anomalies~\cite{pan2013crowd}, generate real-time 
traffic statistics~\cite{waze}, etc.
At the same time, however, large-scale collection of individuals' whereabouts 
prompts serious privacy concerns, as location data may reveal one's 
occupation, lifestyle, as well as political and religious 
beliefs~\cite{krumm2007inference,taxis-rainbows}. 
A possible approach toward mitigating these concerns is to anonymize 
location traces prior to releasing them. Alas, this is 
ineffective, as location data is inherently unique to the user, and the identities of the subjects 
generating the traces can often be recovered~\cite{de2013unique,golle2009anonymity,zang2011anonymization}.

In some cases, mobility models can be trained using only aggregate 
statistics~\cite{chen2012towards,melis2015efficient,pyrgelis2016privacy}, 
e.g., how many people are in a certain location at a given time.
Therefore, a common approach is to consider aggregation as a privacy defense,
and, by using appropriate cryptographic protocols, the aggregation 
can take place in a privacy-preserving way, i.e., removing the need
for a trusted aggregator~\cite{kopp2012privacy,popa2011privacy,pyrgelis2016privacy}.
Moreover, Differential Privacy 
(DP)~\cite{dwork2008differential} can be used to bound the privacy leakage from 
releasing aggregate statistics~\cite{acs2014case, ho2011differential}, using  
output~\cite{chan2011private,dwork2010differential,rastogi2010differentially} or 
input~\cite{erlingsson2014rappor,quercia2011spotme} perturbation.  
However, there is no sound method to reason about the 
privacy lost by single individuals from the release of
raw aggregate time-series. %
Even when using DP, we only get privacy guarantees in terms of 
the theoretical upper-bounds provided by a generic indistinguishability
parameter -- $\epsilon$.
Existing location privacy quantification frameworks~\cite{shokri2011quantify,shokri2011quantifying} 
do not help either, as they typically focus on evaluating single user-centric privacy defense 
mechanisms (e.g., when one user accesses a location-based service).

In this paper, we present a framework geared to address this gap,
and use it to provide a thorough evaluation of aggregation-based location privacy.
We consider an adversary aiming to perform {\em 
localization} attacks, i.e., recovering users' punctual locations, as well as  
{\em profiling}, i.e., inferring their mobility patterns. 
We define appropriate metrics to express the privacy lost from the availability of the aggregates, with respect to the adversary's prior knowledge. 
We propose a few approaches to build such priors, 
parameterized by location and time observations available to her  (e.g., 
users' frequent locations on a Monday morning, or observations 
from the previous week, etc.), and discuss inference strategies, which employ
either Bayesian reasoning or greedy approaches to improve
the knowledge of users' whereabouts, by using the aggregates. 

We then utilize our framework to experimentally measure users' privacy loss when raw aggregate 
time-series are released. We use two mobility datasets obtained from 
Transport for London and the San Francisco Cab network. 
Our comparative analysis shows that, overall, aggregates do improve the adversary's 
prior knowledge about mobility patterns and help her localize users. 
Users' loss of privacy depends not only on the prior knowledge and the inference 
strategy of the adversary, but also on the density of her
observations.
Furthermore, the adversary's inference power is influenced by the nature of the patterns to infer, being regular movements (e.g., 
peak hours/weekdays) better protected than irregular ones (e.g., 
evenings/weekends). 

Next, we study the privacy protection provided by 
DP mechanisms as compared to the release of raw aggregates, vis-\`a-vis the 
utility they provide. 
Although DP ensures an upper bound on the amount of leakage, determined by 
the $\epsilon$ parameter, it is often difficult to interpret
its real-world meaning and to choose appropriate values for it,
despite directly affecting the resulting utility of the data.
Using our framework, we measure the privacy gain provided by using DP techniques, and
find that, in our adversarial model, these mechanisms
only provide meaningful additional privacy protection if the noise 
they introduce is so high that data utility is ultimately destroyed. 
This holds for both input and output perturbation techniques.

Our results demonstrate that, while differential privacy offers
a promising privacy-enhancing solution to several analytics and data mining problems, 
its use in location-oriented applications (including those recently announced by 
Google~\cite{googleTraffic} and  Apple~\cite{AppleDP}) needs to be carefully 
evaluated with respect to the
actual privacy it provides. Overall, our work highlights the need for novel 
defense mechanisms that 
can offer better privacy guarantees to individuals whose location data is part of aggregate time-series releases.

\descr{Paper Organization.} The next section reviews some background 
information. Then, in Section~\ref{sec:analyzing}, we formalize the problem of 
quantifying privacy leakage from aggregate location time-series. 
Section~\ref{sec:experiments} presents an experimental evaluation on two 
real-world mobility datasets, while Section~\ref{sec:counter} analyzes DP 
techniques for protecting aggregates. After reviewing related work in 
Section~\ref{sec:related}, the paper concludes in Section~\ref{sec:conclusion}.

\beforesec
\section{Preliminaries}
\label{sec:preliminaries}
\aftersec

\descr{Kullback-Leibler (KL) Divergence~\cite{kullback1951information}.} Also known as discrimination information, the Kullback-Leibler (KL) divergence captures the ``difference'' between two probability distributions. Specifically, for two discrete probability distributions ${\sf W}$ and ${\sf X}$, the KL-divergence from ${\sf X}$ to ${\sf W}$ is defined as: \reduxB
\begin{equation}
{\sf D_{KL} (W || X) = \sum_{i} W(i) \cdot \log{\frac{W(i)}{X(i)} } }\reduxA
\end{equation}
where ${\sf W}$ usually represents the \textit{true} distribution of data and ${\sf X}$ an approximation of ${\sf W}$. In other words, KL-divergence from ${\sf X}$ to ${\sf W}$ measures the information lost when ${\sf X}$ is used to approximate ${\sf W}$. Note that KL is not a \textit{metric} as it does not satisfy the triangle equality and in general not symmetric in ${\sf W}$ and ${\sf X}$.

\descr{Jensen-Shannon (JS) Divergence~\cite{endres2003new,lin1991divergence}.} It is used to calculate the similarity between two probability distributions. It is based on KL-divergence but it is symmetric and always a finite value. The JS-divergence is a smoothed version of the KL-divergence ${\sf D_{KL}(W || X)}$, defined by:\reduxB
\begin{equation}
{\sf JS(W || X) = \frac{1}{2} \cdot D_{KL}(W || Z) + \frac{1}{2} \cdot D_{KL}(X || Z) }
\reduxA
\end{equation}
where ${\sf Z = \frac{1}{2} \cdot (W + X)}$.
When employing the base $2$ logarithm for calculating KL-divergence, the JS-divergence is bounded by $1$, thus ${\sf 0 \leq JS( W || X ) \leq 1 }$.
Note that the square root of the JS-divergence is a \textit{metric} denoted as Jensen-Shannon distance~\cite{endres2003new} (also bounded by $1$). We use JS-distance to calculate the adversarial error in profiling users. %

\descr{F1 Score.} F1 is often used to evaluate the accuracy of classification/prediction tasks, as it captures overall performance by taking into account both precision and recall. Precision (aka positive predictive value, or ${\sf PPV}$) and recall (aka true positive rate, or ${\sf TPR}$) are defined, respectively, as ${\sf PPV = TP / (TP + FP)}$ and ${\sf TPR = TP / (TP + FN)}$, where ${\sf TP}$, ${\sf FP}$, and ${\sf FN}$ denote, respectively, true positives, false positives, and false negatives. The $\fone$ score is calculated as:\reduxB
\begin{equation}
{\sf F1 = \dfrac{2 \cdot TPR \cdot PPV}{TPR + PPV}}\reduxA
\end{equation}
We use it to quantify adversary's accuracy in localizing users.

\begin{table}[t]
\centering
\small
  \begin{tabular}{ r  l }
  \toprule
    {\bf Symbol} & {\bf Description} \\ 
    \midrule
	\adv & Adversary \\ %
    \Uu & Set of mobile users \\ %
    \Ss & Set of locations (ROIs) \\ %
    \Tt & Time period considered \\ %
	\Tp & Inference period \\ %
    \Th & Observation period \\ %
	\gt & Ground truth \\ %
	\gtP & Ground truth mobility profile \\ 
	\agg & Aggregate time-series \\ %
	\aggP & Aggregate mobility profile \\
	\pagg & Perturbed aggregate time-series\\
	\gs & \adv's prior knowledge \\ %
	\gstar & \adv's inference output \\   
	\bottomrule
	  \end{tabular}
\reduxB\reduxB
	\caption{Notation.} 
  \label{table:notation}
  \vspace{0.2cm}
  \end{table}

  \beforesec
\section{Quantifying Aggregate Location Privacy}
\label{sec:analyzing}

\subsection{Problem Statement}\label{sec:problem}
\aftersec

In the rest of the paper, we use the notation summarized in Table~\ref{table:notation}. We consider a set of users \Uu that move among a set \Ss of regions of interest 
(ROIs) -- e.g., landmarks, neighborhoods, stations -- at time instances 
in the set \Tt. This set represents the time frame in which locations are 
collected (e.g., 1 week, 1 month, 1 year), while locations 
can be aggregated in epochs of different granularity (e.g., 15 mins, 30 mins, 1 
hour).

\descr{Ground truth.} We model the actual locations \Ss of a user $\us\in\Uu$, during $\Tt$, 
using a \textit{ground truth matrix} \gt of size $\Ssize\times\Ttsize$, in which rows are ROIs 
and columns are epochs. \gt is a binary matrix s.t. $\lb_{s,t} \in \gt$ is 1 if 
the user was in $\s\in\Ss$ during epoch $\ti\in\Tt$, and 0 otherwise. We note 
that depending on the time granularity of location reports users can be in more than 
one ROI in the same epoch, thus, there can be more than one $1$ per column.
We also define a \textit{mobility profile}, \gtP, 
where $\lp_{s,t}\in\gtP$ represents the probability that a user is in location $\s$ at 
time slot $\ti$ and is computed as $\lb_{s,t}/\sum_{j \in \Ss} \lb_{j,t}$.

\descr{Aggregates.} The aggregate location time-series is represented by the matrix $\agg$, of 
size $\Ssize \times \Tpsize$.
We call \Tp\ the {\em inference period}, i.e., 
aggregation does not need to happen in the full collection period.
Each item $\ag_{\s,\ti'}\in\agg$ represents the number of users
in $\s$ at epoch $\ti'$, and is calculated as $\ag_{\s,\ti'} = \sum_{\us=1}^{ | \Uu |} \lb_{s,t'}$, where $\lb_{s,t'}$ are the entries of each user's \gt. The aggregation can be performed by a 
trusted aggregator or via cryptographic protocols~\cite{kopp2012privacy, popa2011privacy, 
pyrgelis2016privacy}. 
We also define \aggP, the \textit{aggregate mobility profile}, as a probability distribution matrix whose entries 
$\aP_{s,t'}\in \aggP$ are computed as $\ag_{s,t'} / \sum_{j \in \Ss} 
\ag_{j,t'}$. This represents the probability of users being in a ROI at 
an epoch, while observing the aggregates. For instance, $\aP_{s,t'} = 0.1$ indicates that at time $\ti'$, 
$10\%$ of the user observations are in ROI $\s$.

\descr{Prior knowledge.} We model the prior knowledge the adversary, denoted as 
\adv, may have about a user $\us \in \Uu$ for the inference period \Tp, using a 
matrix $\gs$, of size $\Ssize\times\Tpsize$. \gs can be probabilistic (i.e., 
describing how likely a user is to visit a ROI) or binary (i.e., indicating 
whether a user will visit a ROI or not). We discuss how to build these priors 
in Section~\ref{sec:prior}.

\descr{Quantifying aggregate location privacy.} 
Given the observation of the aggregates (\agg), and the prior knowledge about each user (\gs), \adv aims to infer information about individual users from the time-series. %
We model the output of this inference as a matrix \gstar, for each user, of size 
$\Ssize\times\Tpsize$.
We do so to quantify the privacy loss for individual users given the adversary's prior knowledge  and her capability to exploit the aggregates. Specifically, we measure the 
adversary's \textit{error} vis-\`a-vis the ground truth \gt, after executing inference attacks,
considering two goals: user profiling and user localization.
\descrit{User Profiling:} \adv\ aims to infer the mobility profile of the 
users. Given \gs and \agg\ (or \aggP), \adv outputs a matrix \gstar. This 
matrix contains a probability distribution profile for each user reflecting the 
likelihood that the user is in each ROI at each epoch.
To compute \adv's error, we compare the ground 
truth mobility profile $\gtP$ to \adv's inference $\gstar$ if we consider the result of the 
inference attack, or to the prior $\gs$ if aggregate data is not available. 
For each user and each $\ti' \in \Tp$, we use the JS metric, 
reviewed in Section~\ref{sec:preliminaries}, to measure the distance between 
the probability distributions. For each user profile, we measure \adv's total error 
over the inference period $\Tp$ as: 
\reduxB
\begin{equation}\label{eq:js}
\adverr_{JS} = \dfrac{\sum_{\ti' \in \Tp} \js( \gtP_{t'} || \gs_{t'})}{| \Tp |}
\end{equation}
Intuitively, at each time slot, JS computes the distance between the inferred and the ground truth profile, 
averaged over the ROIs. This captures the adversary's error regarding profile estimation. Eq.~\ref{eq:js} averages the distance per slot over all time slots, i.e., it computes the adversary's mean error on the inference period.

\descrit{User Localization:} \adv\ aims to infer the punctual locations of 
the users over time. More formally, given \gs and \agg\ (or \aggP), \adv outputs 
a binary matrix \gstar for each user, with 1's for ROIs \adv\ predicts the user 
to be in, and 0's elsewhere. 
To measure \adv's performance we compare her 
predictive assignment matrix on each user (either prior $\gs$ or 
posterior $\gstar$) against the ground truth $\gt$. Concretely, we use \adv's 
precision and recall when predicting users' locations to derive the $\fone$ score, 
reviewed in Section~\ref{sec:preliminaries}, and measure the total 
adversarial error as:\reduxB
\begin{equation}\label{eq:f1}
\adverr_{F1} = 1 - \fone \reduxA
\end{equation}
The \fone score captures the distance between the inferred and the ground truth binary matrices, i.e., Eq.~\ref{eq:f1} reflects the adversarial error regarding localization over the inference period.
Note that both adversarial goals have been considered in location privacy literature~\cite{de2008identification,krumm2007inference,shokri2011quantifying,wernke2014classification}, 
although in different contexts, namely, reconstructing traces or recovering a user's location from obfuscated 
individual data.

\descr{Privacy Loss (PL).} For both adversarial goals, we measure the privacy loss for an individual user from 
the aggregate location time-series as the normalized difference between \adv's error using her prior knowledge (\gs), 
with and without $\agg$ ($\adverr_{\gs, \agg}$ and $\adverr_{\gs}$ resp.). More specifically, for each user we 
define the privacy loss (PL) as:
\reduxB
\begin{equation}\label{eq:pl}
{\sf PL} = \begin{cases}
\frac{| \adverr_{\gs, \agg} - \adverr_{\gs} | }{\adverr_{\gs}}~\text{if} \ \adverr_{\gs} \neq 0 \ \wedge \ \\[-.5ex]
	~~~~~~~~~~~~~~~~~~~~~~~~~~~~~~~~~ \adverr_{\gs, \agg} < \adverr_{\gs} \\
0 ~~~~~~~~~~~~~~~~~~~~~~~~~~~\text{otherwise}\reduxA
\end{cases}
\end{equation}
PL is a value between $0$ and $1$ and captures \adv's improvement towards her
goal (either profiling or localizing users). %

\beforesec
\subsection{Adversary's Prior Knowledge}\label{sec:prior}
\aftersec

We now present a few different approaches to build the adversary's prior knowledge, which we divide in  {\em probabilistic} priors, i.e., user profiles averaging location reports, and {\em assignment} ones, i.e., binary matrices representing users' location visits over a certain period of time.
Essentially, they differ in how the \gs matrix is populated, depending on what information 
is assumed to be available to the adversary and the strategy used to extract 
prior knowledge about each user. 

In real life, adversarial prior knowledge may originate from, e.g., social 
networks, data leaks, location traces released by providers, personal knowledge. 
Here we aim to describe a generic quantification framework, comparing different 
adversarial strategies,
hence, we opt to construct priors from a subset of the users' ground truth matrices (\gt),
including epochs in $\Th\subseteq\Tt$, which we call the {\em observation period}.
We follow intuitive strategies, based on a sensible threat model in which \adv obtains information about users' routines and punctual locations (e.g., where one works/lives) over a certain period of time. %
Nonetheless, our framework is generic enough so that new ways of building \adv's priors can be easily incorporated.

\beforesec
\subsubsection{Probabilistic Priors}
\label{sec:probabilistic-prior}
\aftersec
Probabilistic priors model prior information that represent knowledge of user profiles.

\descr{ROI Frequency.} We start by considering that \adv knows the probability
of a user visiting a given ROI during the full observation period. We assume that \adv has access to 
a vector of size \Ssize, indicating how frequently the user visits each ROI during \Th.
\adv then populates \gs by: (i) transforming the vector into a probability
distribution using the total number of user's observations, $\locs$, as normalizing factor,
and (ii) copying the distribution onto \gs, for all time slots of the inference period 
\Tp. More specifically, using the entries $\lb_{s,t}$ in the ground truth \gt, 
we build \gs, $\forall \s\in\Ss, \ti'\in\Tp$, as:\reduxB
\begin{equation}
 \gs_{FREQ\_ROI}(\s, \ti') := \sum_{\ti \in \Th} 
\lb_{s,t} / \locs \reduxA\reduxA\reduxA\reduxA
\end{equation}

\descr{ROI Seasonality.} This prior models the case that \adv knows the seasonal probability
of a user visiting a ROI during the observation period $\Th$, for a given seasonal time period \seas. 
For instance, if \seas corresponds to one day, and epochs are of one hour, we assume that \adv obtains a probability distribution
over the ROIs for every hour in a day. If seasonality is on days of the week, 
the probability distribution over ROIs available to the adversary is for each hour, for each day of the week.
More formally, if $\cy$ denotes the seasonality cycle of \seas (e.g., $\cy=24$ hours for daily or $\cy=7\cdot 24$ hours for weekly seasonalities), then the seasonality profile is, $\forall \s\in\Ss, \forall \ite\in\{1,\ldots,\cy\}$:\reduxB
\begin{equation}
\roiseas_{\s, \ite} = \dfrac{\sum_{\kk=0}^{\Th/\cy-1} \lb_{\s,\ite+\kk\cdot 
\cy}}{\sum_{\jj \in \Ss}\sum_{\kk=0}^{\Th/\cy-1}\lb_{j,\ite+\kk\cdot \cy}} \reduxA
\end{equation}
Then, we build \gs, $\forall \s\in\Ss, \ti'\in\Tp$ as: \reduxB
\begin{equation}
\gs_{ROI\_SEAS}(\s,\ti') := \roiseas_{\s,\ti'\bmod \cy}\reduxA
\end{equation}

\descr{Time Seasonality.} We assume \adv knows the seasonal probability
of a user reporting her location (without any information about which concrete ROIs) during the observation period $\Th$, for a given seasonal time period \seas.
For instance, if \seas corresponds to one day, and the granularity is one hour, \adv learns which hours of a day a user is likely to report locations. 
More formally, if $\cy$ denotes the seasonality cycle of \seas, %
then the time seasonality profile is, $\forall \ite \in \{1,\ldots,\cy\}$:\reduxB
\begin{equation}
\timeseas_{\ite} = \dfrac{\sum_{\jj \in \Ss}\sum_{\kk=0}^{\Th/ \cy-1} \lb_{j,i+\kk \cdot 
c}}{\locs} \reduxA
\end{equation}
where $\locs$ is the total number of user's observations within the period \Th. 
Then, \gs is built, $\forall \s\in\Ss, \ti'\in\Tp$, as:\reduxB
\begin{equation}
\gs_{TIME\_SEAS}(\s,\ti') := 
\begin{cases}
      1/|\Ss|~\text{if}\ \timeseas_{\ti'\bmod c} > 0 \\
      0~~~~~~~\text{otherwise}
\end{cases}\reduxA
\end{equation}
i.e., it is a \textit{uniform} probability distribution over ROIs for the time slots when the user is likely to report locations.

\beforesec
\subsubsection{Assignment Priors}
\label{sec:assignment-prior}
\aftersec

Next, we describe strategies to compute prior information that represents knowledge of users' punctual locations. An assignment prior 
is modeled as a binary matrix that predicts whether or not a user will be in a location $\s \in \Ss$, at time $\ti' \in \Tp$. 

\descr{Most popular prior ROIs.} We model the case that \adv only considers users' favorite locations (\pop). 
Given a probabilistic prior knowledge $\gs$, and a threshold $\delta$ modeling what the adversary considers to be \textit{favorite}, \adv builds a binary location matrix so that $\forall \s\in\Ss,\ti' \in\Tp$: \reduxB
\begin{equation}
\gs_{POP}(s,t') := 
\begin{cases}
1 \text{ if } \gs_{s,t'}\geq \delta\\ 
0 \,\text{ otherwise} 
\end{cases}\reduxA
\end{equation}

\noindent{\bf All prior ROIs.} Next, we consider a scenario where \adv\ considers every location that users visit, but not the frequency (\all). Given a probabilistic prior knowledge $\gs$, \adv builds a binary location matrix so that $\forall \s\in\Ss,\ti' \in\Tp$:\reduxB
\begin{equation}
\gs_{ALL}(s,t') := CEIL( \gs_{s,t'} ) \reduxA
\end{equation}
where CEIL is the ceiling function, thus $\gs_{ALL}(s,t')=1$ iff the probability of visiting $\s$ at $\ti'$ is greater than 0 (i.e., the user has visited that location during time slots of $\Th$). %

\begin{table}[t]
    \centering
\resizebox{0.99\columnwidth}{!}{%
  \begin{tabular}{l  p{5.8cm}}
  \toprule
    {\bf Prior} & {\bf Description} \\ \midrule
    \freqroi & Frequent ROIs, over time \\ %
    \roiday & Most frequent ROIs, for each time instance of a day \\ %
    \roidayweek & Most frequent ROIs, for each time instance of a week \\ %
    \timeday & Most frequent time instances of a day, reporting ROIs \\ %
	\timedayweek & Most frequent time instances of a week, reporting ROIs \\ %
	\lastweek &	Last week's ROIs\\ %
	\lastday & Last day's ROIs \\ %
	\lasthour & Last hour's ROIs \\ 
	\bottomrule
  \end{tabular}
}
\reduxB\reduxB
  \caption{Different ways to build adversarial prior knowledge.}
  \label{table:priors}
  \vspace{0.2cm}
\end{table}

\descr{Last Season.} We assume that \adv has access to the last seasonal information for each user, i.e., the last season \seas constitutes the observation period $\Th$.
For instance, if \seas corresponds to 1 day and time granularity is 1 hour, \adv only knows the locations visited in each hour of the last day.
Formally, if $\cy$ denotes the seasonality cycle, e.g., $\cy = 7 \cdot 24$ hours for weekly seasonality, \gs is built utilizing a sliding window as: \reduxB
\begin{equation}
\forall \s\in\Ss, \ti'\in\Tp\,: \gs_{LAST\_SEAS}(s,t') := \gt_{\s,\ti' - \cy}\reduxA
\end{equation}

\descr{Summary.} Table~\ref{table:priors} summarizes our approaches 
to construct the adversarial prior knowledge. For priors taking 
\textit{seasonality} into account, \seas takes a value indicating the 
seasonal period we consider to build \adv's initial knowledge.

\beforesec
\subsection{Location Inference Strategies}
\label{sec:attacks}
\aftersec

We now describe possible strategies that the adversary can follow to exploit aggregate locations in order to make inferences about individuals.
We present algorithms that, taking as input \adv's prior knowledge about a given user (\gs) and the location aggregate time-series (\agg or \aggP), output an updated matrix \gstar. This matrix represents \adv's posterior knowledge about the user's whereabouts over the inference period $\Tp$ by virtue of the availability of the aggregate time-series. The proposed strategies can be used for both profiling and localization attacks, 
the difference being the nature of the output matrix \gstar, which is 
probabilistic in the former case and binary in the latter.
In the following, we use ${\sf \Theta(:,x)}$ or ${\sf \Theta(x,:)}$ to denote all the instances of a dimension in a matrix ${\sf \Theta}$.

\descr{Bayesian Updating.} The first strategy, summarized in 
Algorithm~\ref{alg:bayesian}, computes the posterior probability of $\us 
\in \Uu$ being in each ROI $\s \in \Ss$, during $\ti' \in \Tp$, given the 
adversarial prior knowledge (\gs) and the aggregate mobility profile $\aggP$.
Let $\E_{\s, \ti'}$ 
denote the event in which the user appears in location $\s \in \Ss$ at time $\ti' \in \Tp$. 
Given \adv's prior information about this event, $\Pr[\E_{\s,\ti'}] = \gs_{s,t'}$, and her observation ${\sf O}$ at time $\ti'$, i.e., a probability distribution of users over all ROIs $\s \in \Ss$ at time $\ti'$, we compute the posterior probability using the Bayes theorem as: %
\begin{equation}
\gstar_{s, t'} = \Pr[E_{s, t'} | O ] = \dfrac{\Pr[O | E_{s, t'}] \cdot \Pr[E_{s,t'}]}{ \Pr[O] } \smallskip
\end{equation}
where $\Pr[\sf O]$ represents the user's un-normalized distribution over ROIs at time $\ti'$ and can be calculated using the law of total probability, i.e.,  %
$\Pr[{\sf O}] = \sum_{\jj \in \Ss} \aggP_{j,t'} \cdot \gs_{j,t'}$, and $\Pr[{\sf 
O} | \E_{s,t'}]=\aggP_{s,t'}$ is given by the released aggregate statistics. 

\begin{figure}[t]
\begingroup
\csname @twocolumnfalse\endcsname
\noindent
\resizebox{.49\textwidth}{!}{%
\begin{algorithm}[H]
\small
\KwIn{$\gs$, $\aggP$}
\For{\textbf{each} $\us \in \Uu$}
{	
\For{\textbf{each} $\ti'  \in \Tp$}
{
$\gstar(:, \ti') = \gs(:, \ti') \times \aggP(:, \ti') $ 	
$\gstar(:, t') = \gstar(:, \ti') / \sum_{\jj \in \Ss} \gstar(\jj, \ti')$ 

	}
	\KwRet $\gstar$;
}
\caption{\bayes}
\label{alg:bayesian}
\end{algorithm}
}%
\endgroup
\end{figure}
\begin{figure}[t]
\begingroup
\csname @twocolumnfalse\endcsname
\noindent
\resizebox{.49\textwidth}{!}{%
\begin{algorithm}[H]
\small
\SetKwComment{Comment}{//}{}
\KwIn{$\gs$, $\agg$}

$\gs_{\Uu}(:,:,:) = \emptyset$

$\loc(:,:,:) = \emptyset$

\For{\textbf{each} $\us \in \Uu$}
{
	$\gs_{\Uu} = \gs_{\Uu} || \gs^{(\us)} $ %
}
\For{\textbf{each} $\s \in \Ss, \ti	' \in \Tp$}
{
	\If{$\agg(\s,\ti') == 0$}
	{
		$ \loc(:, \s, \ti') = 0 $ %
	}
	\Else
	{
		$ {\sf X} = \gs_{\Uu}(:, \s, \ti')$		
		
		$ \Uu^{*} = \sort({\sf X}, \agg(\s, \ti'))$ 
		
		\For{\textbf{each} ${\sf z} \in \Uu^{*}$}
		{
			$\loc({\sf z}, \s, \ti') = 1 $ %
		}
	}
}
\For{\textbf{each} $\us \in \Uu$}
{
	$\gstar = \loc( \us, :, :)$ %
	
	\KwRet $\gstar$;
}
\caption{\maxr}
\label{alg:max-roi}
\end{algorithm}
}
\endgroup
\end{figure}

\descr{Max-ROI.} The Bayesian approach is well-principled but considers users 
independently, thus losing information related to the fact that at most 
$\agg_{\s,\ti'}$ users can be assigned to a location $\s$, at time $\ti'$. We 
now describe a \textit{greedy} alternative that accounts for this constraint. The 
algorithm aims at maximizing the total probability for each ROI by assigning the 
most probable users to each location. It is summarized in 
Algorithm~\ref{alg:max-roi}, where $\sort({\sf V, x})$ denotes a function 
that returns the indexes of the \textit{top} ${\sf x}$ values, of a vector ${\sf 
V}$.
Specifically, \adv first concatenates the probabilistic prior $\gs$ matrices of all users $\us \in \Uu$ and creates a 3-dimensional matrix of size $\Usize \times \Ssize \times \Tpsize$, which we denote as \gsU (lines 3--4, Algorithm~\ref{alg:max-roi}). Additionally, she creates a localization matrix of same size, denoted as \loc. Next, \adv selects all $\s,\ti'$ s.t. $\agg_{\s,\ti'}=0$ and sets the corresponding indexes of \loc to zero, i.e., she discards locations where no users have been observed during the aggregation period (lines 6--7). Then, for all non-zero entries in \agg, she selects the $\agg_{\s,\ti'}$ most probable users according to her prior, \gsU, setting the corresponding indexes in \loc to 1 (lines 9--12). If there are users with equal probability, \adv can use any criterion, e.g., the total number of location reports (as we do in our experiments that are presented in Section~\ref{sec:experiments}) to make a decision. Finally, \adv outputs the location assignment profile of each user as her $\gstar$ matrix (lines 13--15).

\descr{Max-User.} Our final inference attack is similar in spirit to the previous \textit{greedy} strategy but, rather than maximizing the probability over ROIs, it maximizes each user's probability over the ROIs by assigning them to their most likely locations. The algorithm is summarized in Algorithm~\ref{alg:max-user}, where $\ind({\sf V > x})$ denotes a function that returns the indexes of a vector ${\sf V}$, whose values are larger than ${\sf x}$.
More precisely, \adv first sorts users by some criterion, e.g., the total number of locations that they report (as we will do in our experiments in Section~\ref{sec:experiments}). Then, at each time slot $\ti' \in \Tp$, \adv iterates over the users and assigns each of them to their most likely ROIs, provided that each ROI's aggregate $\agg(\s, \ti')$ is still not consumed (lines 3--7, Algorithm~\ref{alg:max-user}). The procedure is repeated until the assignments cover all the revealed aggregate information (lines 8--9).
\begin{figure}[t] 
\begingroup %
\csname @twocolumnfalse\endcsname
\noindent
\resizebox{.49\textwidth}{!}{%
\begin{algorithm}[H]
\small
\SetKwComment{Comment}{//}{}
\KwIn{$\gs$, $\agg$}

$\loc(:, :, :) = \emptyset$

\For{\textbf{each} $\ti' \in \Tp$}
{
	\For{\textbf{each} $\us \in \Uu$}
	{	
		${\sf Idx} = \ind( \gs(:, t') > 0.0 ) $ %
		
		\For{\textbf{each} $\ite \in {\sf Idx}$}
		{
			\If{$\sum_{{\sf v} \in \Uu} \loc({\sf v}, \ite, \ti') < \agg(\ite, \ti')$}
			{
				$\loc(\us, \ite, \ti') = 1$ %
			}
		}
		\If{ $\sum_{{\sf w} \in \Uu, \jj \in \Ss} \loc({\sf w}, \jj, \ti') == \sum_{\jj \in \Ss} \agg(\jj, \ti') $} 
		{	
			$break$; %
		}
	}
}
\For{\textbf{each} $\us \in \Uu$}
{
	$\gstar = \loc( \us, :, :)$ %
	
	\KwRet $\gstar$;
}
\caption{\maxu}
\label{alg:max-user}
\end{algorithm}
}
\endgroup
\end{figure}

\descr{\em Note:} 
Our strategies are suitable for both \adv's inference goals, i.e.,  profiling 
and localization. For instance, if \adv is given a probabilistic prior, she can 
follow \maxr or \maxu strategies and transform their assignment outputs to 
probability distributions that can be used for her profiling goal. Similarly, 
she can run \bayes on the prior and evaluate \pop and \all on its output to 
localize users. 

\beforesec
\section{Privacy Evaluation of Raw\\Aggregates}
\label{sec:experiments}
\aftersec
We now use our framework to experimentally evaluate aggregate location privacy from raw aggregates release. We compare different approaches to build priors (Section~\ref{sec:prior}) as well as strategies to perform inference attacks (Section~\ref{sec:attacks}), using two mobility datasets obtained from London's transportation authority and the San Francisco Cab network.

\beforesec
\subsection{Datasets}
\label{sec:dataset}
\aftersec

\descr{Transport for London (TFL).} We have obtained, from the TFL authority, all Oyster card trips on the TFL network from March 2010 (8GB uncompressed). The Oyster is a personal, pre-paid, RFID-enabled card, and the most common payment system on TFL-operated services.
Each entry in the data describes a unique trip and consists of the following fields: (anonymized) oyster card id, start time, touch-in station id, end time, and touch-out station id. 
(Note that the same dataset has also been used in~\cite{ceapa2012avoiding,lathia2013individuals,pyrgelis2016privacy}).

\descrit{Pre-processing \& Sampling.} We discard trips from March 
29--31, 2010 to obtain exactly four weeks of data, i.e., from Monday March 1st to 
Sunday 28th. This yields $60$ million trips, performed by 
$4$ million unique oyster cards, covering $582$ stations (ROIs). 
Next, we select the top 10,000 oyster ids per total number of trips:
these account for about 6M trips ($10\%$). Considering oyster trips start/end 
stations as ROIs, the top 10,000 users report, $171 \pm 26 $ ROIs in 
total and $19 \pm 9$ unique ROIs. Setting the time granularity to one hour, 
the mean number of \textit{active} time slots for the top 10,000 oysters is 
$115 \pm 21$ out of the $672$ slots ($28$ days$\times24$ hours).

\descrit{Ground Truth.} We use the trips performed by each Oyster card in the 
dataset to populate its ground truth matrix \gt. More specifically, $\lb_{s,t}\in\gt$ 
is 1 if the user touched-in or out at station $\s$, during time slot $\ti \in \Tt$, and 0 
otherwise. When an Oyster card does not report any location at a particular time slot, we 
assign it to a special ROI denoted as \textit{null}. Thus, the ground truth \gt is a matrix of size
$\Ssize \times \Ttsize = 583\times672$.

\descrit{Prior Knowledge (Training data).} We build the probabilistic 
adversarial prior knowledge using the {\em first 3 weeks} of \gt (i.e., $75\%$ of data are used for training). 
Thus, the \textit{observation} period $\Th$ consists of $21 \times 24 = 504$ hourly time 
slots. For the seasonal assignment priors, we utilize a {\em  sliding window} on \gt, as 
described in Section~\ref{sec:assignment-prior}.

\descrit{Testing Data \& Aggregates.} We evaluate \adv's performance in profiling/localizing users against 
the {\em last week} of \gt (i.e., $25\%$ of the data are used for testing). Thus, the 
\textit{inference} period $\Tp$ consists of $7 \times 24 = 168$ hourly time slots. For each station 
$\s \in \Ss$, we count the number of users that report their presence in it (touch-in or touch-out) during 
each epoch $\ti' \in \Tp$, and create the aggregate time-series $\agg$ (of size $\Ssize \times \Tpsize = 583\times168$) whose items $\ag_{\s,\ti'}$ are computed 
as $\sum_{\us=1}^{| \Uu |} \lb_{\s,\ti'}$ (remind that $\lb_{\s,\ti'}$ are the entries of each oyster's \gt). 
During $\Tp$, each station is reported $818 \pm 1,361$ times while stations have commuters touching in/out for 
$71 \pm 54$ out of the $168$ hourly time slots.

\descr{San Francisco Cabs (SFC).} We also use the SFC dataset~\cite{epfl-mobility-20090224}, with  mobility traces recorded by cabs in the San Francisco area from May 17 to June  10, 2008. Each record includes:
cab identifier, latitude, longitude and a time stamp. %

\descrit{Pre-processing \& Sampling.} The dataset consists of approximately 11 million GPS coordinates, 
generated by 536 taxis. To facilitate our experiments, %
we focus on exactly 3 weeks: Monday May 19 to Sunday June 8. We 
restrict to the downtown San Francisco area, %
dividing it into a grid of $10 \times 10=100$ regions (ROIs), each 
covering an area of $ 0.5 \times 0.37$ square miles. We group traces in 
one-hour epochs. We also remove duplicates (e.g., a taxi reporting the same ROI multiple 
times during a time slot). This yields a dataset of over 2 million ROIs reported by 534 taxis, reporting $3,663 \pm 1,116$ locations in total, covering $77 \pm 6$ unique ROIs (out of the 100 we consider). Unlike TFL data, the SFC data is 
less sparse, with cabs reporting more locations. On average, cabs are ``inside the system'' 
for $340 \pm 94$ out of the $504$ ($21$ days$\times24$ hours) time slots.

\descrit{Ground Truth.} For each cab, we build its \gt matrix, by setting $\lb_{s,t} $ 
to 1 if the cab was in the $\s$ ``cell'' during time slot $\ti \in \Tt$, and 0 otherwise. As for TFL data, if a cab 
does not report any location at a time slot we assign it to a special ROI, which we denote as \textit{null}, thus,
the \gt matrix is of size $\Ssize \times \Ttsize = 101\times 504$.

\descrit{Prior Knowledge (Training data).} We build the probabilistic adversarial prior knowledge using the \textit{first 2 
weeks} of \gt (i.e., $\sim66\%$ of the data), thus the \textit{observation} period $\Th$ consists of $14 \times 24 = 
336$ hourly time slots. For the \lastseas assignment priors, we utilize a sliding window on \gt. 

\descrit{Testing Data \& Aggregates.} We quantify \adv's performance in 
profiling/localizing cabs against the \textit{last week} of \gt (i.e., $\sim33\%$ of the data are used for testing), 
thus, the inference period $\Tp$ consists of $7 \times 24 = 168$ hourly time slots. For each $\s \in \Ss$, 
we count the number of taxis reporting it during epoch $\ti' \in \Tp$, and create the aggregate 
time-series $\agg$ (of size $\Ssize \times \Tpsize = 101\times168$) whose items $\ag_{\s,\ti'}$ are computed as $ \sum_{\us=1}^{| \Uu |} \lb_{\s,\ti'}$ (where $\lb_{\s,\ti'}$ are the entries of each cab's \gt). During $\Tp$, each ROI is
reported $6,714 \pm 7,624$ times while ROIs have taxis in them for $135 \pm 61$ out of the 168 time slots.

\beforesec
\subsection{User Profiling}
\aftersec

\begin{figure}[t]
\centering
\begin{subfigure}[b]{0.45\textwidth}
\includegraphics[width=0.99\linewidth]{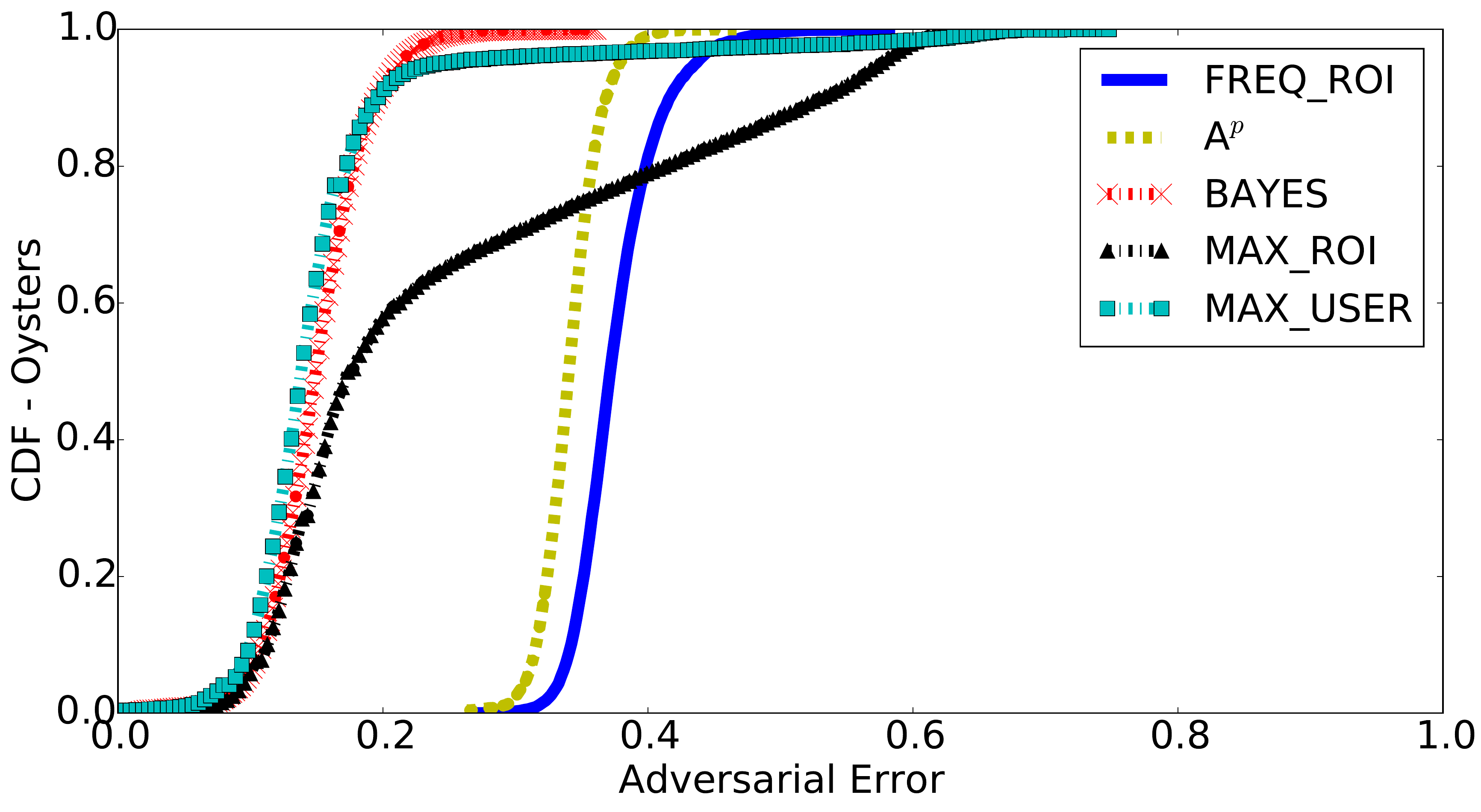}
\caption{TFL}
\label{fig:task1-freq-roi-tfl}
\end{subfigure}
\begin{subfigure}[b]{0.45\textwidth}
\includegraphics[width=0.99\linewidth]{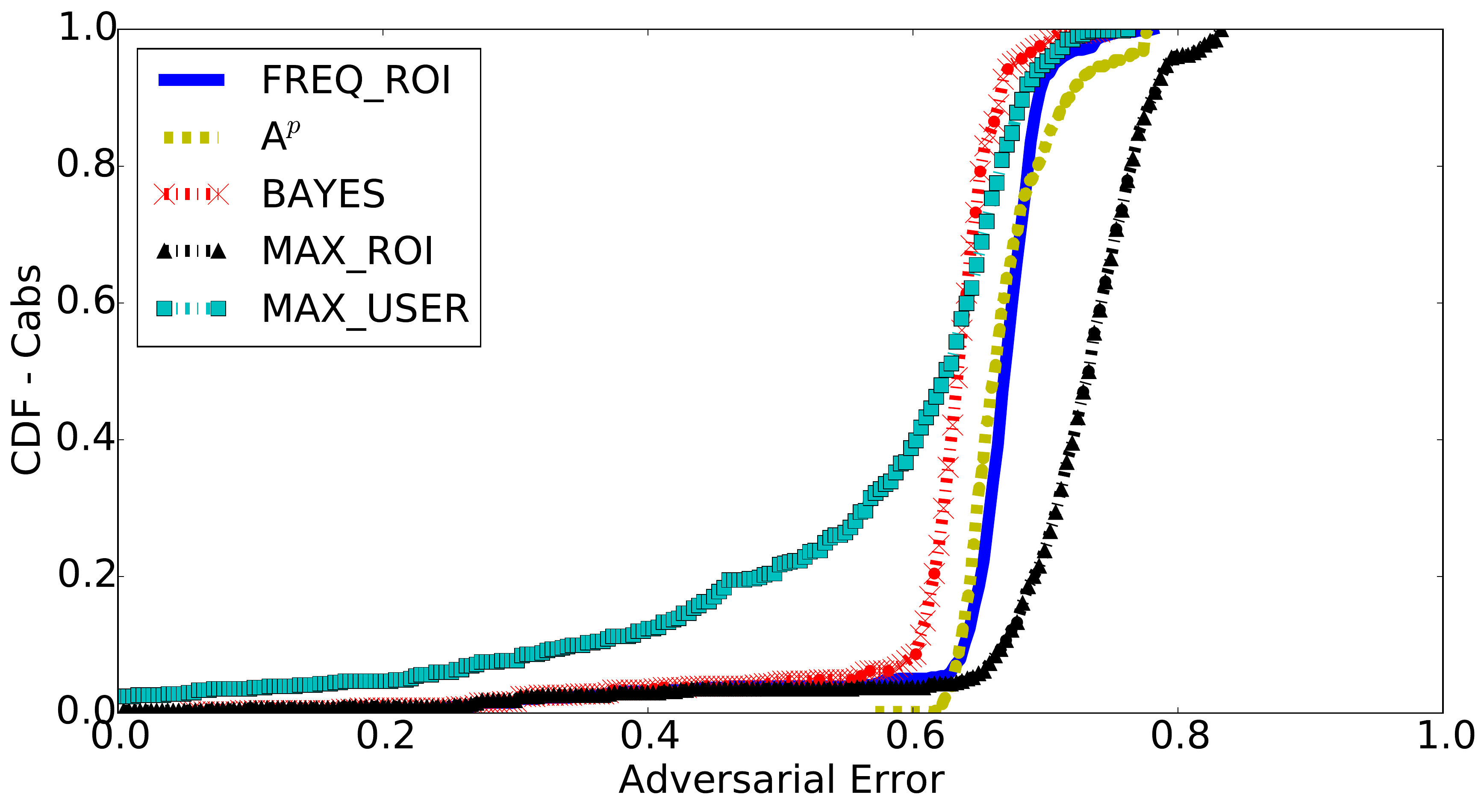}
\caption{SFC}
\label{fig:task1-freq-roi-sfc}
\end{subfigure}
\reduxB\reduxB
\caption{\adv's Profiling Error - \freqroi prior.}
\label{fig:task1-freq-roi}
\end{figure}

We start our experimental evaluation by quantifying aggregate location privacy 
against \textit{user profiling} (cf.~Section~\ref{sec:problem}). We study the impact of the information used to 
build \adv's prior vis-\`a-vis the strategy used to exploit aggregate data. 
Specifically, we measure \adv's performance using the JS distance from the ground truth (Eq.~\ref{eq:js}),
and use this metric in our plots (Figures~\ref{fig:task1-freq-roi}--\ref{fig:task1-time-day-week}).
During our analysis, we also discuss the privacy loss (PL, Eq.~\ref{eq:pl}),
allowing us to better understand the effect of aggregate data publication on privacy, independently of the prior mobility pattern of the user, as PL reflects how much the adversary has learned with respect to her initial knowledge.

\beforesec
\subsubsection{Probabilistic Priors}
\aftersec

\noindent{\bf FREQ\_ROI.} In \figurename~\ref{fig:task1-freq-roi}, we plot the CDF (over the user population) of \adv's error over 
the testing week when building priors using \freqroi, i.e., each user's frequent ROIs over time, 
for both datasets, using different inference strategies.
Using the TFL dataset (\figurename~\ref{fig:task1-freq-roi-tfl}), a baseline attack where
\adv uses only her prior (blue line) has an average error of $0.37$, while \aggP (i.e., the population profile extracted from the aggregates) reduces her error to $0.34$. When \adv uses both her prior and the aggregates for the inference, the error is notably reduced, yielding
average errors amounting to $0.15, 0.25$ and $0.15$, respectively, with \bayes, \maxr and \maxu. More specifically, this corresponds to an average privacy loss of, resp., $0.6$, $0.41$, and $0.59$ for individuals whose locations are included in the aggregate time-series.

We also observe that inferences affect users in different ways, i.e., 
with \bayes, the adversarial error is reduced for all users, while with \maxr and \maxu for $77\%$ and $95\%$ of all users, respectively. This confirms that \maxr and \maxu are somewhat greedy strategies and may end up selecting users that either report few (\maxr) or many (\maxu) ROIs overall, to ``consume'' the aggregates. %

With the SFC data (cf.~\figurename~\ref{fig:task1-freq-roi-sfc}), the adversarial error only relying on cabs' frequent ROIs prior (\freqroi) is higher compared to that of TFL -- $0.65$ on average, and in this case it is quite similar to that owing to the aggregates (\aggP). It drops to $0.62$ with the Bayesian updating (corresponding to $0.06$ privacy loss) and to $0.56$ with \maxu ($0.16$ PL), indicating that taxis reporting the most locations are regular within them and end up losing more privacy. We also observe that, unlike in the TFL experiments, the greedy strategy \maxr actually deteriorates \adv's mean error ($0.71$), owing to the bias introduced by taxis visiting few ROIs (i.e., cabs having high probability to appear in a ROI). Overall, we find that profiling commuters based on their frequent ROIs is more effective than profiling cabs, as cabs report more locations and follow variable routes during their shifts.

\begin{figure}[t]
\centering
\begin{subfigure}[b]{0.44\textwidth}
\includegraphics[width=0.99\linewidth]{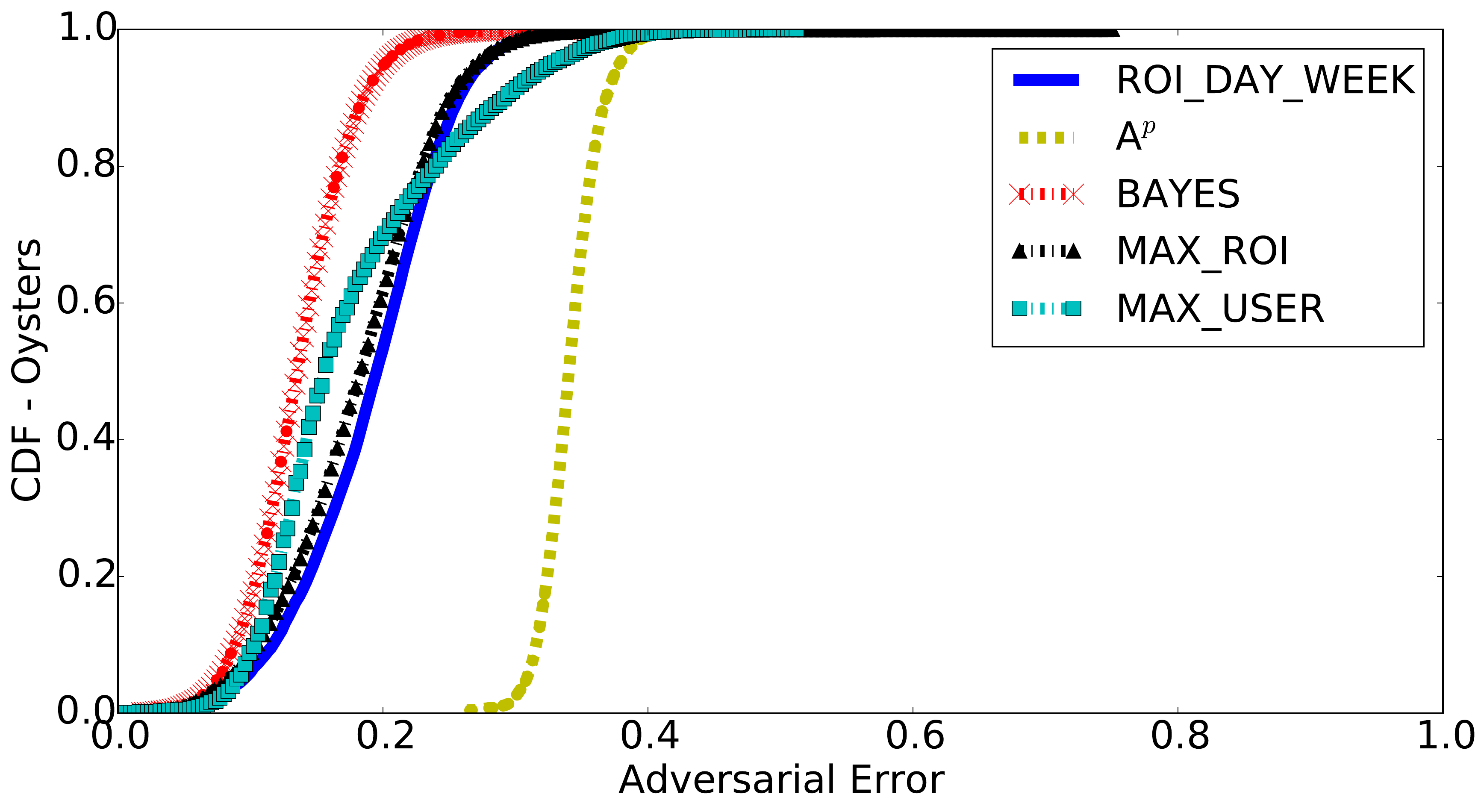}
\caption{TFL}
\label{fig:task1-roi-day-week-tfl}
\end{subfigure}
\begin{subfigure}[b]{0.44\textwidth}
\includegraphics[width=0.99\linewidth]{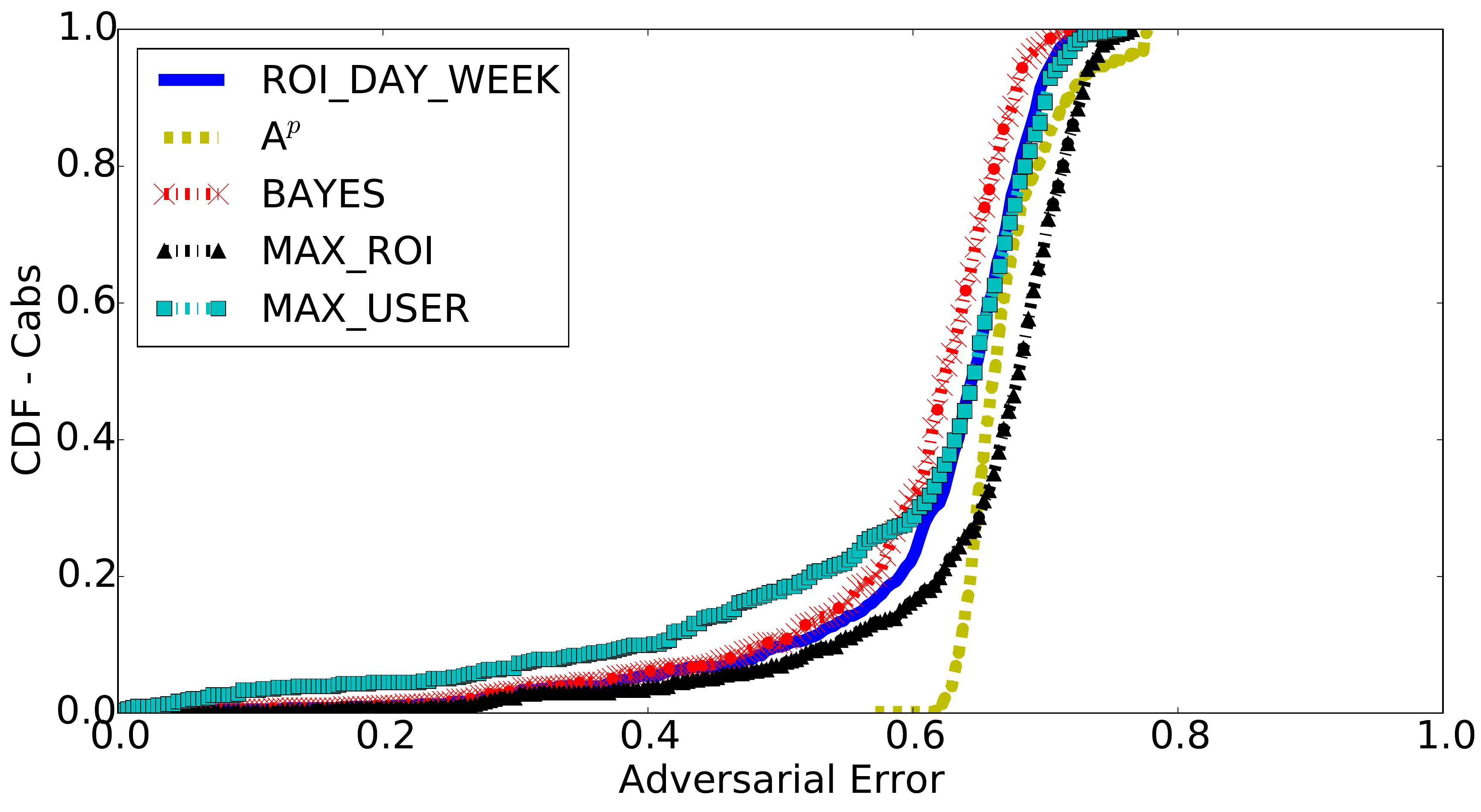}
\caption{SFC}
\label{fig:task1-roi-day-week-sfc}
\end{subfigure}
\reduxB\reduxB
\caption{\adv's Profiling Error - \roidayweek Prior.}
\label{fig:task1-roi-day-week}
\end{figure}

\descr{ROI\_DAY\_WEEK.} Next, we report \adv's error when
using the \roidayweek as her prior, i.e., a weekly profile 
that takes into account location frequency as well as time and day semantics (e.g., users' locations on Mondays, 3pm). The results are plotted in \figurename~\ref{fig:task1-roi-day-week}, for both datasets.
We have also experimented with location frequency and time only (and not day) semantics
to build the prior (\roiday), which yields larger errors, as less information is considered.
To ease presentation, we defer details to Appendix~\ref{sec:app-prof-prob}.

With the TFL data (\figurename~\ref{fig:task1-roi-day-week-tfl}), it is clear that commuters' most frequent 
ROIs for the time instances of a week (\roidayweek) are a more informative prior than their frequent 
ROIs (\freqroi), with an average prior error as low as $0.19$. This shows how time and day semantics help \adv 
profile tube commuters. \maxr and \maxu strategies slightly enhance \adv's posterior knowledge and result in, resp., $0.08$ and $0.14$ mean privacy loss. Whereas, the Bayesian inference significantly improves \adv's performance towards her 
profiling goal, yielding an average of $0.27$ privacy loss for the users. With the SFC dataset (\figurename~\ref{fig:task1-roi-day-week-sfc}), the average prior error is lower than with \freqroi as
time and day semantics enhance \adv's performance, but it still remains
relatively high ($0.61$). Two of the inference strategies reduce \adv's error, 
although not dramatically: \bayes and \maxu help \adv to profile cabs' mobility and yield, resp., $0.03$ and $0.07$ privacy loss. 
In contrast, \maxr actually deteriorates \adv's performance and does not harm the cabs' privacy. Overall, we notice that profiling cabs using their weekly profiles as prior knowledge is more challenging than profiling commuters whose mobility patterns are more regular.

\begin{figure}[t]
\centering
\begin{subfigure}[b]{0.44\textwidth}
\includegraphics[width=0.99\linewidth]{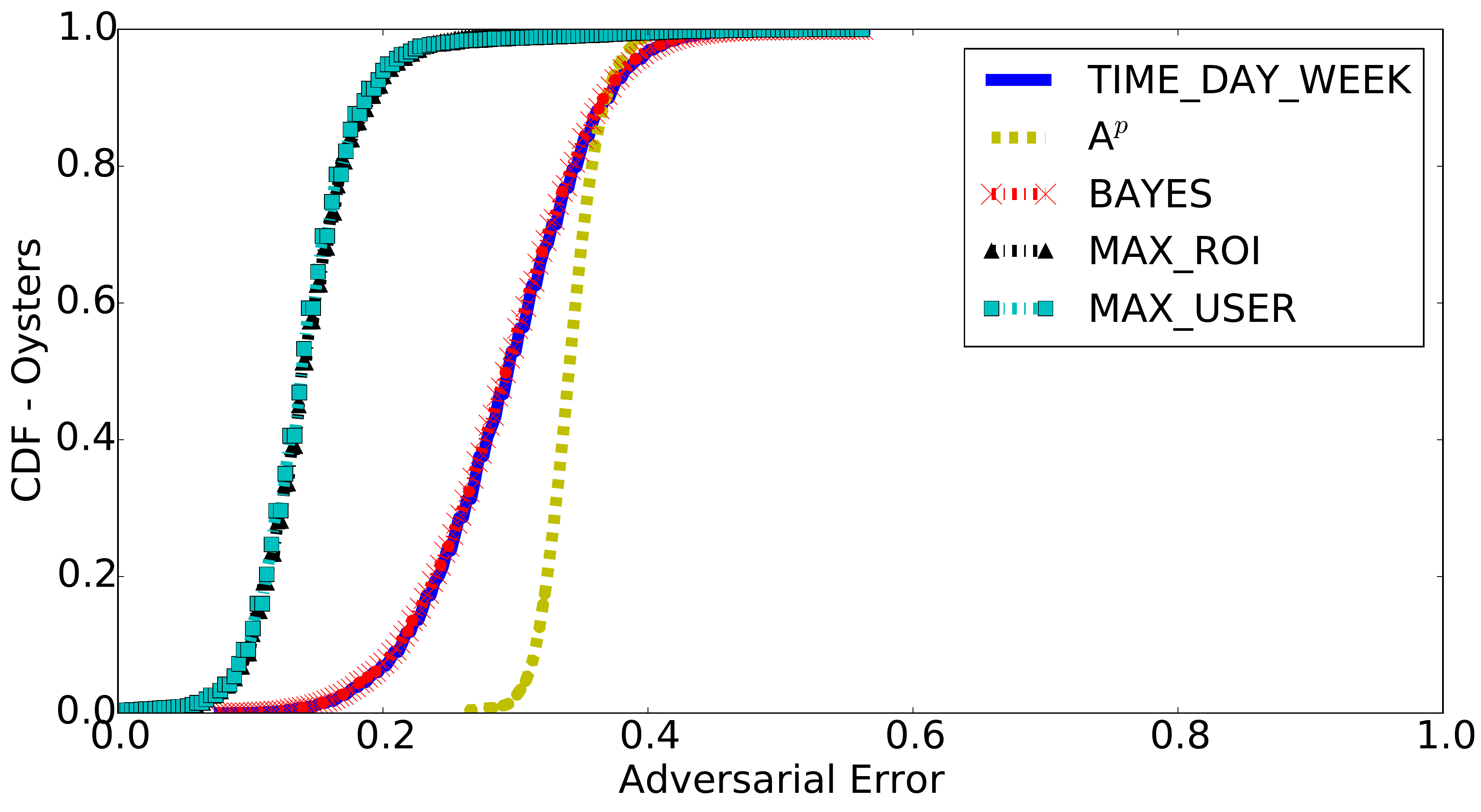}
\caption{TFL}
\label{fig:task1-time-day-week-tfl}
\end{subfigure}
\begin{subfigure}[b]{0.44\textwidth}
\includegraphics[width=0.99\linewidth]{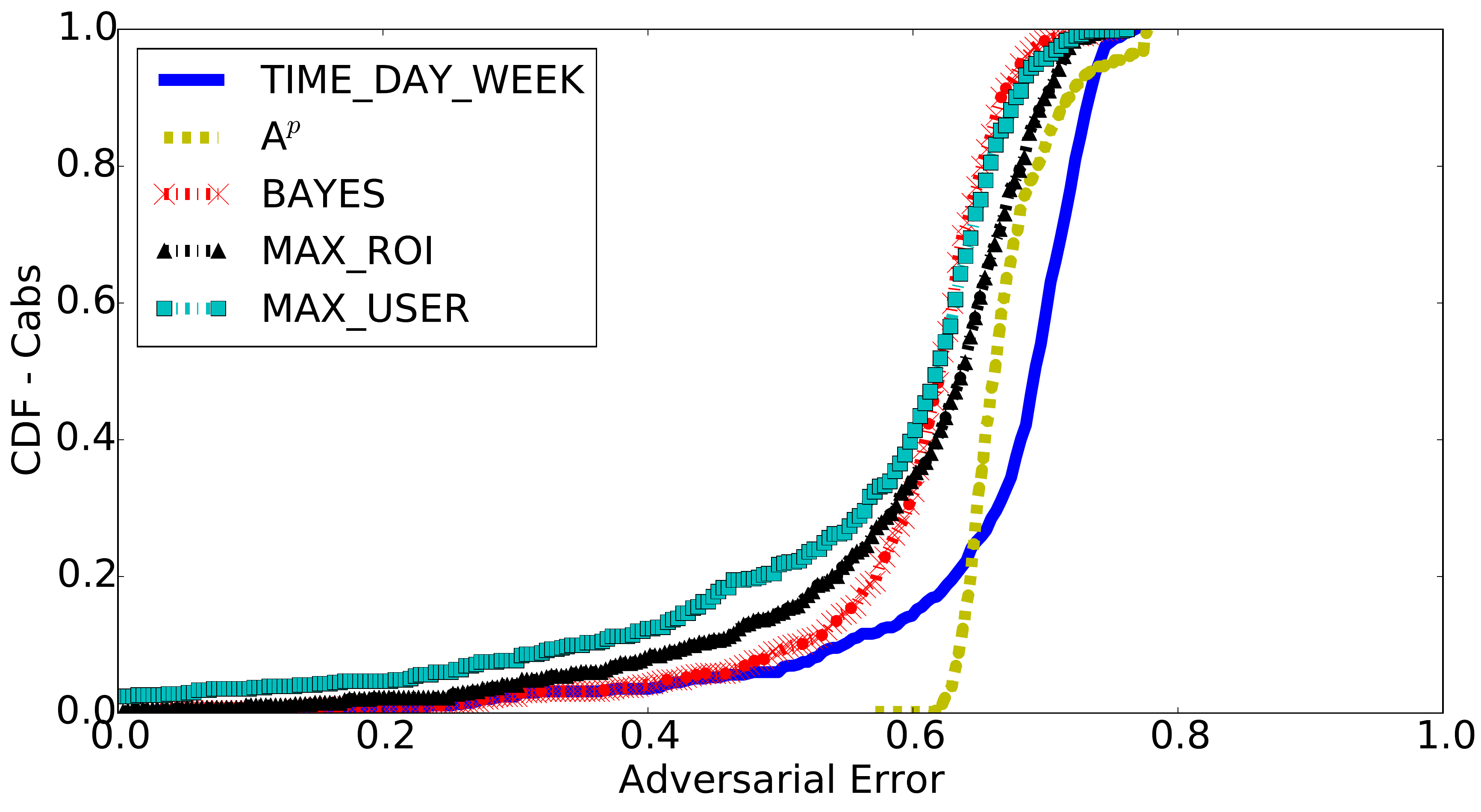}
\caption{SFC}
\label{fig:task1-time-day-week-sfc}
\end{subfigure}
\reduxB\reduxB
\caption{\adv's Profiling Error - \timedayweek Prior.}
\label{fig:task1-time-day-week}
\end{figure}

\descr{TIME\_DAY\_WEEK.} Our last experiments with probabilistic priors measure \adv's error
(see \figurename~\ref{fig:task1-time-day-week}) when her prior knowledge consists only of time information for the users, i.e., she knows which time slots of the inference week a user is likely to report ROIs, but not which ROIs. Similar experiments in which \adv knows which time slots of \textit{any} day a user reports ROIs (\timeday) result in larger error and are discussed in Appendix~\ref{sec:app-prof-prob}. With the TFL data (\figurename~\ref{fig:task1-time-day-week-tfl}), the error based on this prior is larger than with \roidayweek, namely, $0.3$. ``Greedy'' strategies (\maxr and \maxu) remarkably improve \adv's performance (i.e., they result in $0.5$ privacy loss on average), as in this case the users reporting the most ROIs are chosen to consume the aggregates (due to the prior, users have equal probability to appear in ROIs). On the other hand, the Bayesian inference only slightly decreases the adversarial error, due to the small probabilities of her prior, which consists of a uniform distribution over the tube stations for the users' most frequent time slots of a week.

With the SFC data (\figurename~\ref{fig:task1-time-day-week-sfc}), when \adv knows the cabs' most frequent time slots reporting ROIs, her prior error is larger compared to that of cabs' frequent ROIs over the time instances of a week (\roidayweek), i.e., $0.66$. However, exploiting the aggregate knowledge the error is reduced and \bayes, \maxr and \maxu inferences yield $0.09, 0.1$ and $0.2$ mean privacy loss, respectively. Overall, we point out that due to the different nature of the datasets (sparse TFL vs.~dense SFC), user profiling with time information as prior knowledge yields different amounts of privacy leakage.

\beforesec
\subsubsection{Assignment Priors}
\label{sec:exp-prof-ass-prior}
\aftersec
Next, we evaluate \adv's performance with assignment priors, i.e., when she obtains a historical location profile as her prior knowledge for the users. We experiment with \lastweek, \lastday and \lasthour, described in Section~\ref{sec:assignment-prior}. Unlike probabilistic ones, the privacy loss from aggregates with assignment priors is very small, as the sliding window on the ground truth of commuters/cabs already yields highly informative priors.
Since the CDF plots are less illustrative in this setting, we defer them to Appendix~\ref{sec:app-prof-ass} to ease presentation %
(Figures~\ref{fig:task1-ass-priors-tfl}--\ref{fig:task1-ass-priors-sfc}).

With TFL, when \adv knows users' last week's whereabouts (\lastweek), her baseline mean error is $0.17$, indicating that commuters are fairly regular in their weekly patterns. \bayes, \maxr and \maxu inferences somewhat reduce \adv's error and achieve only little privacy loss ($0.01$, $0.03$ and $0.05$ resp.). When the users' last day's ROIs are available to \adv (\lastday), her initial error is comparable to \lastweek but smaller ($0.15$ on average). \bayes and \maxr only slightly reduce the adversarial error, causing, resp., $0.02$ and $0.05$ privacy loss. On the contrary, \maxu does not harm commuters' privacy as it actually increases \adv's error, indicating the most mobile users might not follow the patterns of their previous day. \lasthour generates larger error compared to the previous ones ($0.19$), as passengers do not exhibit as strong hourly seasonality. Once again, all inferences yield negligible privacy loss. In general for TFL, we remark that seasonal historic profiles are more instructive priors than probabilistic ones (e.g., \freqroi or \timedayweek), thus, the privacy loss for individuals from the aggregate time-series is actually small compared to that of probabilistic priors.

Our experiments on the SFC data show that, unlike TFL, \lasthour is the most ``revealing'' among the assignment priors, with a mean error of $0.53$ (vs.~$0.63$ for \lastday and $0.67$ for \lastweek). Interestingly, \adv profiles cabs more efficiently knowing their last hour's ROIs than with probabilistic priors, e.g., their most frequent ROIs (\freqroi) or their most frequent ROIs for the time slots of a week (\roidayweek). That is, cabs of San Francisco are more likely to appear in those ROIs they visited during the last hour, while their daily/weekly patterns are less regular. In all assignment prior cases, \bayes and \maxu reduce \adv's error by little, while \maxr increases it, thus, the privacy loss from the aggregates is again quite low.

\beforesec
\subsubsection{Take Aways}
\aftersec

Overall, our experiments show that
aggregates do help the adversary on the profiling inference goal.
The actual degree of privacy loss for the users depends on the prior: assignment ones yield smaller privacy leakages, as they are already quite informative for the adversary compared to probabilistic ones. We also observe that inferring the mobility profiles of commuters from aggregates is significantly easier than profiling cabs. In other words, cabs' patterns are not as regular as those of tube passengers, who exhibit high seasonality. As a consequence, commuters lose much more privacy than cabs from aggregate locations.

\beforesec
\subsection{User Localization}
\aftersec

We now measure privacy loss in the context 
of \textit{localization} attacks, i.e., as \adv attempts to predict users' 
future locations. Our experimental setup is the same as with profiling. 
However, \adv's output is not a probability distribution, but a binary 
localization matrix, and \adv's main performance metric (error) is now computed 
as $1-\fone$ (see Eq.~\ref{eq:f1}).
\begin{figure}[t]
\centering
\begin{subfigure}[b]{0.45\textwidth}
\includegraphics[width=0.99\linewidth]{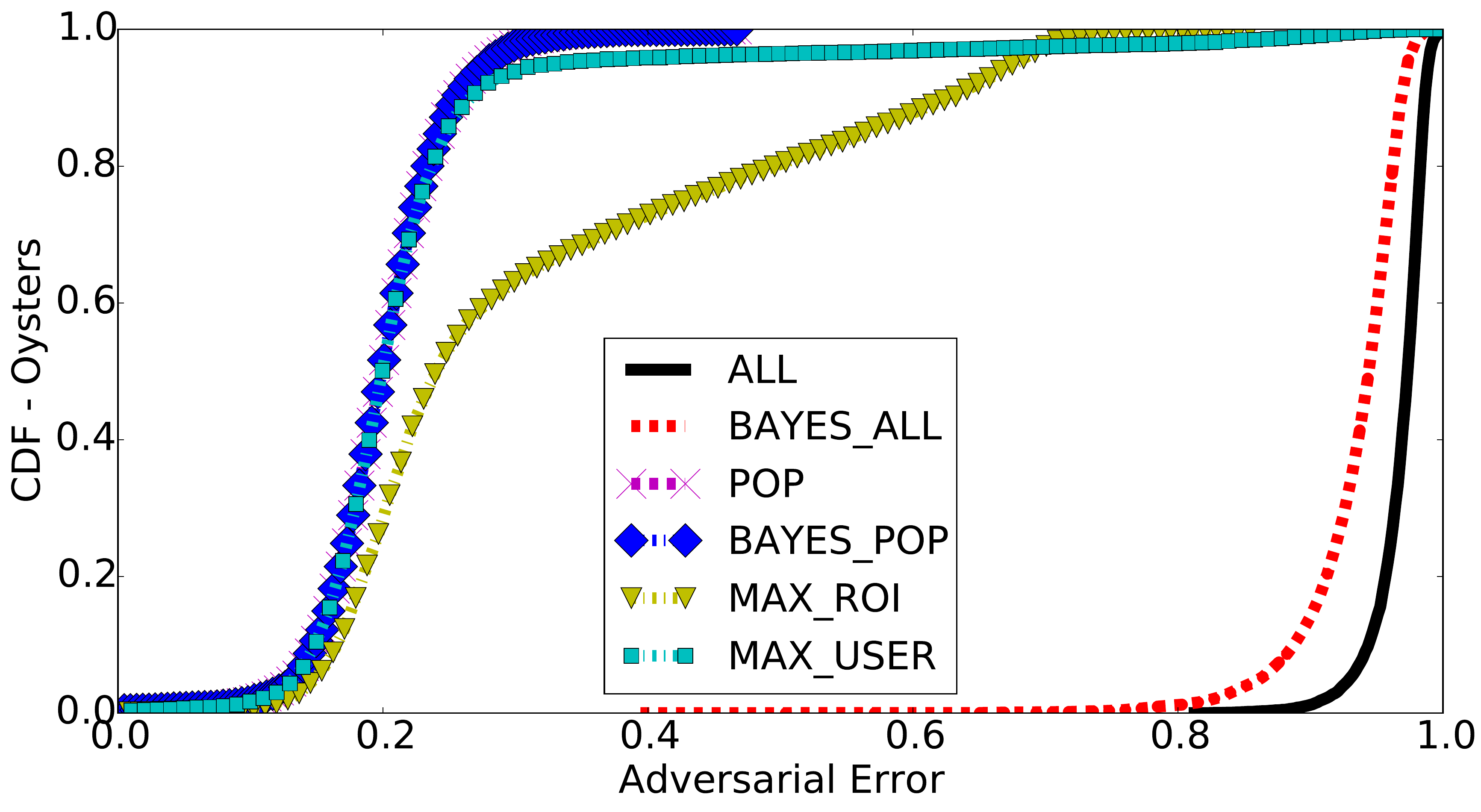}
\caption{TFL}
\label{fig:task2-freq-roi-tfl}
\end{subfigure}
\begin{subfigure}[b]{0.45\textwidth}
\centering
\includegraphics[width=0.99\linewidth]{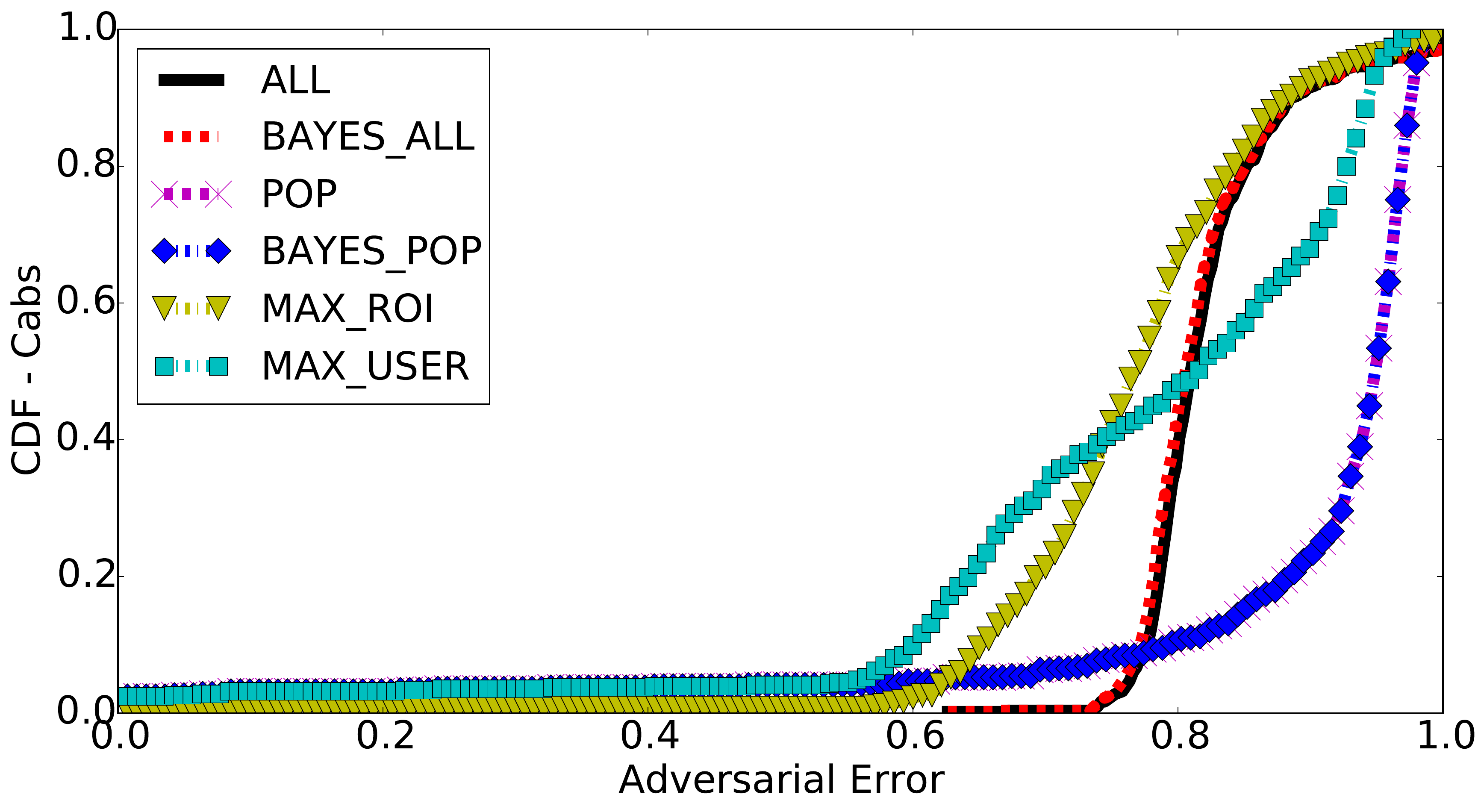}
\caption{SFC}
\label{fig:task2-freq-roi-sfc}
\end{subfigure}
\reduxB
\caption{\adv's Localization Error - \freqroi Prior.}
\label{fig:task2-freq-roi}
\vspace{0.1cm}
\end{figure}

\beforesec
\subsubsection{Probabilistic Priors}
\aftersec

\begin{figure}[t]
\centering
\begin{subfigure}[b]{0.45\textwidth}
\includegraphics[width=0.99\linewidth]{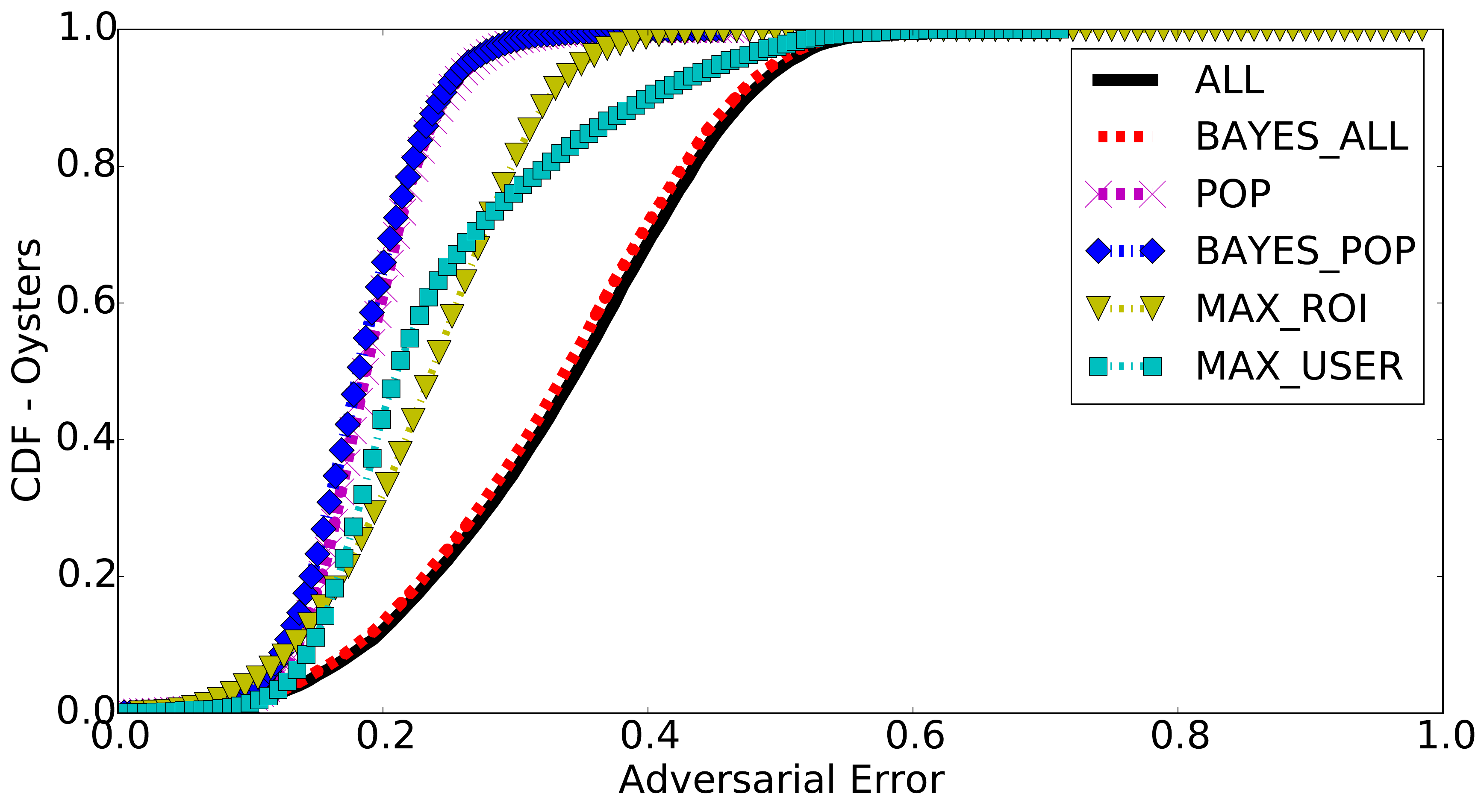}
\reduxB\reduxB\caption{TFL}
\label{fig:task2-roi-day-week-tfl}
\end{subfigure}
\begin{subfigure}[b]{0.45\textwidth}
\includegraphics[width=0.99\linewidth]{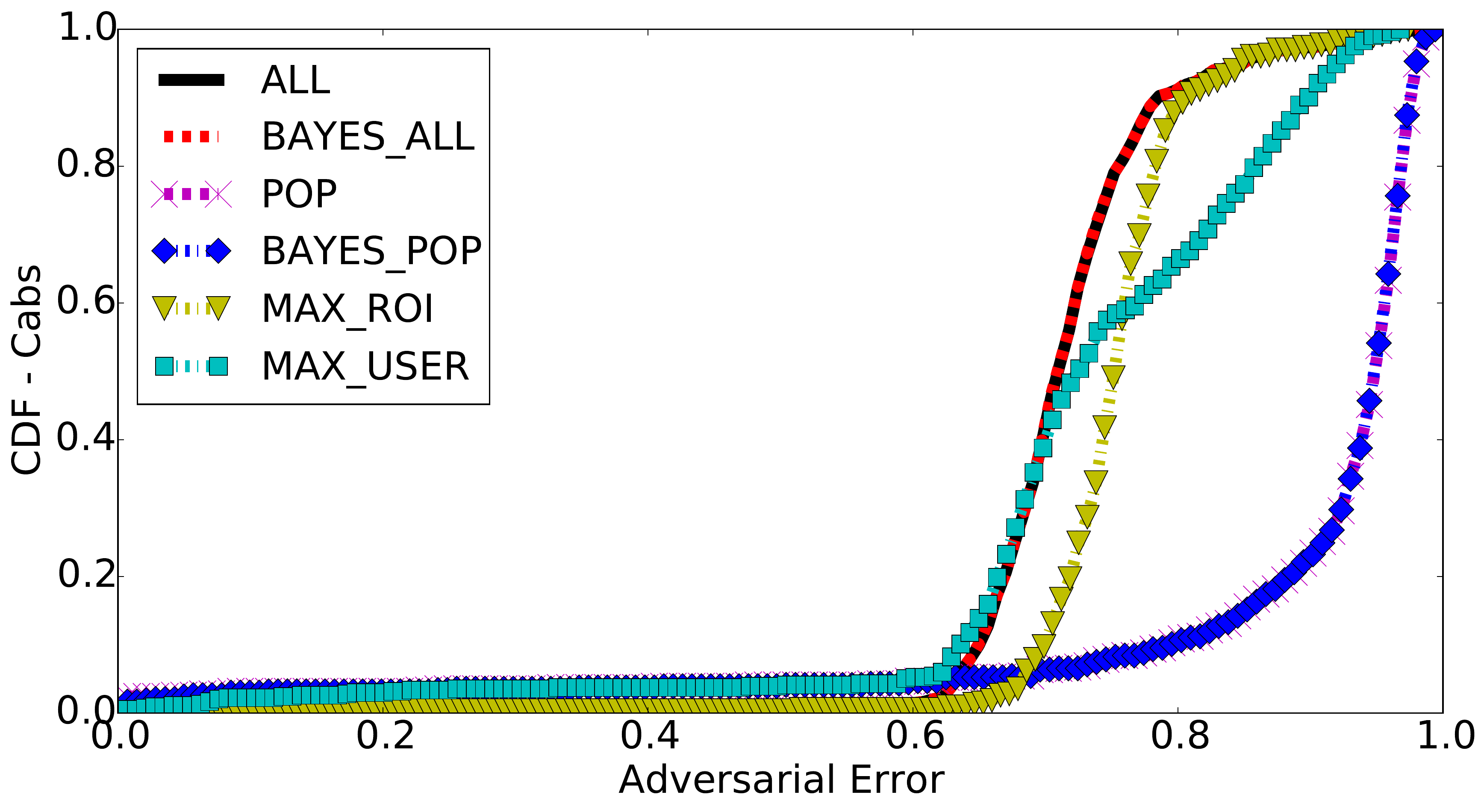}
\reduxB\reduxB\caption{SFC}
\label{fig:task2-roi-day-week-sfc}
\end{subfigure}
\reduxB\reduxB
\caption{\adv's Localization Error - \roidayweek Prior.}
\vspace{0.1cm}
\label{fig:task2-roi-day-week}
\end{figure}

We quantify \adv's error in localizing users when the prior knowledge matrix \gs 
is built according to users' most frequent ROIs over time (\freqroi) and their 
most frequent ROIs for each time slot of a week (\roidayweek). Since her prior is 
a probability distribution over ROIs for each time slot of the inference period, 
\adv's baseline prediction is to extract the users' most popular prior ROIs 
(\pop) or all prior ROIs (\all) (cf. Section~\ref{sec:assignment-prior}). For 
\pop, we set the threshold $\delta$ to $0.5$, i.e., we consider users' favorite 
ROIs those with more than $50\%$ chance of visiting. As part of her inference 
strategy, \adv (i) applies \bayes and evaluates \pop and \all on its output, and 
(ii) employs \maxr and \maxu. 
Figures~\ref{fig:task2-freq-roi}--\ref{fig:task2-roi-day-week} plot the 
corresponding results, while additional experiments with \adv knowing users' 
most frequent time slots of a week reporting ROIs are deferred to 
Appendix~\ref{sec:app-loc-prob}.

\descr{FREQ\_ROI.} \figurename~\ref{fig:task2-freq-roi-tfl} plots the CDF of \adv's error in localizing TFL passengers with their frequent ROIs over time as prior knowledge. Using only the prior, i.e., predicting that users will appear in all their frequent ROIS (\all), we get a very large average error ($0.97$); evaluating \all after applying the Bayesian inference slightly reduces the adversarial error ($0.93$) and yields very small privacy loss (on average, $0.04$).
When predicting that commuters will appear in their most popular ROIs (\pop), \adv's mean error drops to $0.21$. Again, \bayes does not improve \adv's performance, as the prior probabilities are so small that, after updating, they do not exceed $\delta$. We observe that with \pop, \adv predicts users to be \textit{out} of the transportation system during the time slots of $\Tp$. Interestingly, such a conservative strategy yields a small adversarial error overall, however, this occurs due to the fact that the TFL dataset is relatively \textit{sparse}. With the greedy inference strategies (\maxr and \maxu), \adv's mean error is much smaller than \all, respectively, $0.32$ and $0.23$. Their error patterns are different as they select different sets of users to cover the aggregates. \maxr achieves an error of $0.5$ or less for $70\%$ of the users, while \maxu for $90\%$. In both cases, \adv's error is reduced notably in comparison with the \all baseline strategy, and we find that the aggregates do indeed yield substantial privacy loss (resp., $0.66$ and $0.77$).

In \figurename~\ref{fig:task2-freq-roi-sfc}, we plot the CDF of \adv's error while attempting to localize SFC cabs over $\Tp$, again given their most frequent ROIs as prior. Similar to TFL experiments, when \adv extracts cabs' most popular prior ROIs (\pop), she predicts all of them to be \textit{outside} the network, since the prior probabilities are smaller than the threshold ($\delta = 0.5$), and the Bayesian inference updates them negligibly. However, unlike TFL, \adv's error with \pop is $0.9$ on average, proving it to be a bad strategy for localizing cabs. Predicting that cabs will show up in all their prior ROIs (\all) slightly improves her predictive power as the mean error drops to $0.83$, while \bayes negligibly reduces it further. Both \maxr and \maxu inferences improve \adv's predictions compared to the \all baseline, and they yield, resp., $0.08$ and $0.11$ privacy loss. However, we observe that \maxr behaves more consistently than \maxu (which reduces \adv's error only for $50\%$ of the cabs), indicating ROI regularity. Overall, it is clear that localization strategies behave quite differently on datasets of dissimilar characteristics.

\descr{ROI\_DAY\_WEEK.} \figurename~\ref{fig:task2-roi-day-week} displays the CDF of \adv's error localizing users with their most frequent ROIs for each time slot of a week (\roidayweek) as prior knowledge. For TFL, we notice that all prior ROIs yield a mean adversarial error of $0.34$ and Bayesian updating slightly reduces it and yields insignificant privacy loss ($0.03$). In this case, users' most popular prior ROIs (\pop) reduce \adv's error to $0.19$. The \bayes and \pop inference results in a negligible mean privacy loss ($0.06$). In contrast, compared to \all, \maxr and \maxu generate a notable privacy loss ($0.29$ and $0.26$ on average). \maxu yields larger errors for users that are selected to cover the aggregates, while the error gets smaller for those users that were not (because the aggregates were consumed). On the other hand, \maxr predicts better than \maxu for $25\%$ of users, who are highly regular in the ROIs they visit. In comparison to \freqroi, \roidayweek enables \adv to localize commuters more efficiently, proving it to be a more informative prior.

With the SFC data (\figurename~\ref{fig:task2-roi-day-week-sfc}), localizing cabs via all their prior ROIs (\all) yields a mean error of $0.71$, while \bayes reduces it insignificantly. Interestingly, with \roidayweek being an instructive prior, \all proves to be the best strategy. Extracting the cabs most popular ROIs (\pop) results in an average error of $0.9$ confirming once again that this strategy does not perform well on the dense cab data. Furthermore, \maxr and \maxu yield significant privacy loss (resp., $0.17$ and $0.18$), compared to the baseline \pop. 

Overall, our experiments demonstrate that \adv is more effective in localizing commuters/cabs with \roidayweek than \freqroi, however, the privacy loss for individuals is smaller due to the more revealing prior knowledge.

\subsubsection{Assignment Priors}

Finally, we assume \adv obtains a historical assignment prior for the users, i.e., we experiment with \lastweek, \lastday, and \lasthour priors in the context of the localization inference task. We defer the details of the corresponding results (and plots) to Appendix~\ref{sec:app-loc-ass}. We find that TFL commuters are best localized with their last week's ROIs (average error is $0.24$ with \lastweek, $0.27$ with \lastday, and $0.31$ with \lasthour), whereas, SFC cabs with their last hour's ROIs (average error is $0.73$ with \lastweek, $0.71$ with \lastday and $0.64$ with \lasthour). Moreover, as in the profiling case, the availability of aggregates yields limited privacy loss when the adversarial prior knowledge is built via assignments. This indicates that, since assignment priors are already quite instructive, the aggregates do not significantly improve \adv's knowledge of individual users' whereabouts.

\beforesec
\subsubsection{Take Aways}
\aftersec

Similar to profiling, localization inferences performed using the aggregates yield different degrees of loss in privacy for individual users, depending on \adv's prior knowledge. Assignment priors are more revealing than probabilistic ones, thus aggregates end up leaking less privacy overall. We also observe that commuters are best localized via their popular ROIs (\pop), while cabs by their most recent ROIs (\lasthour). Once again, localizing commuters is easier than localizing cabs as the former ones exhibit seasonality, while the latter ones have irregular patterns.

\begin{figure}
\centering
\begin{subfigure}[b]{0.475\textwidth}
\centering\includegraphics[width=0.9\linewidth]{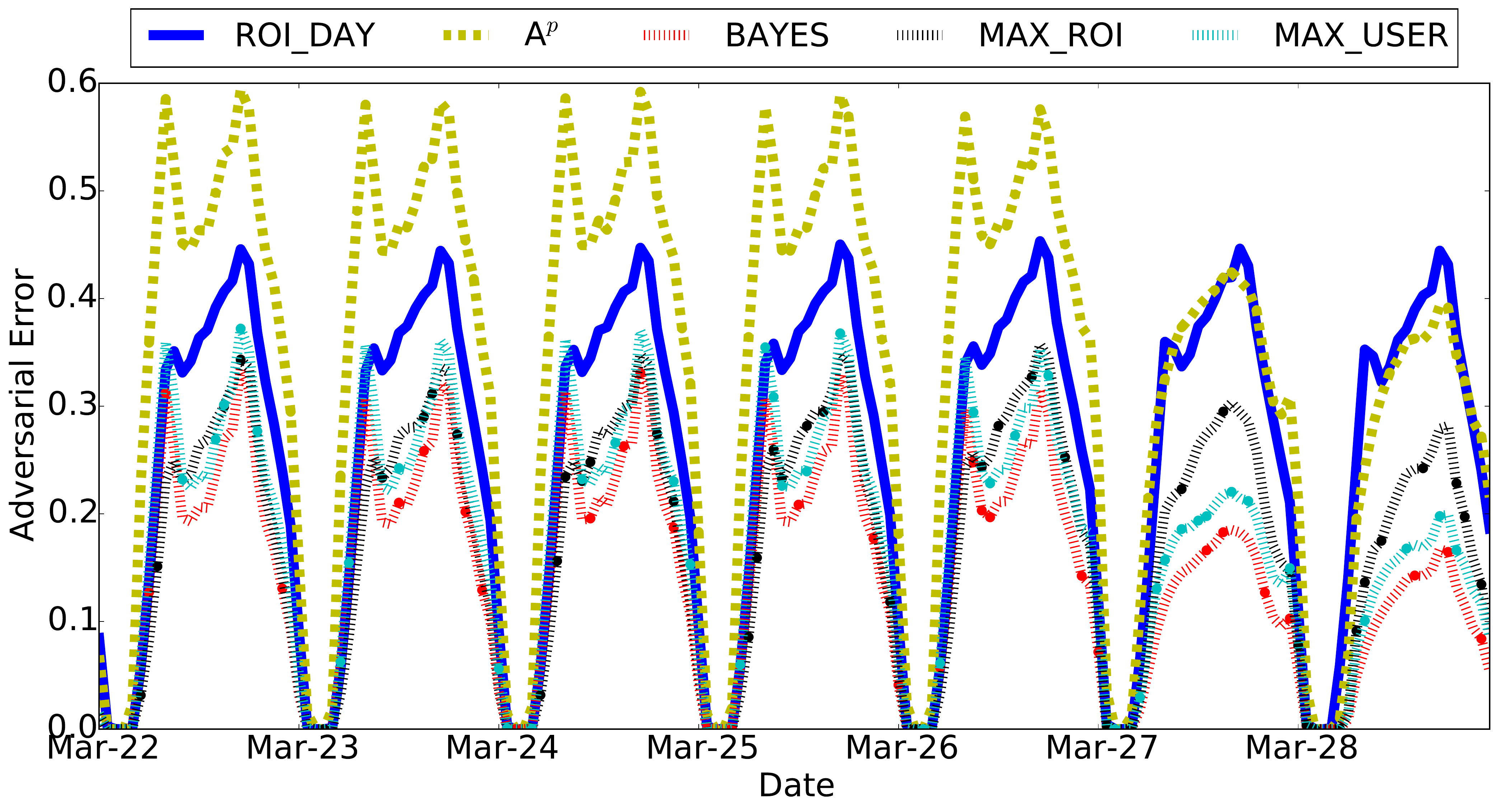}
\reduxB\reduxB\caption{TFL}
\label{fig:task1-roi-day-adv-err-time-tfl}
\end{subfigure}
\begin{subfigure}[b]{0.475\textwidth}
\centering\includegraphics[width=0.9\linewidth]{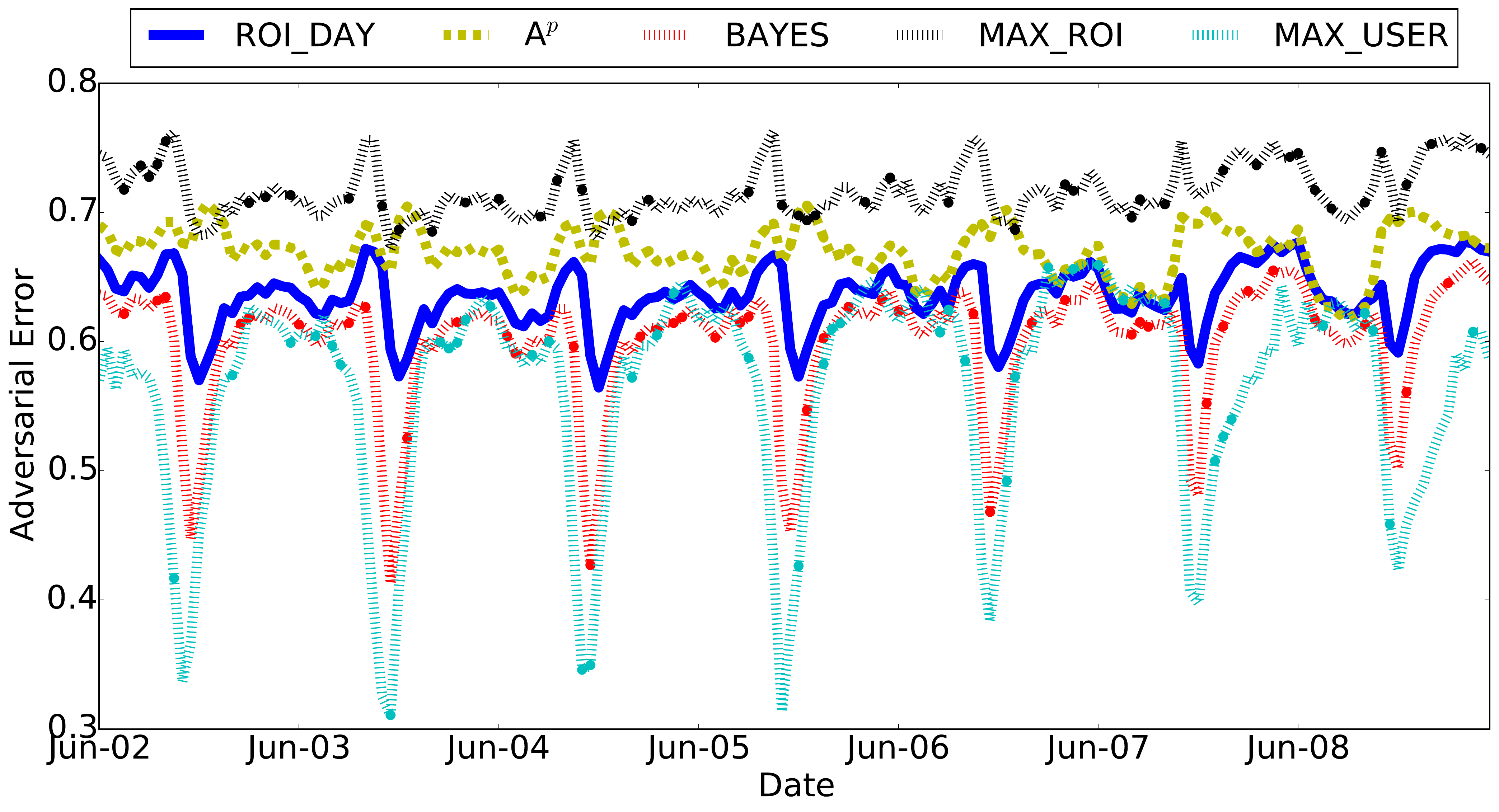}
\reduxB\reduxB\caption{SFC}
\label{fig:task1-roi-day-adv-err-time-sfc}
\end{subfigure}
\reduxB\reduxB\reduxB
\caption{\adv's Hourly Profiling Error - \roiday Prior.}
\label{fig:task1-roi-day-adv-err-time}
\end{figure}

\beforesec
\subsection{Privacy Implications of Regular Mobility Patterns}
\aftersec

Experimenting with our framework also provides some interesting considerations about \adv's error over the time slots of the inference week. 
One would expect the leakage to vary according to time of the day (e.g., peak hours vs.~night) or days of the week (e.g., weekdays vs.~weekends) since the number of users in the system, and their concentration, varies significantly. In this case, users would have variable levels of privacy protection over time. In order to validate this intuition, we pick a case-study out of our experimental setup and examine the patterns in \adv's mean error during the hourly time slots of the inference week. \figurename~\ref{fig:task1-roi-day-adv-err-time} plots the evolution of \adv's mean error over time for tube passengers (TFL) and taxis (SFC), when \adv obtains their most frequent ROIs for the time slots of day (\roiday) as prior knowledge.

For TFL (\figurename~\ref{fig:task1-roi-day-adv-err-time-tfl}), we observe different patterns w.r.t. hours of the day and weekdays, as expected. Only considering the prior (\roiday), \adv's error is smaller in the morning hours than mid-day or evening hours, likely because tube passengers are regular in their commuting routines to work, while in the evening they might go to the gym, meet friends, or go shopping before traveling back home. As the aggregate time-series is available to \adv, her error is reduced during morning hours not nearly as much as in mid-day and evening hours. In other words, commuters lose more privacy if they travel during mid-day, as there are fewer users in the transportation system, or in the evening hours, because the aggregates reflect their irregular mobility pattern. Similarly, we observe that the aggregates give \adv a much more significant advantage during the weekends than on weekdays, as  commuters more likely follow variable routes.

Likewise, for SFC (\figurename~\ref{fig:task1-roi-day-adv-err-time-sfc}), we observe distinct patterns in \adv's error w.r.t hours of the day and weekdays. With the prior (\roiday), \adv's error has a spike in the morning peak hours of weekdays indicating that cabs follow variable routes at these times and are not highly predictable. We find that \adv's prior error is smaller ($0.57$) during mid-day hours (i.e., 12pm--4pm) as cabs might be parked waiting for clients, or fewer routes might be performed during that shift. Indeed, the availability of the aggregate time-series harms cabs' privacy more during mid-day time slots as \bayes and \maxu reduce \adv's error significantly (higher privacy loss). Finally, we note that, among the inference strategies, \maxu gives \adv remarkable advantage in profiling cabs during weekends as the cabs reporting the most ROIs are likely to follow routes that are reflected by the aggregates.

\beforesec
\section{Privacy Evaluation of Defense Mechanisms}
\label{sec:counter}
\aftersec

In the previous section, we have shown that aggregate location time-series leak information about individuals' whereabouts, and have evaluated how, based on different priors and inferences. Next, we study whether mechanisms supporting the release of aggregate information in a privacy-respecting manner are effective at avoiding such privacy leakage, and to what extent. Specifically, we focus on the protection offered by Differential Privacy (DP)~\cite{dwork2008differential}, using either output or input perturbation techniques. The former add noise to the output of the aggregation process, whereas, with the latter, noise is added to users' inputs before aggregation. We do not consider other defense mechanisms, e.g., based on k-anonymity, as they have already been shown to be ineffective~\cite{shokri2010unraveling}.

In theory, one can assess the level of privacy provided by DP mechanisms as it is
configured by the parameter $\epsilon$, which determines the privacy 
risk incurred when releasing statistics computed on sensitive data (providing 
an upper bound). While $\epsilon$ expresses the relation between the level of privacy 
before and after the release, and provides protection against arbitrary risks, it
is not an absolute measure of privacy and it is often not clear how to 
interpret, in practice, the actual level of privacy enjoyed by individuals in the dataset, 
nor is how to choose the value of $\epsilon$ to obtain the desired protection.

In the rest of this section, we use our framework to measure to which extent DP mechanisms reduce the privacy leakage compared to the release of raw aggregates, vis-\`a-vis the resulting utility of the data. That is, we quantify the protection that these mechanisms provide to users in presence of an adversary that, as in Section~\ref{sec:experiments}, has access to the aggregates (now perturbed via a DP mechanism) and uses that information to improve her prior knowledge about users' whereabouts.

\beforesec
\subsection{Metrics} 
\aftersec

\noindent{\bf Privacy Gain.} We quantify the protection
provided by DP techniques in terms of the ``privacy gain'' they yield, which we define 
to denote the difference in \adv's error when using her prior (\gs) with the noisy aggregates $\pagg$ ($\adverr_{\gs, \pagg}$) minus that with the raw aggregates $\agg$ ($\adverr_{\gs, \agg}$), normalized by the maximum gain the mechanism can provide. That is, we measure privacy gain (PG) as: \reduxB
\begin{equation}\label{eq:pg}
{\sf PG}=\begin{cases}
\frac{\adverr_{\gs, \pagg} - \adverr_{\gs, \agg} }{1 - \adverr_{\gs, \agg}} ~ \text{if} \ \adverr_{\gs, \agg} \neq 1~\wedge\\[-1.5ex]
	~~~~~~~~~~~~~~~~~~~~~~~~~~~~~~~~~~~~~~~~~~~ \adverr_{\gs, \pagg}>\adverr_{\gs, \agg}\\[0.5ex]
0 ~~~~~~~~~~~~~~~~~~~~~~~~~~~~~~~~\text{otherwise}\reduxA
\end{cases}
\end{equation}
PG is a value between 0 and 1 capturing Adv's deterioration
towards her goal (e.g., profiling users) owing to the noise added by the DP techniques. 

\descr{Mean Relative Error (MRE).} We also use the MRE to measure utility, specifically, to capture the error between an original time-series ${\sf Y}$ of ${\sf n}$ time points and its \textit{noisy} version ${\sf Y'}$, which comes as the result of perturbation. More precisely:\reduxB\reduxB
\begin{equation}\label{eq:MRE}
{\sf MRE(Y, Y') = (1/n) \sum_{i=0}^{n} \frac{| Y'_{i} - Y_{i}|}{max(\beta, Y_i)}}\reduxA\reduxA
\end{equation}
where ${\sf \beta}$ is a sanity bound mitigating the effects of very small counts. As done in previous work~\cite{acs2014case}, we use MRE to measure the utility loss when a privacy mechanism is applied to an aggregate time-series, and we adjust ${\sf \beta}$ to $0.1\%$ of ${\sf \sum_{i=0}^{n} Y_{i}}$.

\beforesec
\subsection{Output Perturbation}
\label{sec:counterOutput} 
\aftersec

We first evaluate differentially private mechanisms based on output perturbation, 
in which an entity adds noise to the 
statistics prior to their release. This entity can be trusted with the 
individual users' data~\cite{acs2014case,fan2012real} or only be allowed
to compute aggregate statistics, e.g., using cryptographic protocols
for private aggregation~\cite{popa2011privacy,brown2013haze,pyrgelis2016privacy}. 
We evaluate two specific approaches: the Simple Counter 
Mechanism~\cite{chan2011private,dwork2010differential} and the Fourier 
Perturbation Algorithm~\cite{rastogi2010differentially}.

\descr{Simple Counter Mechanism (SCM)~\cite{chan2011private,dwork2010differential}.} 
SCM is a straightforward extension of the Laplace mechanism proposed by Dwork et 
al.~\cite{dwork2008differential} for time-series. It answers a
new query at each time slot (e.g., how many users are in a ROI at that time) 
and randomizes the answer with fresh independent noise. Given $\epsilon$, for a ROI 
$\s \in \Ss$, for each time slot $\ti' \in \Tp$, SCM samples a fresh random value from the 
Laplace distribution ${\sf \gamma_{t'} \sim Lap( 1 / \epsilon)}$ (recall that each user is counted 
at most \emph{once} in $\agg_{\s\ti'}$) and releases the perturbed 
aggregate $\pagg_{\s\ti'} = \agg_{\s\ti'} + {\sf \gamma_{t'}}$, where $\agg_{\s\ti'}$ is the true aggregate value.

Due to the composition theorem~\cite{dwork2010differential}, and given that the 
number of locations and time slots in the inference period for which data is released, 
are $| \Ss |$ and $| \Tp |$ respectively, the mechanism is overall 
${\sf O( |\Ss| \cdot | \Tp | \cdot \epsilon)}$ differentially private. 
Thus, the privacy leakage increases linearly with the number of ROIs and the length 
of the inference period. This version of SCM only guarantees \textit{event-level} 
privacy~\cite{chan2011private,dwork2010differential} for the users, i.e., it protects 
whether or not a user was in a ROI at a specific time slot. If one desires to achieve stronger 
privacy guarantees with SCM, then the noise can be distributed according to 
${\sf Lap( | \Tp | / \epsilon)}$ (i.e., users are protected within the 
aggregates of a region, during the whole period $\Tp$) and SCM becomes ${\sf O( |\Ss| \cdot \epsilon)}$ differentially private. 
Alternatively, to guarantee $\epsilon$-DP (i.e., users are protected within the aggregates of all regions, 
during $\Tp$) the noise must be distributed according to ${\sf Lap(| \Ss | \cdot | \Tp | / \epsilon )}$, 
increasing privacy at the cost of utility.

\descr{Fourier Perturbation Algorithm 
(FPA)~\cite{rastogi2010differentially}.} FPA improves the privacy/utility 
trade-off offered by SCM by reducing the amount of noise needed 
to obtain the same level of privacy. This reduction is based on performing the noise 
addition in the compressed domain as follows. First the time-series is 
compressed using the Discrete Fourier Transform (DFT) and the first 
${\sf k}$ Fourier coefficients, ${\sf F_k}$, are kept. Then, ${\sf F_k}$ is perturbed with noise 
distributed according to ${\sf Lap(\sqrt{k  \cdot | \Tp |} / \epsilon)}$, and 
padded with zeros to the size of the original time-series. Finally, the 
inverse DFT is applied to obtain the perturbed aggregates to be released. This version of FPA 
guarantees $\epsilon$-DP for each ROI (thus, overall it's ${\sf O( |\Ss| \cdot \epsilon)}$ differentially private) 
with better utility than SCM. Note that a mechanism similar to FPA has also been applied in~\cite{acs2014case}.
\descr{Evaluation.}
We present the results of our evaluation on two case-studies: (i) user profiling on the TFL 
dataset with \adv obtaining \freqroi as her prior knowledge and following the greedy \maxr 
strategy, and (ii) user profiling on the SFC data when \adv knows \freqroi and employs \maxu. 
Although we restrict to two cases, %
their choice is reasonable
as our analysis in Section~\ref{sec:experiments} shows that, in these settings, 
the aggregates yield significant privacy loss for individual users.

We parameterize SCM and FPA perturbation mechanisms with $\epsilon \in \{ 0.001, 0.01, 0.1, 1.0 \}$. 
For SCM, we experiment with variable magnitude of Laplacian noise to demonstrate the actual protection 
it offers with respect to its theoritical privacy guarantees. Since SCM with ${\sf Lap( |\Ss| \cdot | \Tp | / \epsilon)}$ is expected to yield unnecessarily huge error in the aggregates (i.e., it is practically impossible for commuters/cabs to appear in \emph{all} ROIs in \emph{every} time slot of the inference period), we also report SCM with noise distributed according to ${\sf Lap(\Delta / \epsilon)}$, where ${\sf \Delta}$ is the sensitivity of users within the aggregates $\agg$, i.e., the maximum number of location reports by a user/cab during $\Tp$ in the TFL and SFC datasets (224 and 2,687 resp.). Furthermore, for FPA, as done in~\cite{rastogi2010differentially}, we experiment with the parameter ${\sf k}$ to minimize its total error, finding that 
${\sf k}=25$ yields the best results on TFL and ${\sf k}=20$ on the SFC data.

\descr{Utility.} Tables~\ref{table:utility-tfl} 
and~\ref{table:utility-sfc} report the utility for both datasets, in 
terms of MRE, of the mechanisms for different values of $\epsilon$. 
Overall, as expected, for all mechanisms the higher the privacy (i.e., lower $\epsilon$ values), the lower the utility (i.e., bigger MRE). 

In our first case study (Table~\ref{table:utility-tfl}), SCM-${\sf Lap( |\Ss | \cdot |\Tp| / \epsilon)}$ 
yields the worse utility, with perturbed 
aggregates being about 700 times worse estimates than raw ones,
for all values of $\epsilon$. Moreover, SCM-${\sf Lap( \Delta / \epsilon)}$ and SCM-${\sf Lap( | \Tp | / \epsilon)}$ still 
yield very high errors, even for a mild level of privacy ($\epsilon=0.01$). The highest utility is 
provided by SCM-${\sf Lap(1/ \epsilon)}$, followed by FPA. Nonetheless, with the former,
the utility is at least 8 times worse than the raw aggregates (MRE$=$7.8) for small
$\epsilon$ values ($0.01$ or less). %
In our second case study (Table~\ref{table:utility-sfc}), we observe that SCM-${\sf Lap( |\Ss | \cdot |\Tp| / \epsilon)}$ 
and SCM-${\sf Lap( \Delta / \epsilon)}$ result in very large errors (MRE$\geq 24$), while SCM-${\sf Lap( | \Tp | / \epsilon)}$ 
follows closely. FPA and SCM-${\sf Lap(1/ \epsilon)}$ yield the best utility, although for 
sensible levels of (expected) privacy (i.e., $\epsilon=0.01$) the perturbed aggregates are 
about 8 and 5 times worse estimates than the raw ones, respectively.

\begin{table}[t]
\centering
\small
  \begin{tabular}{  l   r r r r }
 \toprule
 \hfill $\mathbf{\epsilon}$ & {\bf 0.001} & {\bf 0.01} &  {\bf 0.1} & {\bf 1.0} \\ 
	\midrule
	SCM - ${\sf  Lap( | \Ss | \cdot | \Tp | / \epsilon ) }$ & 739.9 & 743.2 & 735.8 & 709.4 \\
	SCM - ${\sf  Lap( \Delta / \epsilon )}$ & 720.1 & 605.1& 168.9 & 16.7 \\
	SCM - ${\sf  Lap( | \Tp | / \epsilon)}$ & 719.8 & 549.6 & 123.5 & 12.8  \\    
    FPA & 117.1 & 11.7 & 1.3 & 0.3 \\ 
    SCM - ${\sf  Lap(1 / \epsilon)}$ & 74.4 & 7.8 & 0.9 & 0.1 \\
    \bottomrule
	\end{tabular}
	\reduxB	\reduxB
	\caption{TFL: MRE (Utility) of output perturbation mechanisms.}
	  \label{table:utility-tfl}
	  \vspace{0.3cm}
  \end{table}

\begin{table}[t]
\centering
\small
  \begin{tabular}{  l  r r r r }
 \toprule
 \hfill $\mathbf{\epsilon}$ & {\bf 0.001} & {\bf 0.01} &  {\bf 0.1} & {\bf 1.0} \\ 
	\midrule
	SCM - ${\sf  Lap( | \Ss | \cdot | \Tp | / \epsilon ) }$ & 26.8 & 26.3 & 26.2 & 26.2 \\
	SCM - ${\sf  Lap( \Delta / \epsilon )}$ & 26.9 & 26.5& 25.9 & 24.3 \\
	SCM - ${\sf  Lap( | \Tp | / \epsilon)}$ & 26.1 & 25.9 & 22.4 & 8.3  \\    
    FPA & 24.1 & 8.8 & 1.1 & 0.3 \\ 
    SCM - ${\sf  Lap(1 / \epsilon)}$ & 19.9 & 5.1 & 0.6 & 0.1 \\
    \bottomrule
	\end{tabular}
		\reduxB	\reduxB
	\caption{SFC: MRE (Utility) of output perturbation mechanisms.}
	  \label{table:utility-sfc}
	  \vspace{0.1cm}
  \end{table}

\begin{figure}[t]
\includegraphics[width=0.99\linewidth]{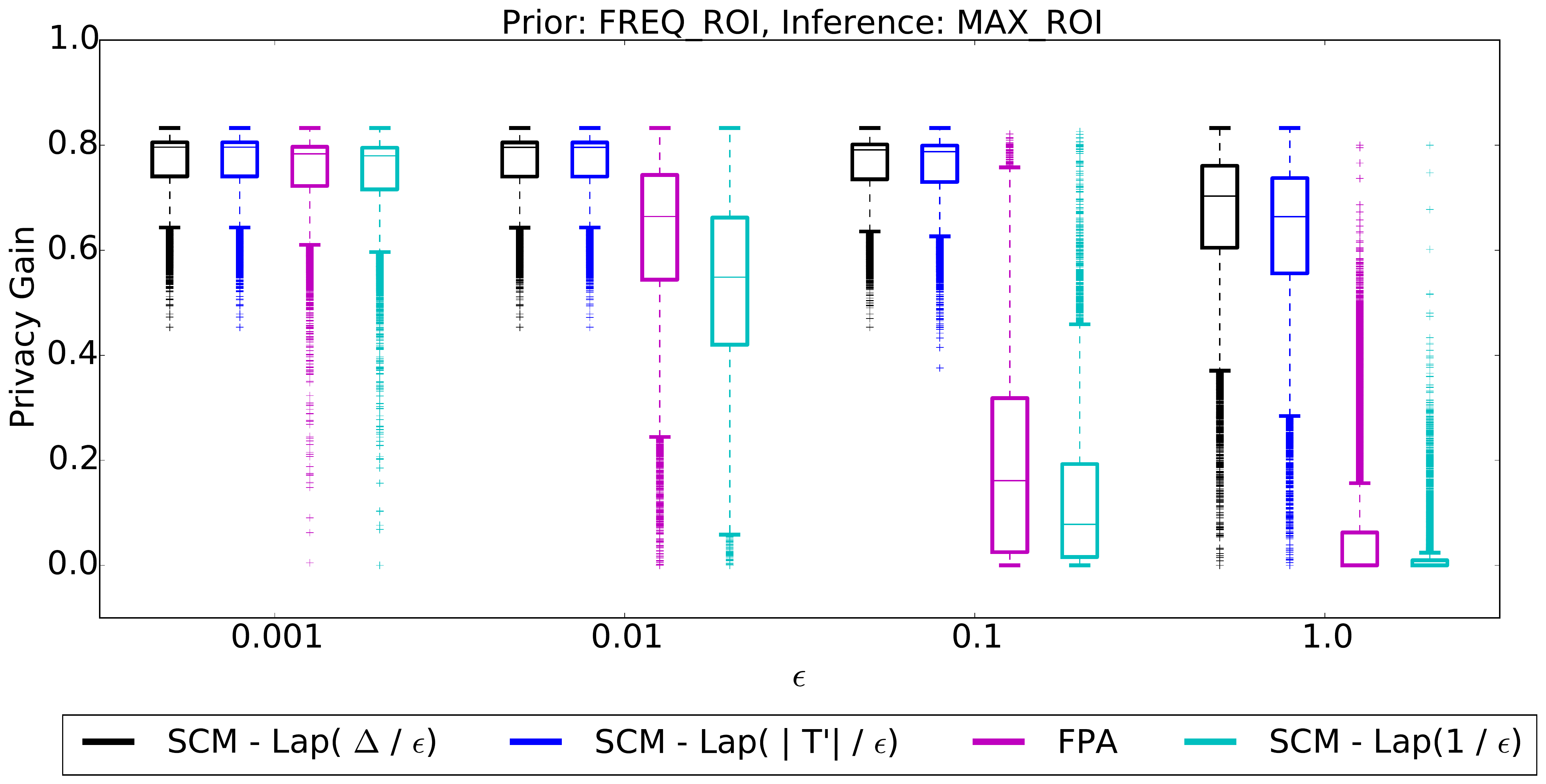}
\reduxB\reduxB
\caption{TFL: Privacy gain for output perturbation DP mechanisms.} %
\label{fig:task1-freq-roi-privacy-tfl}
\end{figure}

\begin{figure}[t]
\includegraphics[width=0.99\linewidth]{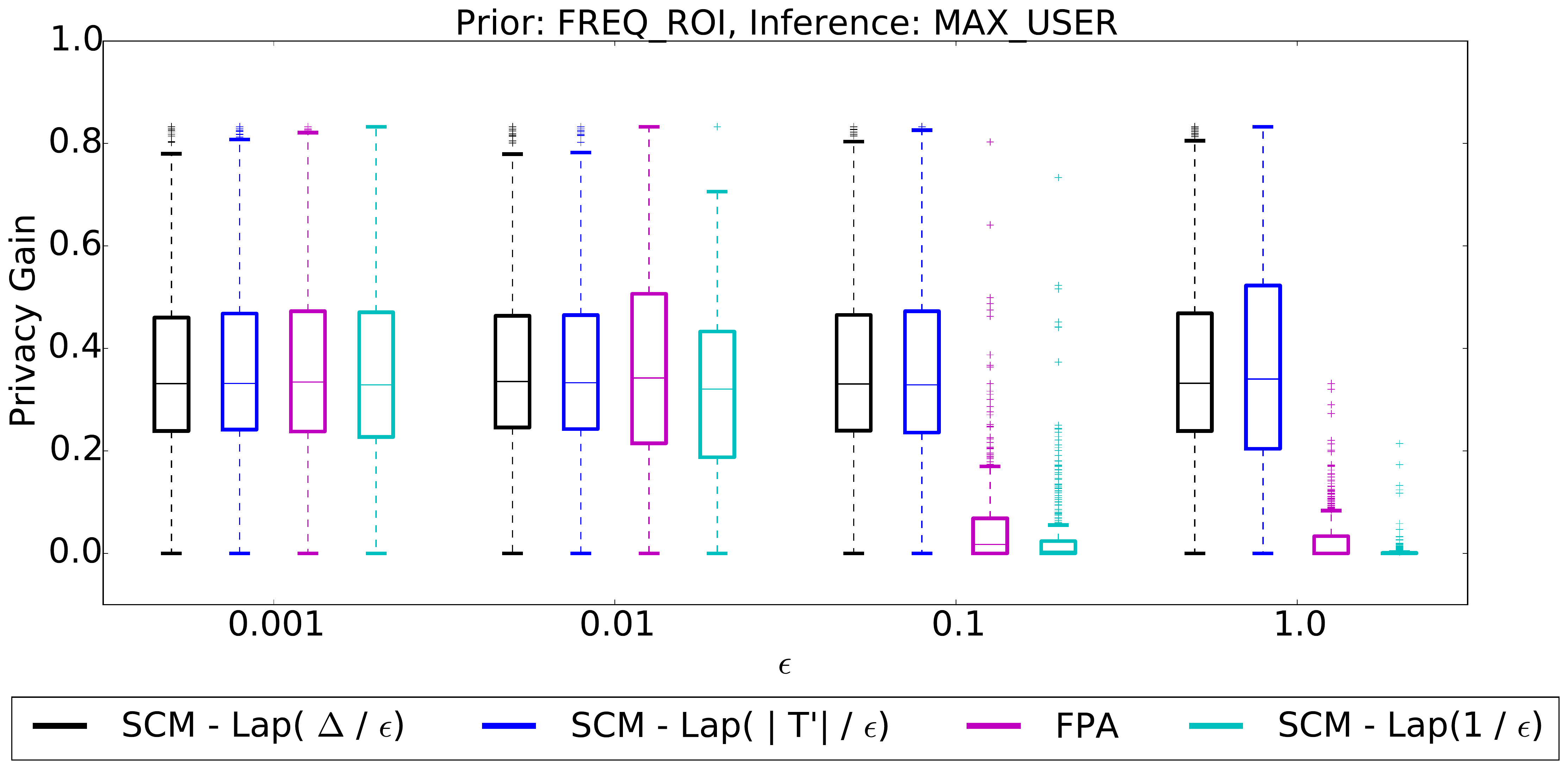}
\reduxB\reduxB
\caption{SFC: Privacy gain for output perturbation DP mechanisms.} %
\label{fig:task1-freq-roi-privacy-sfc}
\end{figure}

\descr{Privacy Quantification.} Figs.~\ref{fig:task1-freq-roi-privacy-tfl} and~\ref{fig:task1-freq-roi-privacy-sfc} 
display box-plots of the privacy gain (PG, see Eq.~\ref{eq:pg}) enjoyed by individual users in both datasets 
thanks to the perturbation mechanisms, for increasing values of $\epsilon$. To ease presentation, the plots 
do not include SCM-${\sf Lap( |\Ss | \cdot |\Tp| / \epsilon)}$, which, as discussed earlier, completely destroys utility.
In the TFL case study (\figurename~\ref{fig:task1-freq-roi-privacy-tfl}), the four mechanisms exhibit very different behaviors. SCM-${\sf Lap(\Delta / \epsilon)}$ and 
SCM-${\sf Lap(| \Tp | / \epsilon)}$ offer the best privacy protection, with an average privacy gain as 
high as $0.77$ for $\epsilon\leq 0.1$, and $0.65$ and $0.62$, resp., for $\epsilon=1.0$. However, as 
discussed above (and shown in Table~\ref{table:utility-tfl}), this protection comes with very poor utility. 
We also find that SCM-${\sf Lap( 1 / \epsilon)}$ and FPA offer similar protection 
(PG=$0.74$ on average) for $\epsilon = 0.001$, while, as $\epsilon$ grows, the  
gain drops significantly, being negligible when $\epsilon=1.0$. While this is somewhat expected for SCM-${\sf Lap( 1 / \epsilon)}$, it is quite surprising for FPA, which in theory should provide as much protection as SCM-${\sf Lap(| \Tp | / \epsilon)}$.

In the SFC case (\figurename~\ref{fig:task1-freq-roi-privacy-sfc}), we observe that 
SCM-${\sf Lap(\Delta / \epsilon)}$ and SCM-${\sf Lap(| \Tp | / \epsilon)}$ provide the best
privacy gain ($0.36$ on avg.) for all values of $\epsilon$. FPA and SCM-${\sf Lap( 1 / \epsilon)}$
behave similarly to the previous two for $\epsilon \leq 0.01$, however, as $\epsilon$ increases the 
privacy gain approaches zero.

\descr{Remarks.} Our evaluation not only highlights a possible gap between theory and practice w.r.t.
privacy guarantees offered by DP mechanisms, but also
shows that these struggle to offer strong privacy under continual observation 
(as in the case of aggregate location time-series) without destroying 
utility. For instance, FPA with $\epsilon=0.01$ provides reasonably high gain in privacy (PG=$0.62$) for TFL commuters, however, the MRE of the published aggregates is approximately 11. For instance, if there are 100 people in an underground station, the system will report that there are instead 1,200. Similarly,
on the SFC dataset, when FPA provides good level of privacy for cabs (i.e., PG $=0.36$ 
with $\epsilon=0.01$), the MRE is almost 9.

\beforesec
\subsection{Input Perturbation}
\label{sec:counterInput} 
\aftersec

We now look at input perturbation-based DP techniques, whereby users add noise 
to their inputs  prior to the aggregation process. In particular, we focus on Randomized 
Response (RR)~\cite{erlingsson2014rappor,quercia2011spotme,warner1965randomized}.
We do not consider geo-indistinguishability~\cite{andres2013geo}, 
a mechanism to provide individual users with differential privacy guarantees
while using location-based services, since there is no scheme that uses such 
approach to collect or release aggregate locations. (In fact, we consider this as an
interesting open problem for future work.)

\descr{Randomized Response (RR)}  can be used to privately collect statistics 
from users participating in surveys~\cite{warner1965randomized}, crowdsourcing 
statistics from client software~\cite{erlingsson2014rappor}, sharing  
historical traffic data~\cite{googleTraffic}, as well as privately aggregating 
user locations in real-time~\cite{quercia2011spotme}. In particular, the SpotMe 
system~\cite{quercia2011spotme} lets users perturb their location at each 
time instance $\ti' \in \Tp$ by claiming to be in a ROI $\s \in \Ss$ (a ``yes'' response) with some 
probability ${\sf p}$, or report the truth (i.e., whether they are or not in location $\s$)
with probability ${\sf 1 - p}$.
The aggregator collects the perturbed user inputs and computes the aggregation 
estimating the number of individuals in each location $\s \in \Ss$ and every time slot $\ti' \in \Tp$, 
via ${\sf \agg_{\s,\ti'} = total_{\s,\ti'} \cdot \frac{Pyes_{\s,\ti'} - p}{1-p}}$, where ${\sf total_{s,t'}}$ is the total number 
of responses received for ROI $\s$ at time $\ti'$ and ${\sf Pyes_{\s,\ti'} = \frac{yes_{\s,\ti'}}{total_{\s,\ti'}} }$ depicts the proportion of ``yes'' responses. This mechanism is ${\sf \ln \frac{ | \Ss | - ( | \Ss | - 1) \cdot p}{p}}$-DP at each 
time slot~\cite{waseda2016analyzing}, thus, overall it guarantees ${\sf O( |\Tp | \cdot \ln \frac{ | \Ss | - ( | \Ss | - 1) \cdot p}{p})}$ differential privacy due to the composition theorem.

\descr{Evaluation.} We evaluate SpotMe~\cite{quercia2011spotme}, as a representative for RR input perturbation mechanisms, using our framework. In 
this context, \adv is assumed to obtain the estimated perturbed aggregates $\pagg$ that
result as users apply the RR mechanism on their inputs.
As in the output perturbation case, we focus on two user profiling case-studies: (i) TFL data with 
\freqroi adversarial prior knowledge and \maxr inference, and (ii) SFC dataset with \freqroi prior and
\maxu strategy.

\descr{Utility.} Table~\ref{table:spotme-utility} shows the MRE of the perturbed 
aggregates, highlighting that, as ${\sf p}$ grows (i.e., as commuters/cabs perturb 
their inputs with higher probability) the utility of the aggregates 
declines. For TFL, with ${\sf p}=0.1$, the MRE over all stations is $2.1$,
and $17.6$ with ${\sf p}=0.9$.
For SFC, the MRE over all ROIs is $0.4$ for ${\sf p}=0.1$, while for ${\sf p}=0.9$ the perturbed 
aggregates are approximately $3$ times worse than the raw ones. 

\begin{figure}[t]
\centering
\includegraphics[width=0.9\linewidth]{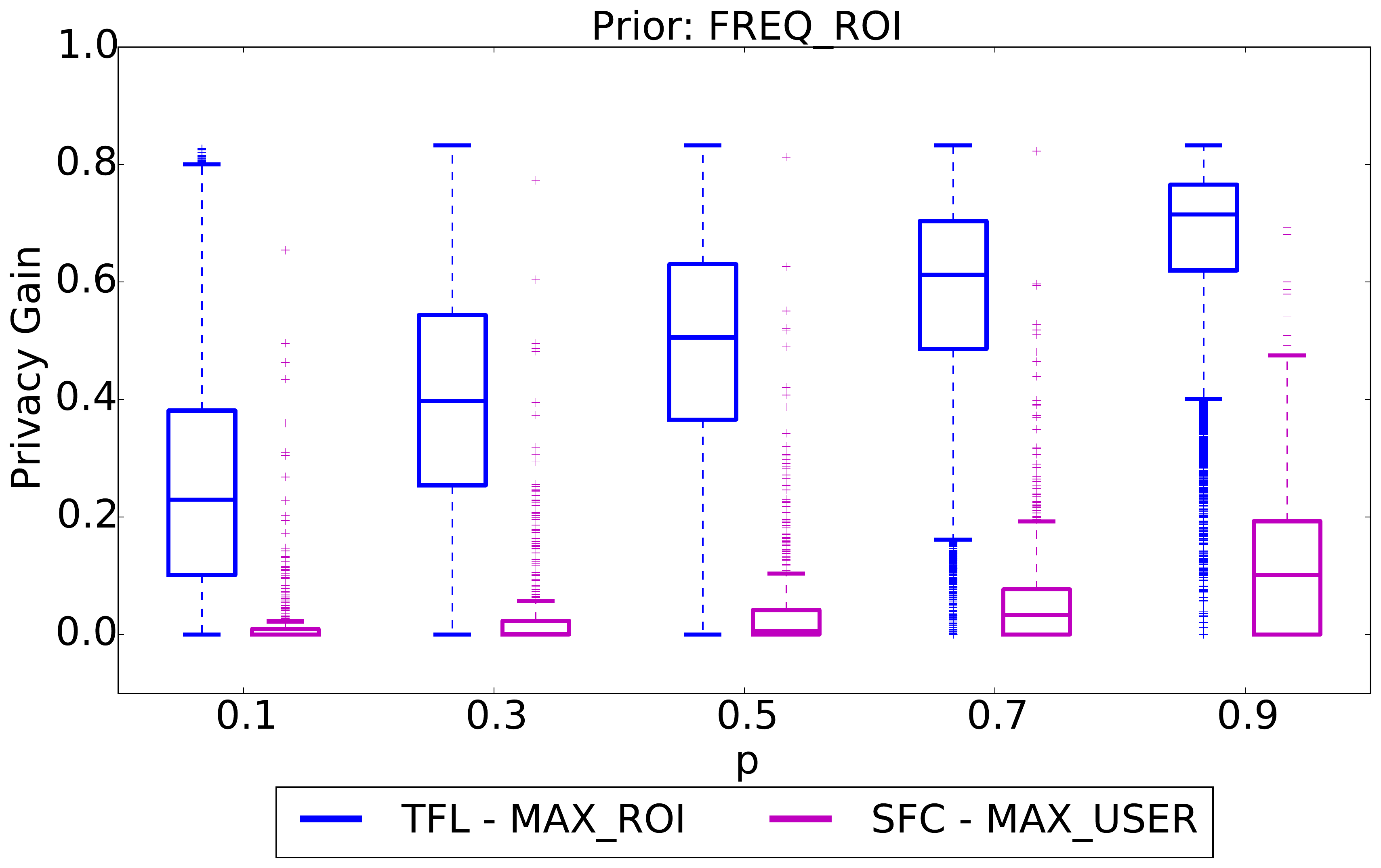}
\reduxB
\caption{SpotMe~\cite{quercia2011spotme}: Privacy gain for increasing values of ${\sf p}$, on TFL and SFC datasets.}
\label{fig:task1-spot-me-privacy}
\end{figure}

\descr{Privacy Quantification.} \figurename~\ref{fig:task1-spot-me-privacy}
plots the privacy gain provided by the RR mechanism w.r.t.~the parameter ${\sf p}$, 
for users in both TFL and SFC datasets. For TFL, we observe that, as ${\sf p}$ increases,
PG also increments, reaching up to $0.6$ with the 
most conservative parameterization (${\sf p}=0.9$). In comparison to the output perturbation
mechanisms applied on TFL data, SpotMe yields smaller privacy gains while keeping the 
utility levels higher.
Interestingly, for SFC, we observe that, as ${\sf p}$ grows, the privacy gain only increases negligibly. 
For ${\sf p}=0.5$, the average PG is $0.04$, while it's only $0.1$ when ${\sf p}=0.9$. Recall that, with
output perturbation mechanisms on the SFC data, privacy gain reaches $0.36$, although yielding lower 
utility. This highlights the challenges of using RR mechanisms, such as SpotMe, on dense datasets with few users.

\begin{table}
\centering
\small
  \begin{tabular}{  l | c  c  c  c  c }
  \toprule
    \hfill ${\sf  p}$ & {\bf 0.1} & {\bf 0.3} &  {\bf 0.5} & {\bf 0.7} & {\bf 0.9} \\ \midrule
	{ TFL - MRE} & 2.1 & 3.9 & 6.1 & 9.3 & 17.6  \\ 	
	{ SFC - MRE} & 0.4 & 0.7 & 1.1 & 1.6 & 2.9  \\ 	
	\bottomrule
	\end{tabular}
\reduxB\reduxB
	\caption{SpotMe~\cite{quercia2011spotme}: MRE (Utility) for increasing values of ${\sf  p}$, on TFL and SFC datasets.}
  \label{table:spotme-utility}
  \vspace{0.3cm}
  \end{table}
  
\beforesec
\subsection{Discussion}
\label{sec:discussion} 
\aftersec

Our evaluation of defense mechanisms based on differential privacy (DP) highlights the difficulty to fine-tune the trade-off between privacy and utility. More specifically, our case studies show that using existing DP mechanisms in a straightforward manner yields poor utility in the context of aggregate location time-series in the settings considered in this paper, i.e., mobility analytics over transport data. 
As expected, we observe that the performance of DP mechanisms in terms of privacy and utility is highly dependent on the intrinsic characteristics of the datasets used in our experiments. For instance, in the sparse TFL dataset containing thousands of users moving among a relatively large 
number of ROIs (583), output and input perturbation achieve reasonable levels of privacy, with the latter performing better than the former in terms of utility. On the other hand, on the denser SFC dataset, which includes fewer users and ROIs (101), output perturbation does not yield
significant privacy protection, and input perturbation only a negligible one. Moreover, our analysis shows that it is challenging to achieve good utility while applying DP on continuous data, such as aggregate location time-series, and mechanisms 
that reduce the required amount of noise (e.g., FPA) still do not provide acceptable privacy guarantees. 

Moreover, data pre-processing techniques~\cite{acs2014case}, like sub-sampling and clustering, could theoretically be used to improve the utility of DP mechanisms (e.g., by reducing the number of locations reported by the users or merging sparse ROIs together), however, such an approach is application dependent and cannot be considered a generalizable solution.

Finally, although the generic framework of differential privacy abstracts from adversarial prior knowledge, our analysis indicates that the concrete nature of this prior should be taken into account when evaluating defense mechanisms. While some priors may not help the adversary, our experiments show that realistic approaches of building adversarial prior knowledge, for example considering users' frequent locations, can help an adversary when performing inference attacks to extract knowledge, even from aggregates perturbed with DP. 

\beforesec
\section{Related Work}
\label{sec:related}
\aftersec

\descr{Attacks on Location Privacy.} Prior work presenting attacks on location 
privacy mostly focuses on inferring users' whereabouts from 
access to individuals' location data, whether obfuscated or not. 
Some show that both anonymization 
and k-anonymity-based mechanisms are ineffective at protecting privacy~\cite{golle2009anonymity,taxis-rainbows,shokri2010unraveling,shokri2011quantifying,zang2011anonymization}. (Also see surveys by 
Krumm~\cite{krumm2009survey} and Ghinita~\cite{ghinita2013privacy}). 
More recently, researchers analyzed the protection provided by location proximity
schemes adopted by social networks~\cite{PolakisAPSK15,WangWWNZZ14,XueBLNWRQ16},
confirming that mechanisms like cloaking or naive perturbation are also unsuccessful.

Independently of our work, Xu et al.~\cite{xu2017trajectory} have recently presented an attack 
that recovers individual users' trajectories from aggregate mobility data, by exploiting the uniqueness 
and the regularity of human mobility. Although their setting is somewhat similar to ours, 
the adversarial task they consider is quite different. Moreover, our work 
introduces a methodology to reason about the effect of releasing location aggregates on individuals'
privacy---with and without DP protection.

\descr{Privacy-Preserving Aggregation.} There are two main privacy-enhancing strategies to 
collect location data and compute aggregate time-series. 
(1) Cryptographic protocols for private aggregation can let a server obtain aggregates without learning users' individual records~\cite{popa2011privacy,pyrgelis2016privacy,melis2015efficient},
but make no consideration about the privacy loss from learning and/or releasing
exact statistics. We have evaluated this scenario in Section~\ref{sec:experiments}.
(2) Perturbation techniques can be used to hide individual inputs
rather than encrypting them. Ho et  al.~\cite{ho2011differential} use 
quadtree spatial decomposition and density based clustering for privately mining location 
databases, while Kopp et al.~\cite{kopp2012privacy}'s framework enables the collection of quantitative 
visits to sets of locations following a distributed approach. Chen et al.~\cite{chen2016private} focus on
spatial data aggregation in the local setting and propose a framework that allows an untrusted server 
to learn the user distribution over a spatial domain relying on a personalized 
count estimation protocol and clustering. As discussed earlier, 
SpotMe~\cite{quercia2011spotme} uses an algorithm based on Randomized 
Response~\cite{warner1965randomized} to estimate the number of people in 
geographic locations. We have evaluated this kind of solutions, specifically, 
SpotMe~\cite{quercia2011spotme}, in Section~\ref{sec:counterInput}.

\descr{Private Location Data Publishing.} Machanavjjhala et 
al.~\cite{machanavajjhala2008privacy} use synthetic data generation techniques 
to publish commuting patterns in a differentially private way, while Acs and 
Castelluccia~\cite{acs2014case} describe a differentially private scheme to 
release the spatio-temporal density of Paris regions using records provided by 
a telco operator. To et al.~\cite{todifferentially} focus on releasing location 
entropy for ROIs under differential privacy guarantees: they study the bounds 
of location entropy and show that $\epsilon$-differential privacy requires an 
excessive amount of noise, so they use weaker notions achieving better utility.
Besides specific location-oriented private publishing, differential privacy has 
been proposed as a solution for releasing generic time-series of aggregate 
statistics. Examples are the various differentially private counting mechanisms 
by Chan et al.~\cite{chan2011private}, or Fan et al.'s adaptive 
system~\cite{fan2012real} that uses a combination of filtering and 
sampling to increase the utility of differentially private aggregates.
Rastogi and Nath~\cite{rastogi2010differentially} use an algorithm based on 
Discrete Fourier Transform to privately release aggregate time-series, while 
Shi et al.~\cite{shi2011privacy} combine encryption with data randomization to 
achieve differential privacy for time-series data. We have evaluated the 
privacy provided by this approach in Section~\ref{sec:counterOutput}, 
using the schemes in~\cite{chan2011private,rastogi2010differentially}.

\descr{Quantifying Location Privacy.} Previous work on privacy 
quantification has studied the privacy loss incurred when 
disclosing obfuscated traces of individual users, e.g., when using location-based services. 
The main work in this area is the quantification framework by Shokri et 
al.~\cite{shokri2011quantify,shokri2011quantifying}, which
considers a strategic adversary that has prior 
information about users' mobility patterns, knows the location 
privacy-protection mechanism they use, and deploys inference attacks 
based on this information and the observation of the obfuscated traces.

This framework is conceived for evaluating privacy-preserving 
mechanisms applied to individuals' traces, therefore, the techniques used in their 
work are not applicable in the context of location 
privacy-preserving mechanisms based on aggregation. 
Nonetheless, if we were to compare our framework to Shokri et al.'s,
we would observe that it does not only differ in the modeling of the adversary's prior 
knowledge, observation, and goal, but it is also driven by the definition of new metrics to model the 
adversary's error in this scenario. Moreover, we introduce new inference attacks 
tailored to the aggregate scenario and evaluate the impact on privacy of: 
(i) priors of different nature -- specifically, both assignment and probabilistic, while only probabilistic are considered in~\cite{shokri2011quantify,shokri2011quantifying}, (ii) priors based on more or less complete information, 
and (iii) sparsity of the location data that should be protected.

\beforesec
\section{Conclusion}
\label{sec:conclusion}
\aftersec

Publishing aggregate location information is often considered a privacy-friendly 
strategy to support mobility analytics applications, especially if 
the aggregation itself is performed in a privacy-preserving way~\cite{kopp2012privacy,popa2011privacy,pyrgelis2016privacy} (i.e., without the need for trusted aggregators),
and/or Differential Privacy (DP) is used to perturb aggregates~\cite{chan2011private,dwork2010differential,rastogi2010differentially,quercia2011spotme}.
However, as opposed to privacy-preserving 
mechanisms for single users' traces %
\cite{shokri2011quantify,shokri2011quantifying}, 
there has been very little work on understanding the privacy threat 
that releasing aggregate location time-series poses on individuals whose locations are part of such aggregates.

This paper presented a first-of-its-kind analysis of aggregate location privacy. 
We introduced appropriate metrics
to reason about privacy in the presence of an adversary 
aiming to localize and/or profile individual users,
and proposed strategies to model the adversary's prior knowledge 
as well as to exploit aggregate information to perform inference attacks. 
We used two real-world mobility datasets with different mobility 
characteristics to evaluate both the case in which raw aggregates are released,
and when aggregates are perturbed to achieve Differential Privacy (DP) guarantees. Our experiments show 
that aggregates do help the adversary  
uncover mobility patterns and localize users,
and that DP only improves privacy when adding so much noise that the utility
of the time-series is destroyed.

We believe that our work will encourage further research on inference
attacks as well as on the adversary's capability to obtain 
useful priors, also aiming to gain a better understanding of their dependence. 
Moreover, our results motivate future work towards the design of new differential privacy techniques that take into account temporal as well as spatial correlations, such as those discussed in~\cite{andres2013geo,cao2017quantifying} which may provide a promising direction. Overall, we highlight the need for novel defense mechanisms that 
can offer better privacy guarantees to individuals whose location data is part of aggregate time-series 
releases, including in the context of ``privacy-friendly'' applications recently announced by 
Google~\cite{googleTraffic} and Apple~\cite{AppleDP}.

\descr{Acknowledgments.} We wish to thank Mirco Musolesi and Gordon Ross for useful feedback and comments, as well as Rinku Dewri for shepherding the paper. This research is partially supported by a Xerox University Affairs Committee grant on ``Secure Collaborative Analytics.''

{
\bibliographystyle{abbrv}
\bibliography{bibfile}
}

\aftersec
\appendix

\begin{figure}[t]
\centering
\begin{subfigure}[b]{0.45\textwidth}
\includegraphics[width=0.99\linewidth]{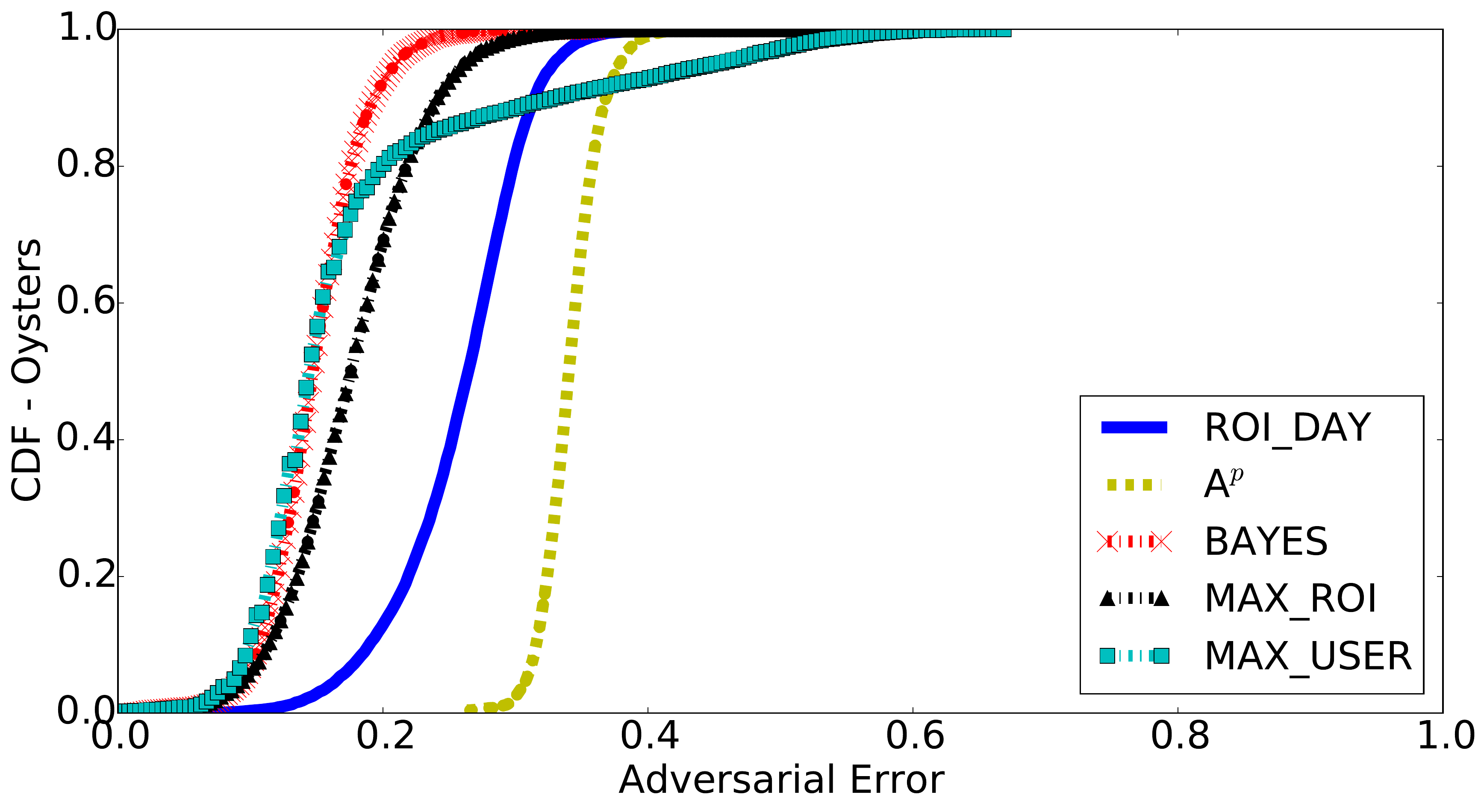}
\caption{TFL}
\label{fig:task1-roi-day-tfl}
\end{subfigure}
\begin{subfigure}[b]{0.45\textwidth}
\includegraphics[width=0.99\linewidth]{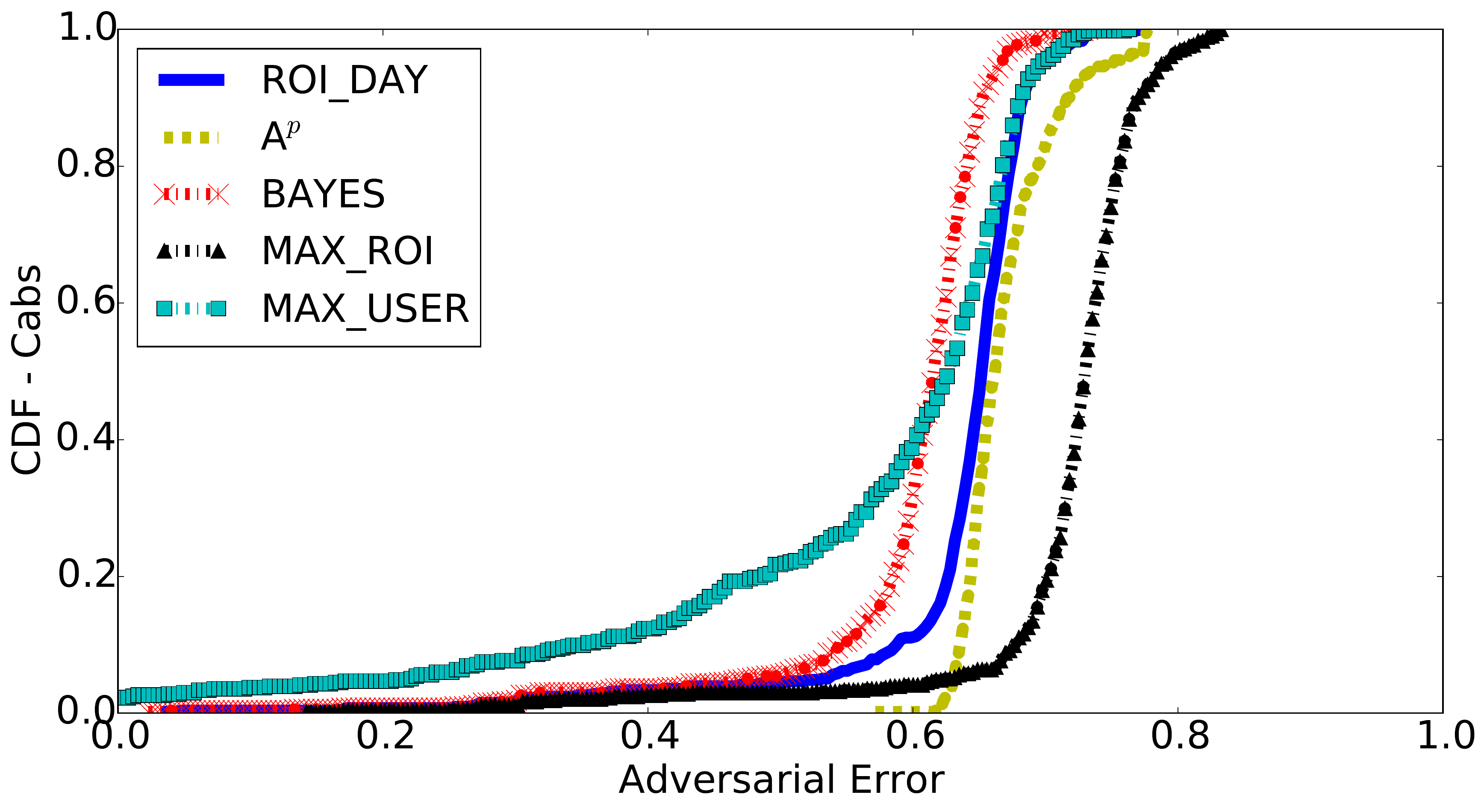}
\caption{SFC}
\label{fig:task1-roi-day-sfc}
\end{subfigure}
\vspace{-0.2cm}
\caption{\adv's Profiling Error - \roiday Prior.}
\label{fig:task1-roi-day}
\end{figure}

\begin{figure}[t]
\centering
\begin{subfigure}[b]{0.45\textwidth}
\includegraphics[width=0.99\linewidth]{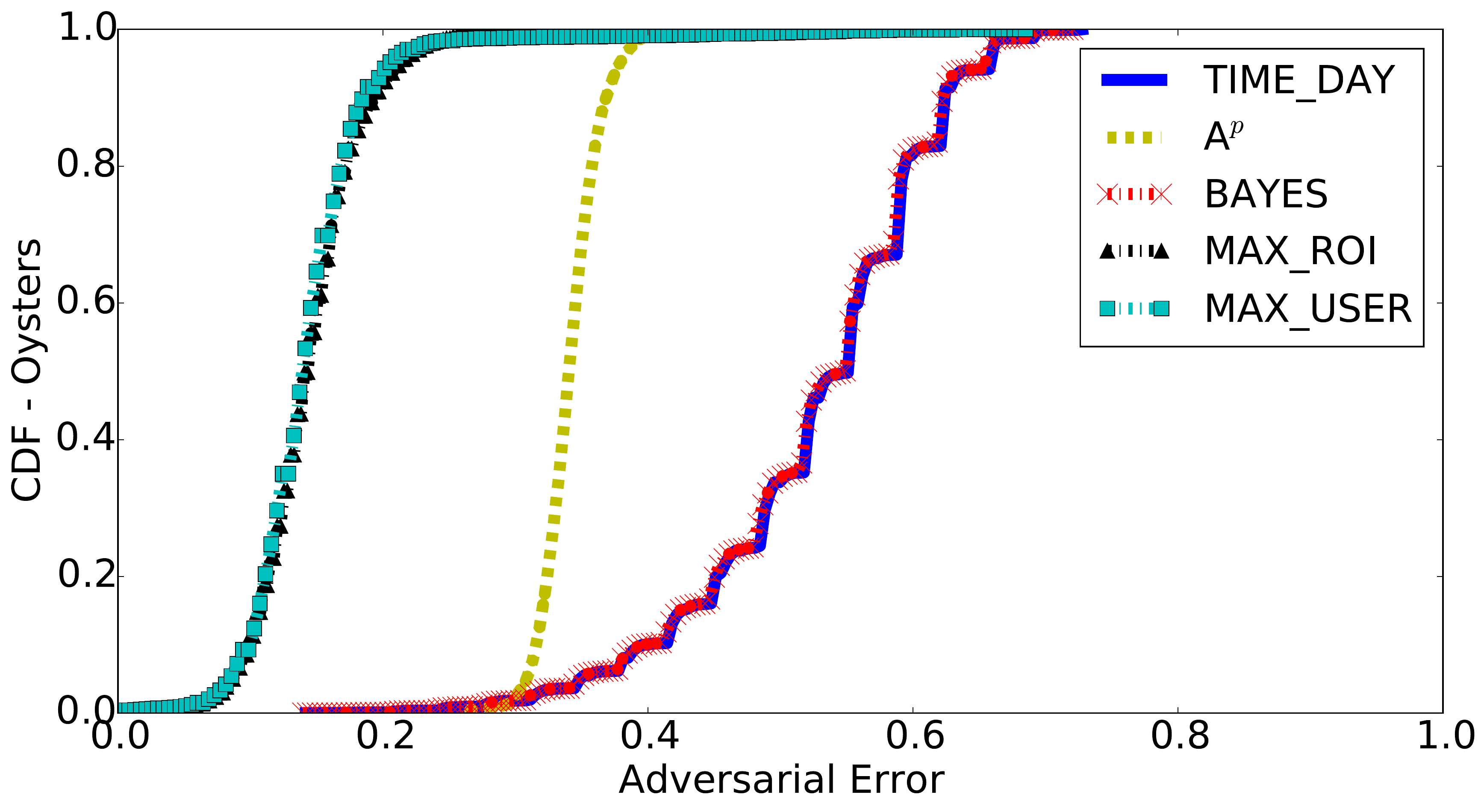}
\caption{TFL}
\label{fig:task1-time-day-tfl}
\end{subfigure}
\begin{subfigure}[b]{0.45\textwidth}
\includegraphics[width=0.99\linewidth]{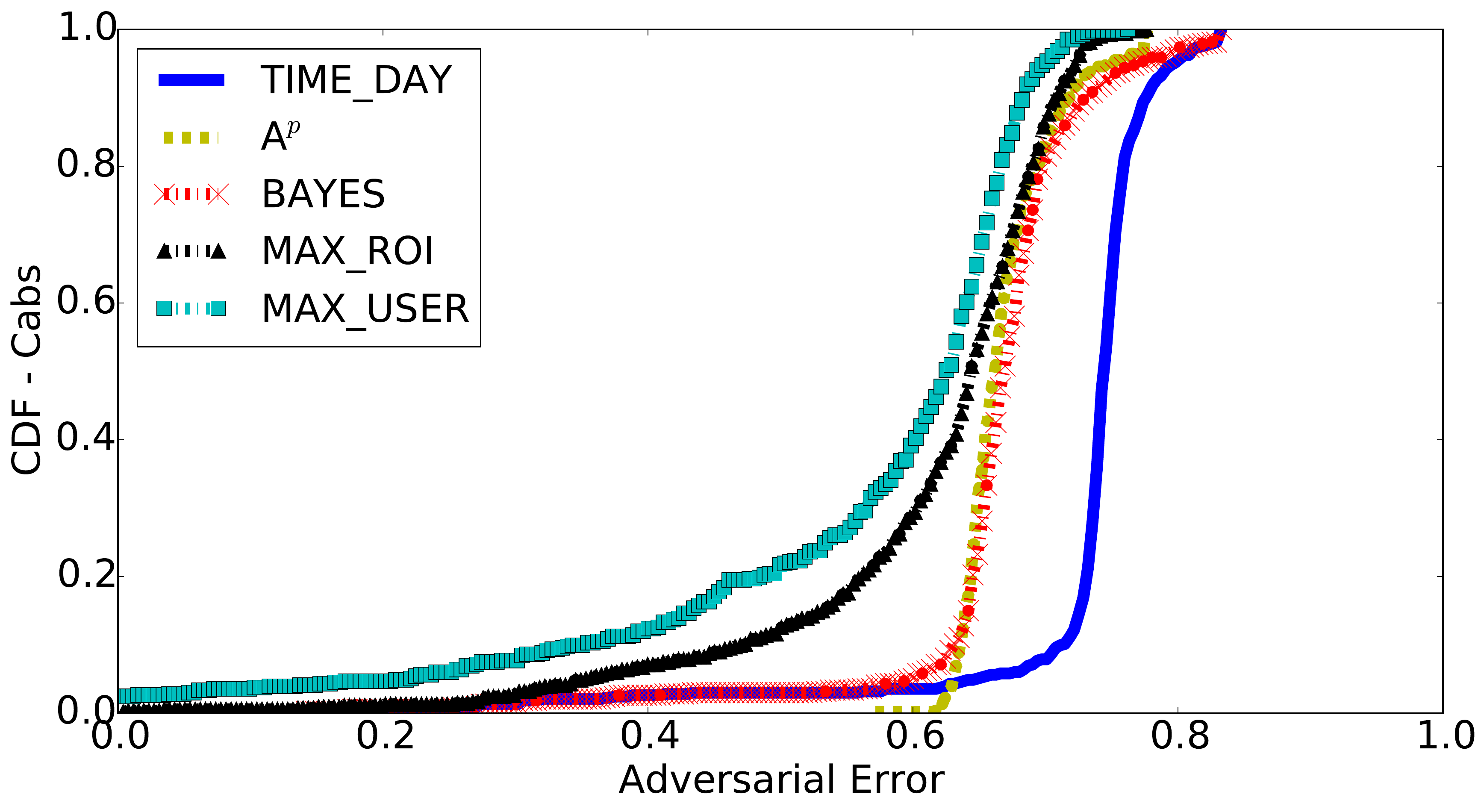}
\caption{SFC}
\label{fig:task1-time-day-sfc}
\end{subfigure}
\vspace{-0.2cm}
\caption{\adv's Profiling Error - \timeday Prior.}
\label{fig:task1-time-day}
\end{figure}

\section{Additional Experiments}
\label{sec:appendix}

We now report  additional details about experimental results on inference tasks based on other approaches of prior knowledge.
\subsection{User Profiling}

\subsubsection{Probabilistic Priors}
\label{sec:app-prof-prob}

\descr{ROI\_DAY.} Recall that, with \roiday, Adv knows for the users, a
profile for each hour of \textit{any} day (e.g., user's frequent locations at 4pm). For TFL (\figurename~\ref{fig:task1-roi-day-tfl}), 
we observe that this is a more instructive prior than commuters' frequent ROIs (\freqroi), with an average prior error of $0.25$. 
Moreover, we note that \bayes and \maxr inferences remarkably improve \adv's profiling accomplishment for all users, 
yielding $0.41$ and $0.31$ average privacy loss, respectively. \maxu improves \adv's predictions for $\sim 80\%$ of 
the users and achieves $0.37$ average loss in privacy. Similarly for SFC (\figurename~\ref{fig:task1-roi-day-sfc}), \roiday ($0.63$ avg. error) is a more revealing prior knowledge than cabs' frequent ROIs (\freqroi \,-- $0.65$) for \adv. \bayes and \maxu give advantage to \adv in profiling users (resulting in $0.06$ and $0.13$ privacy loss, resp.) while \maxr does not, once again, indicating the \textit{bias} of this strategy towards less active cabs.

\descr{TIME\_DAY.} \figurename~\ref{fig:task1-time-day-tfl} plots the CDF of \adv's total error in 
profiling TFL commuters, with the \timeday prior knowledge, i.e., a time profile 
indicating which hours of day a user is likely to report ROIs. We observe that 
\adv's performance is worse ($0.52$ mean error) compared 
to priors containing location information (i.e., \freqroi, \roiday or \roidayweek). This is expected, since \timeday prior contains 
only time information for the users, and it is a \textit{uniform} 
distribution over all ROIs, for the time slots that they are likely to be 
inside the transportation system. Indeed, \figurename~\ref{fig:task1-time-day-tfl} shows 
that profiling only with the aggregate profile (\aggP), \adv achieves smaller error ($0.34$). 
Among the inference strategies, we note that \bayes negligibly improves \adv's error in 
profiling users due to the very small prior probabilities. \maxr and \maxu attacks exhibit similar performance, as in both cases the
users who are more likely to be \textit{inside} the system, are selected to cover the aggregate values (in this case both strategies pick users based on their total number of ROIs). With these strategies, \adv's performance increases significantly and there is notable privacy loss for the users ($0.72$ on average).

\begin{figure}[t]
\centering
\begin{subfigure}[b]{0.45\textwidth}
\includegraphics[width=0.99\linewidth]{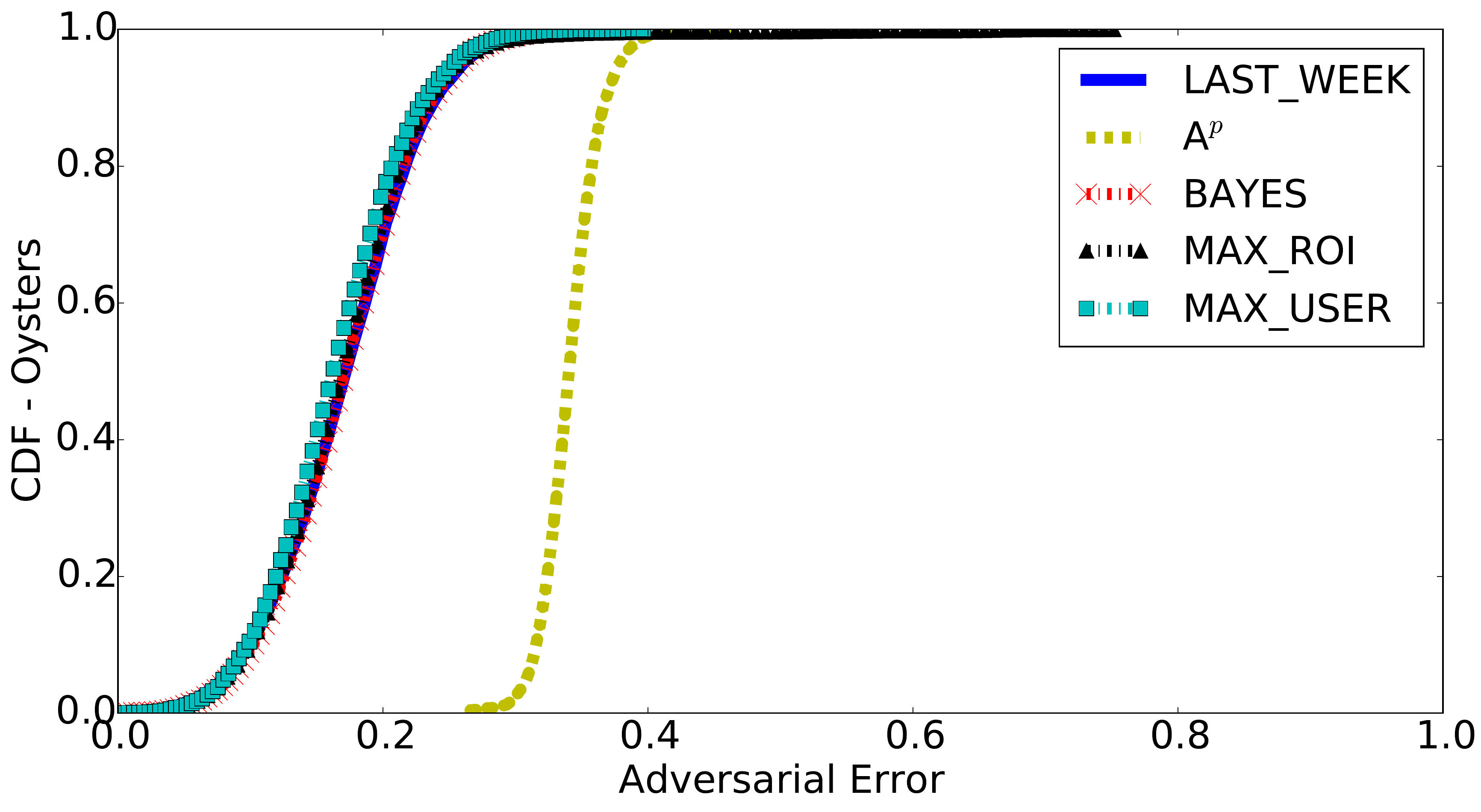}
\label{fig:task1-last-week-tfl}
\end{subfigure}
\begin{subfigure}[b]{0.45\textwidth}
\includegraphics[width=0.99\linewidth]{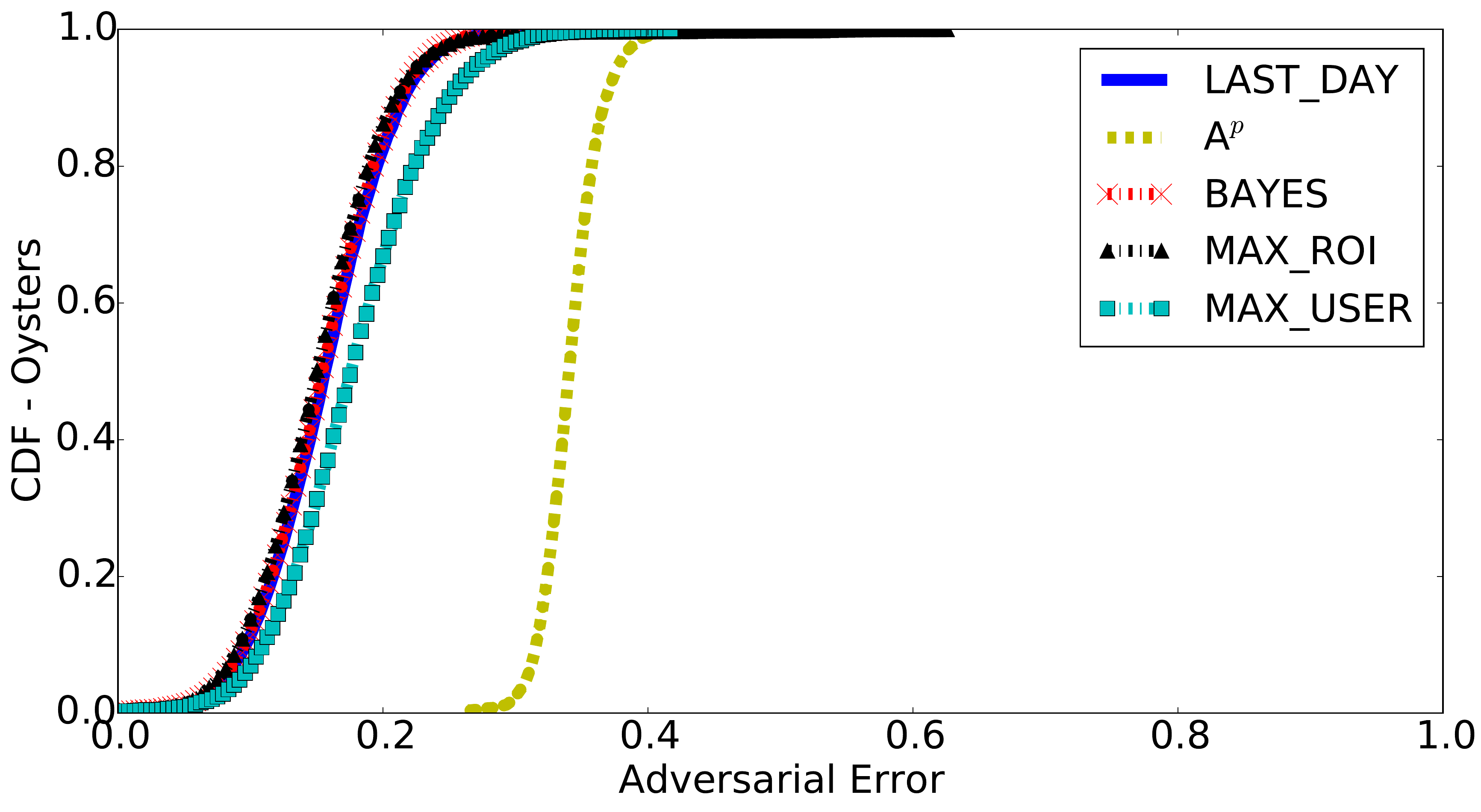}
\label{fig:task1-last-day-tfl}
\end{subfigure}
\begin{subfigure}[b]{0.45\textwidth}
\includegraphics[width=0.99\linewidth]{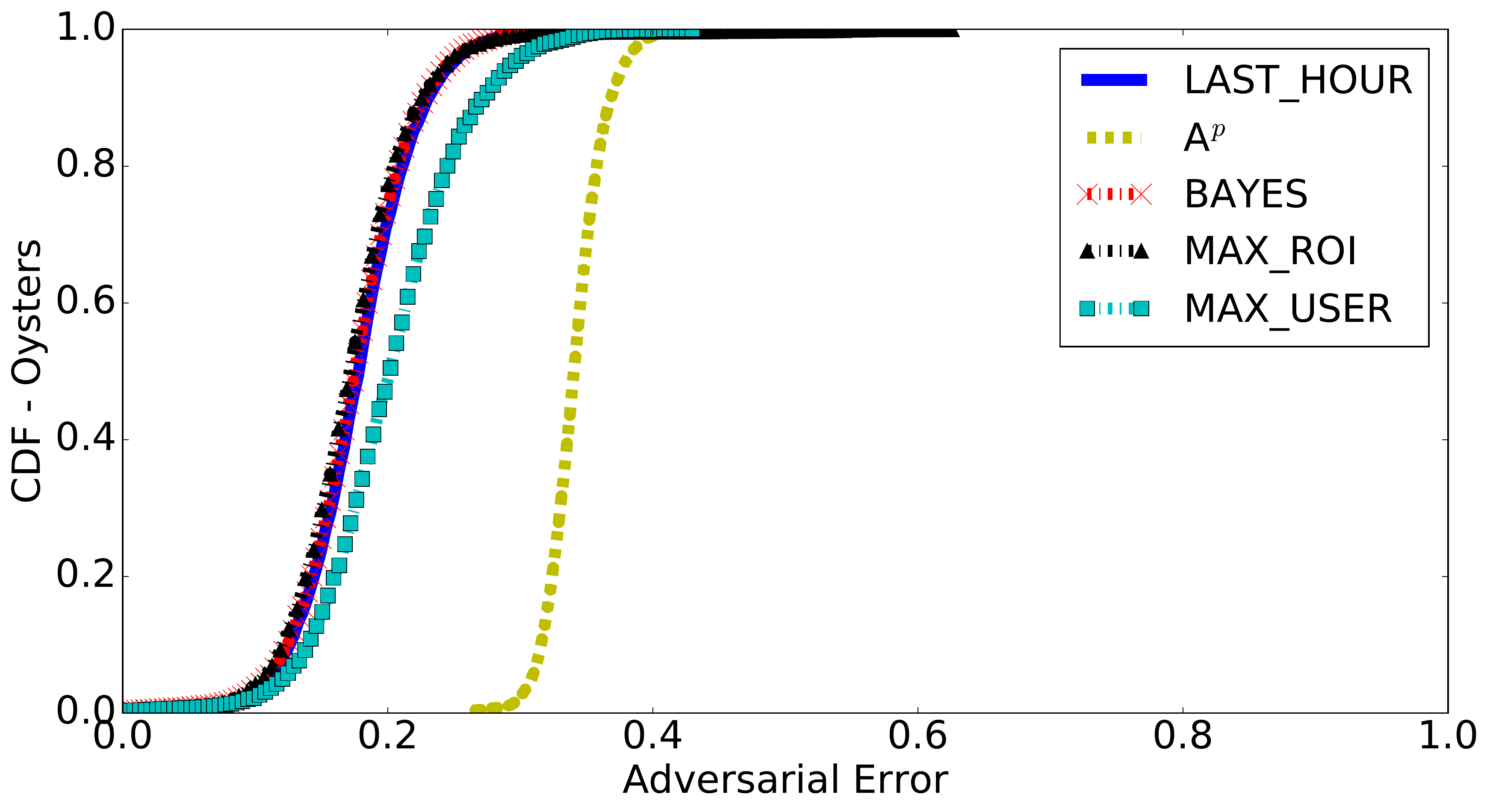}
\label{fig:task1-last-hour-tfl}
\end{subfigure}
\vspace{-0.4cm}
\caption{\adv's Profiling Error - \lastweek, \lastday and \lasthour Priors - TFL.}
\label{fig:task1-ass-priors-tfl}
\end{figure}

\begin{figure}[t]
\centering
\begin{subfigure}[b]{0.45\textwidth}
\includegraphics[width=0.99\linewidth]{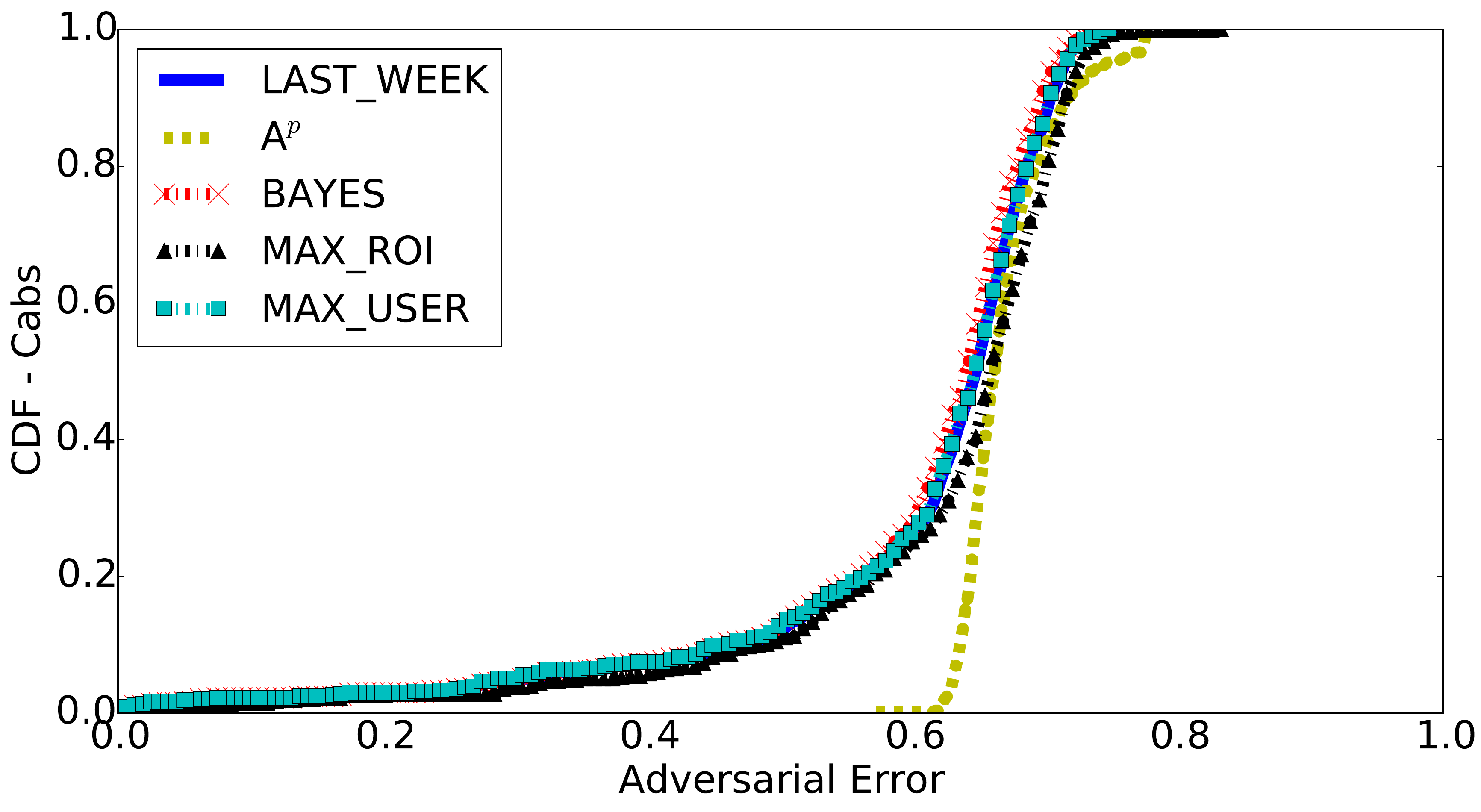}
\label{fig:task1-last-week-sfc}
\end{subfigure}
\begin{subfigure}[b]{0.45\textwidth}
\includegraphics[width=0.99\linewidth]{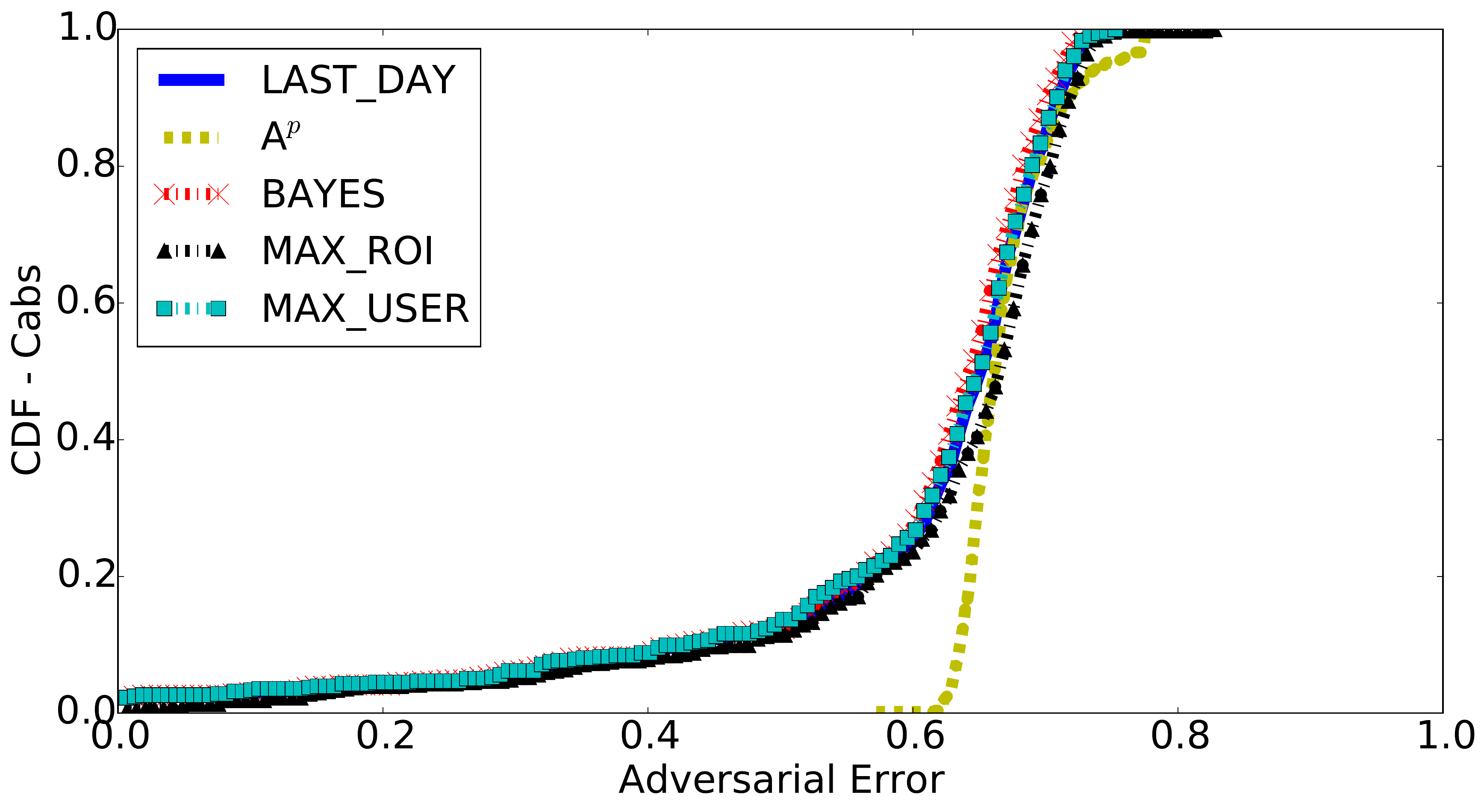}
\label{fig:task1-last-day-sfc}
\end{subfigure}
\begin{subfigure}[b]{0.45\textwidth}
\includegraphics[width=0.99\linewidth]{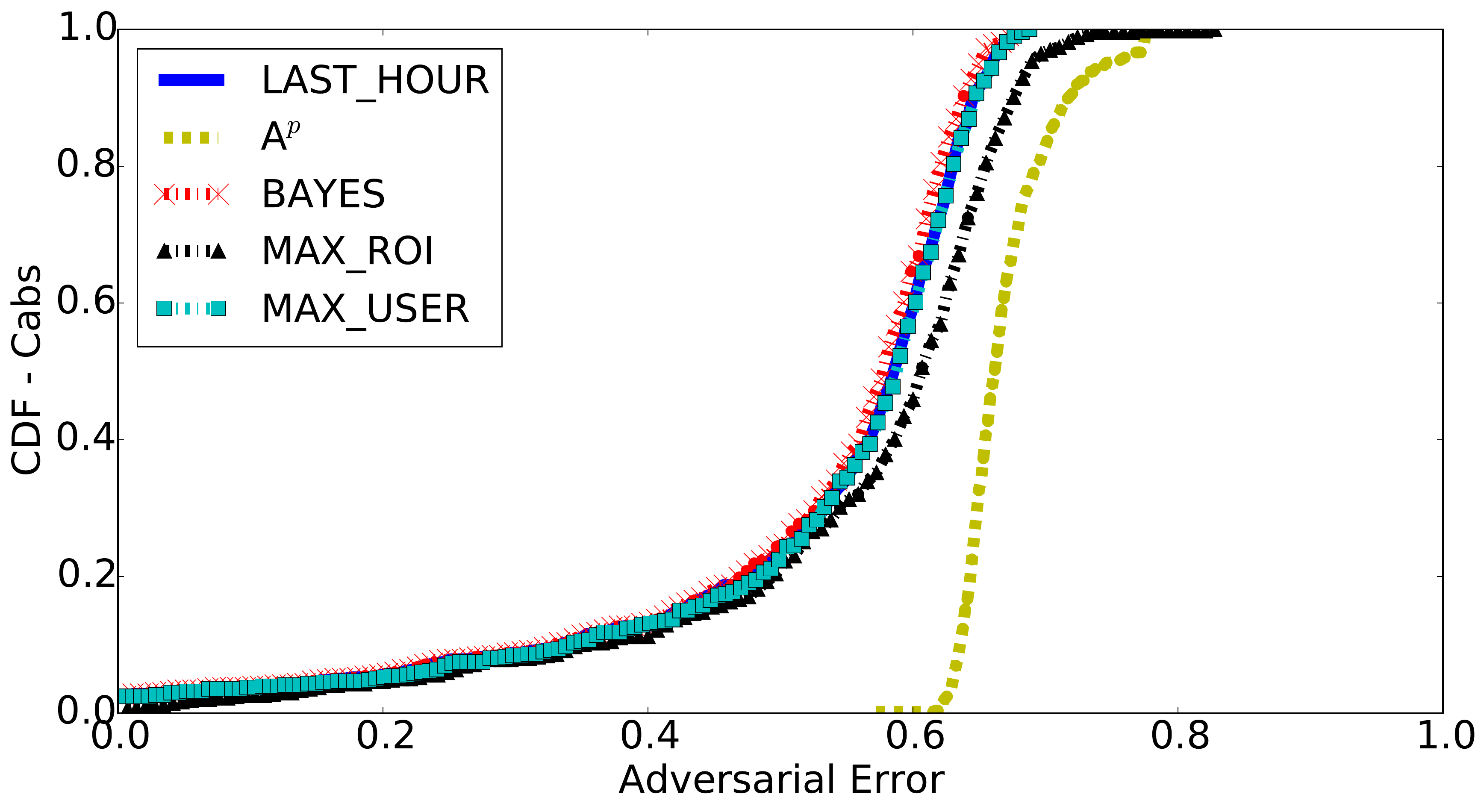}
\label{fig:task1-last-hour-sfc}
\end{subfigure}
\vspace{-0.4cm}
\caption{\adv's Profiling Error - \lastweek, \lastday and \lasthour Priors - SFC.}
\label{fig:task1-ass-priors-sfc}
\end{figure}

\begin{figure}[t]
\centering
\begin{subfigure}[b]{0.45\textwidth}
\includegraphics[width=0.99\linewidth]{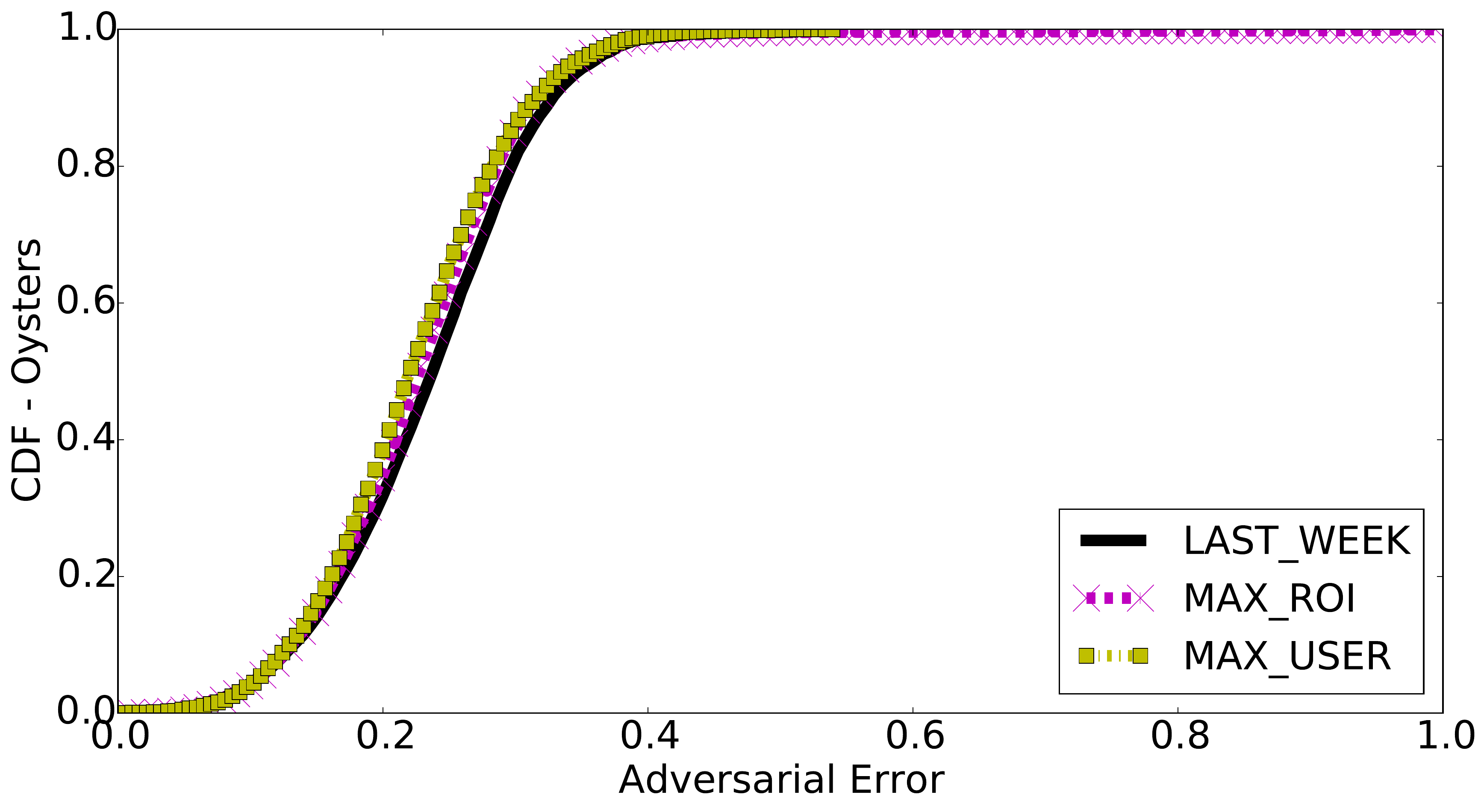}
\label{fig:task2-last-week-tfl}
\end{subfigure}
\begin{subfigure}[b]{0.45\textwidth}
\includegraphics[width=0.99\linewidth]{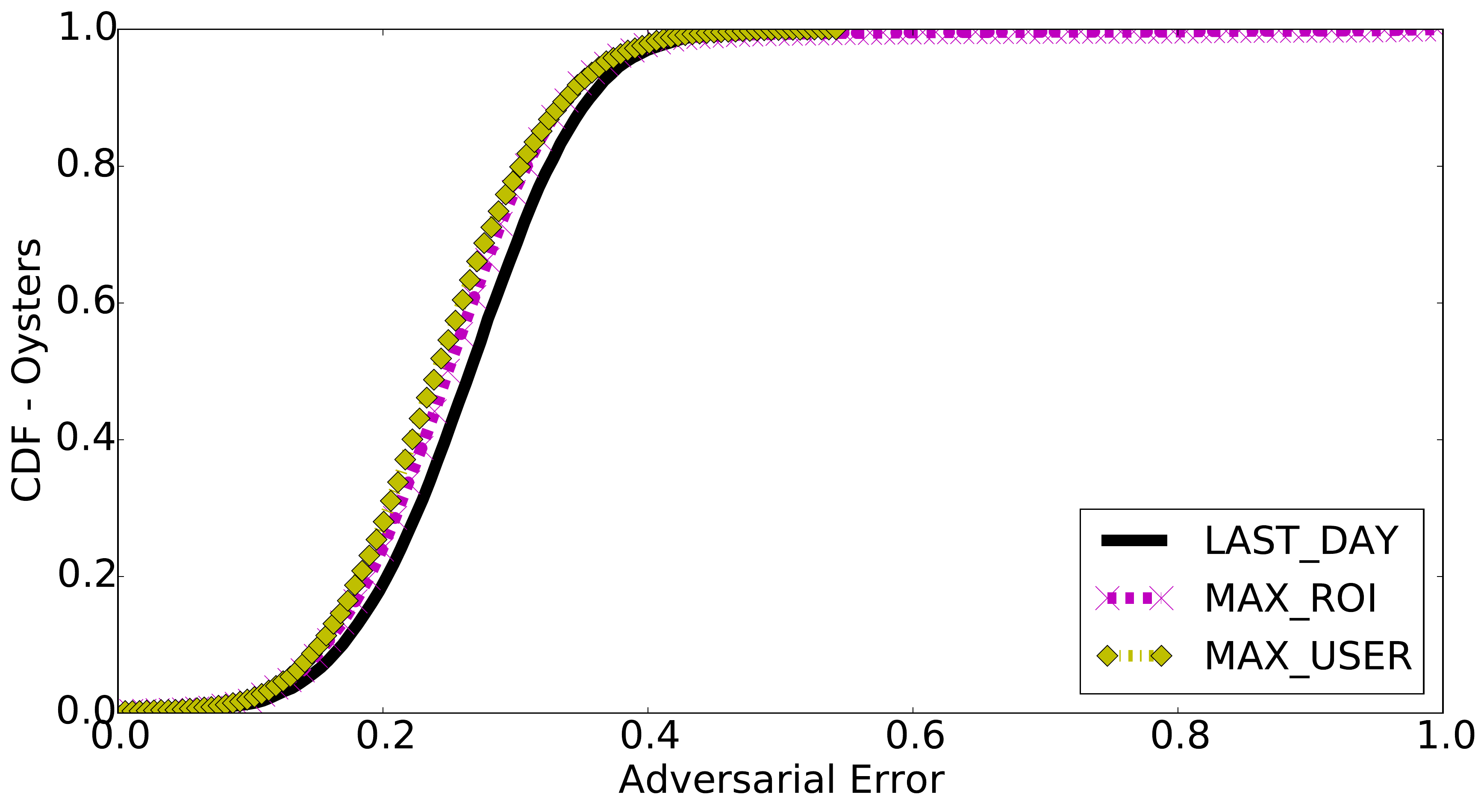}
\label{fig:task2-last-day-tfl}
\end{subfigure}
\begin{subfigure}[b]{0.45\textwidth}
\includegraphics[width=0.99\linewidth]{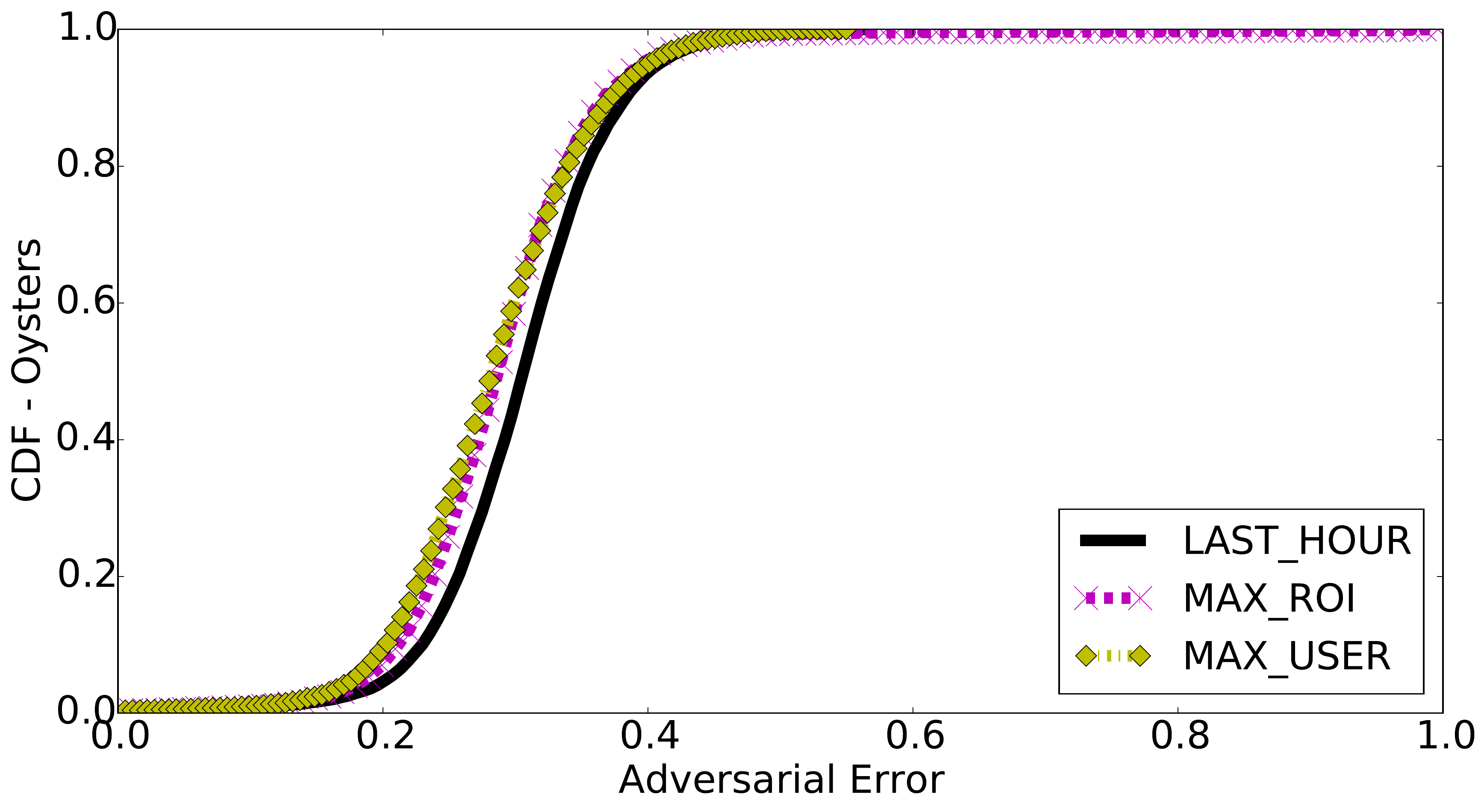}
\label{fig:task2-last-hour-tfl}
\end{subfigure}
\vspace{-0.4cm}
\caption{\adv's Localization Error - \lastweek, \lastday and \lasthour Priors - TFL.}
\label{fig:task2-ass-priors-tfl}
\end{figure}

\begin{figure}[t]
\centering

\begin{subfigure}[b]{0.45\textwidth}
\includegraphics[width=0.99\linewidth]{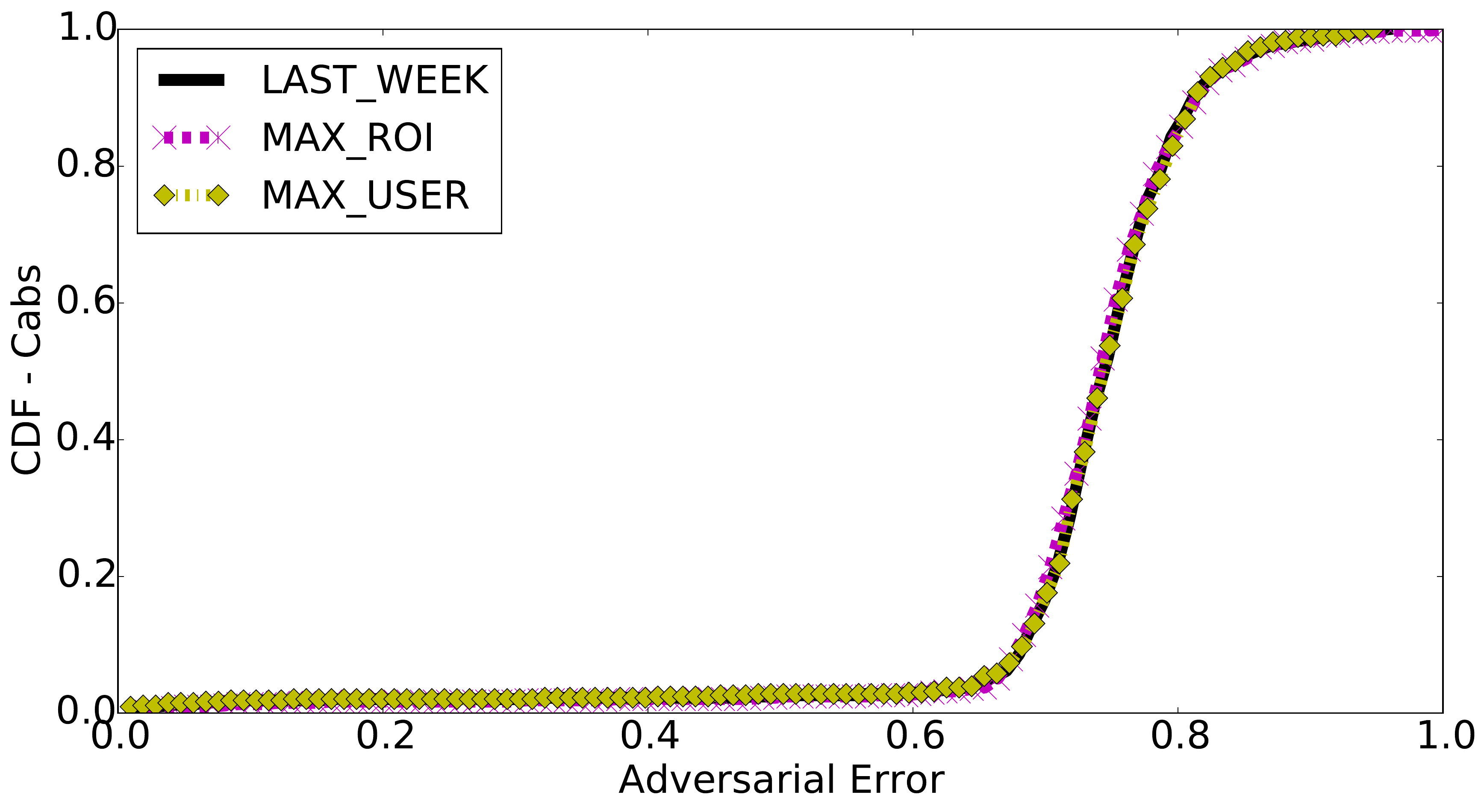}
\label{fig:task2-last-week-sfc}
\end{subfigure}
\begin{subfigure}[b]{0.45\textwidth}
\includegraphics[width=0.99\linewidth]{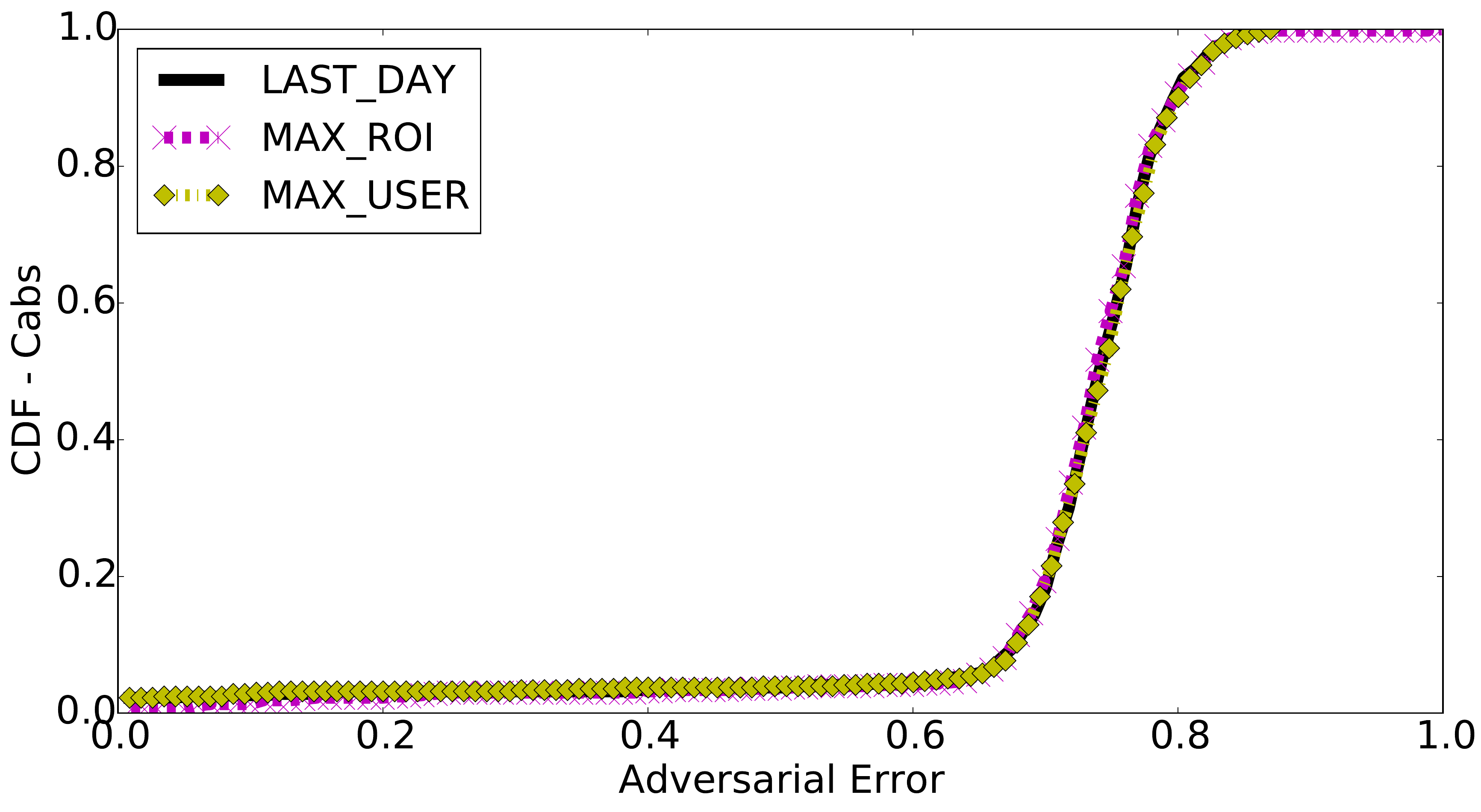}
\label{fig:task2-last-day-sfc}
\end{subfigure}
\begin{subfigure}[b]{0.45\textwidth}
\includegraphics[width=0.99\linewidth]{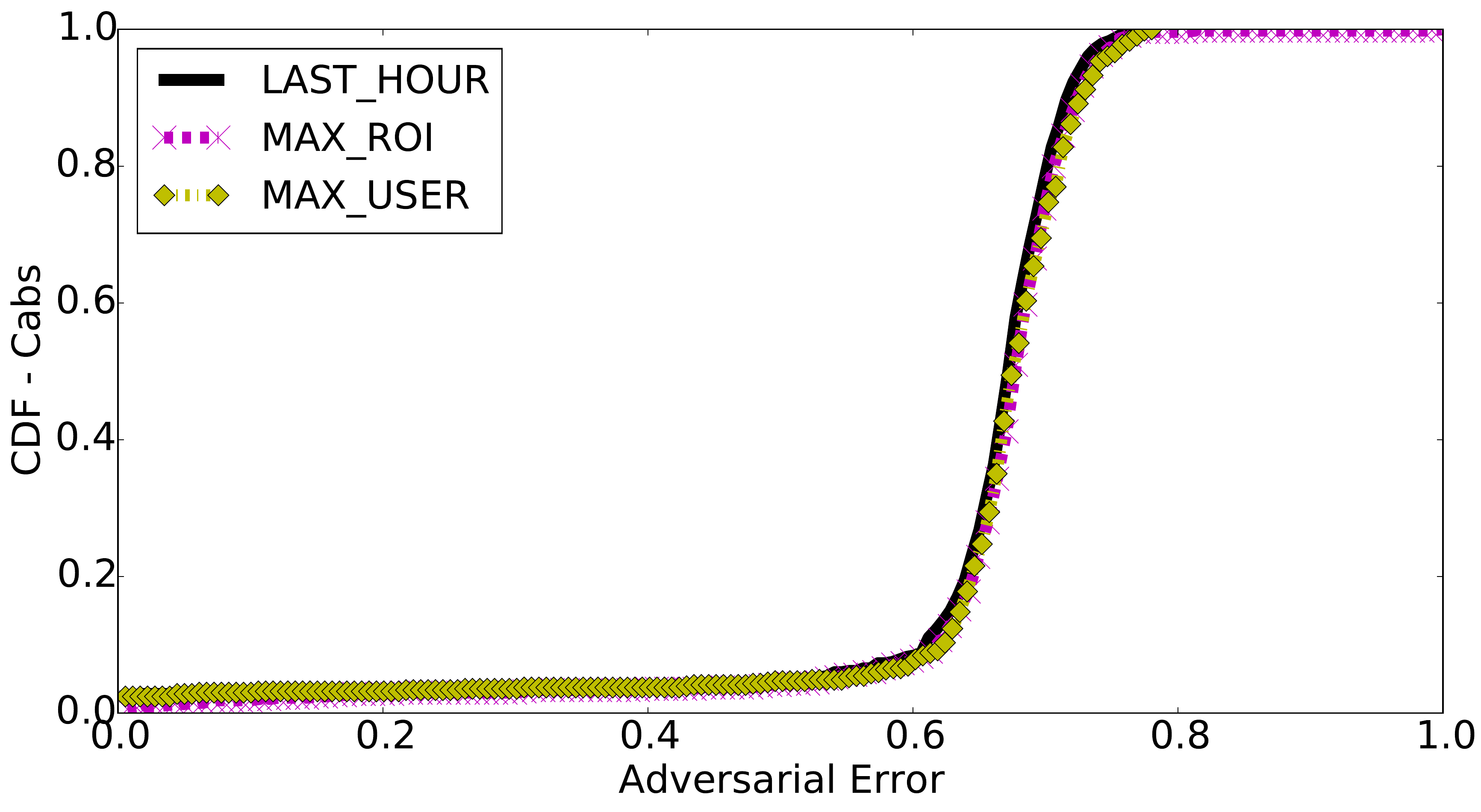}
\label{fig:task2-last-hour-sfc}
\end{subfigure}
\vspace{-0.4cm}
\caption{\adv's Localization Error - \lastweek, \lastday and \lasthour Priors - SFC.}
\label{fig:task2-ass-priors-sfc}
\end{figure}

On the SFC dataset (\figurename~\ref{fig:task1-time-day-sfc}), we observe that when \adv knows the cabs' most frequent time slots of day (\timeday), she obtains a worse prior ($0.73$ mean error) compared to cabs' most frequent ROIs (\freqroi \,-- $0.65$) or cabs' most frequent ROIs with time and day semantics (\roidayweek \,-- $0.61$). Unlike TFL, Bayesian updating yields a $0.1$ privacy loss (as with fewer ROIs in the SFC data, \bayes affects significantly the posterior probabilities), while the greedy strategies perform even better. More precisely, with \maxr the mean privacy loss is $0.16$ and with \maxu $0.22$.

\beforesec
\subsubsection{Assignment Priors}
\label{sec:app-prof-ass}
\aftersec

To ease presentation, the figures that show the results of user profiling based on assignment priors (i.e., \lastweek, \lastday and \lasthour) are presented here. Figures~\ref{fig:task1-ass-priors-tfl} and~\ref{fig:task1-ass-priors-sfc} plot the results that are discussed in Section~\ref{sec:exp-prof-ass-prior}.

\subsection{User Localization}
\label{sec:app-loc}
\aftersec

\subsubsection{Probabilistic Priors}
\label{sec:app-loc-prob}
\aftersec

\begin{figure}[]
\centering
\begin{subfigure}[b]{0.45\textwidth}
\includegraphics[width=0.99\linewidth]{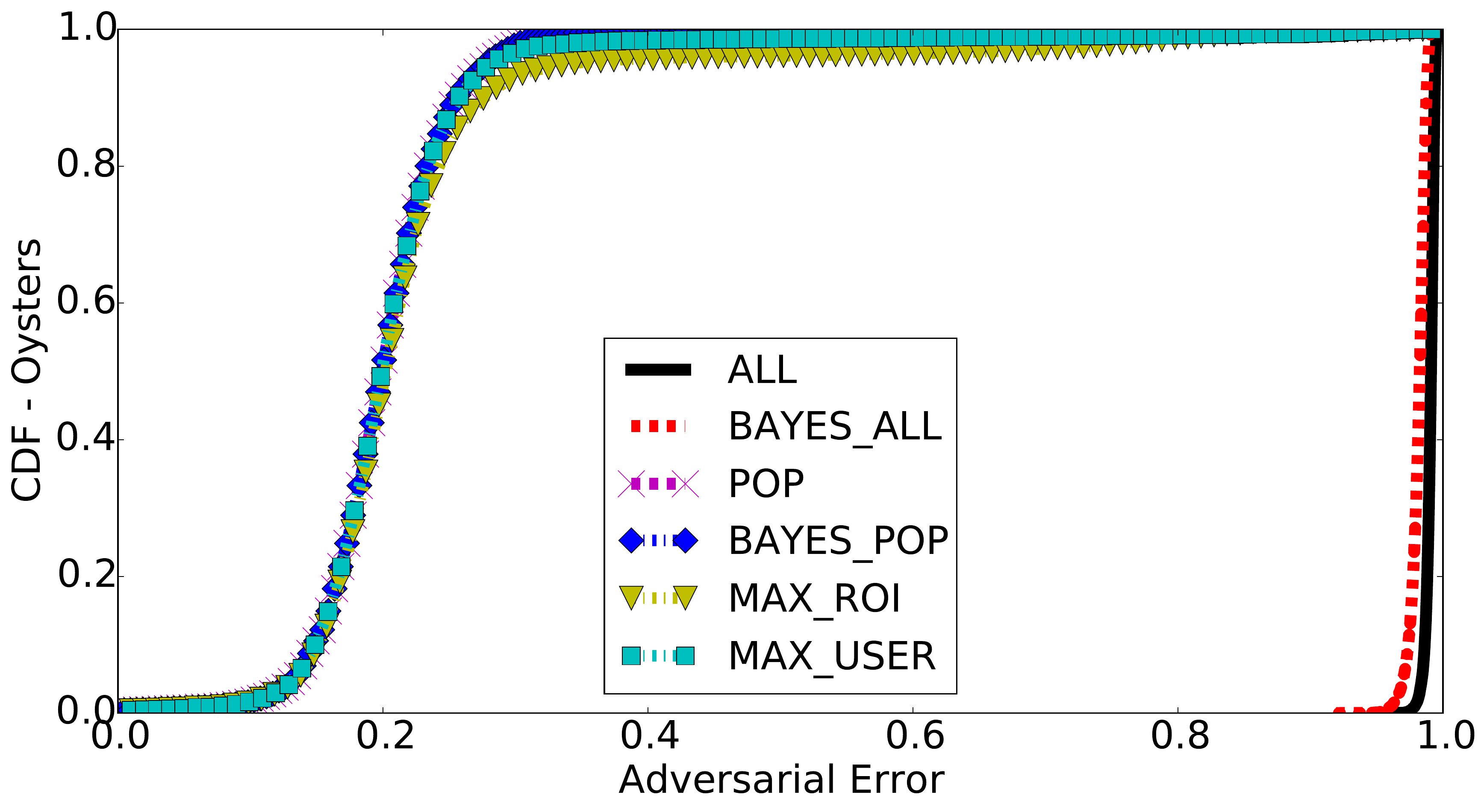}
\caption{TFL}
\label{fig:task2-time-day-week-tfl}
\end{subfigure}
\\
\begin{subfigure}[b]{0.45\textwidth}
\includegraphics[width=0.99\linewidth]{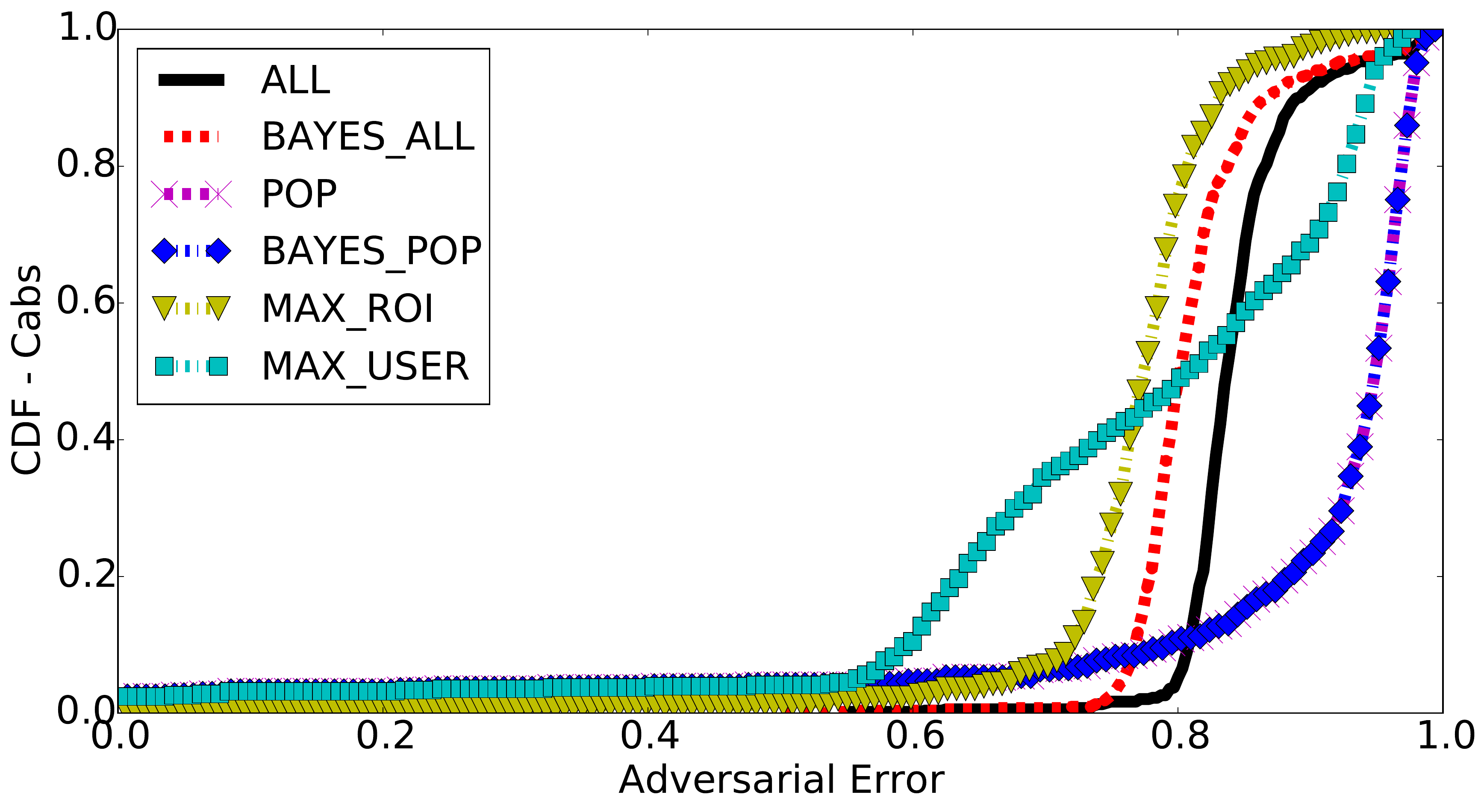}
\caption{SFC}
\label{fig:task2-time-day-week-sfc}
\end{subfigure}
\caption{\adv's Localization Error - \timedayweek Prior.}
\label{fig:task2-time-day-week}
\end{figure}

\descr{TIME\_DAY\_WEEK.} \figurename~\ref{fig:task2-time-day-week} displays \adv's error when localizing users with the \timedayweek prior. For TFL (\figurename~\ref{fig:task2-time-day-week-tfl}), we observe that \all results in a very large error ($0.99$ on average). This is not surprising since \timedayweek is a uniform distribution over ROIs, for the time slots that users are likely to be \textit{in} the transportation system. \all after \bayes achieves negligible privacy loss ($0.03$), while we observe no adversarial advantage between \pop and \pop after \bayes due to the very small prior probabilities. Furthermore, both \maxu and \maxr improve remarkably \adv's performance compared to \all and they yield $0.79$ and $0.77$ average privacy loss respectively. We note that \maxr achieves error larger than $0.25$ for $20\%$ of the users while \maxu yields error larger than $0.25$ for only $5\%$ of the users, i.e., those users that report the most locations and always get assigned to locations to consume the aggregates.

For the SFC data (\figurename~\ref{fig:task2-time-day-week-sfc}) we observe that \all yields $0.84$ average error, while \all after \bayes results in small privacy loss ($0.04$). Once again, \pop is the worst inference strategy as the small probabilities of the prior do not exceed the threshold $\delta$ and cabs are predicted to be outside the network. Moreover, unlike the case of \roidayweek, \maxr now improves \adv's performance for localizing all the cabs, yielding $0.09$ privacy loss. \maxu achieves a similar mean loss in privacy, however, \adv's knowledge is only improved for $60\%$ of the cabs compared to the baseline \all. Once again, we remark how localization strategies result to different amount of privacy leakage on sparse (TFL) and dense (SFC) datasets.

\subsubsection{Assignment Priors}
\label{sec:app-loc-ass}

We evaluate \adv's performance against user localization, i.e., predicting users' future locations with a seasonal part of their ground truth as prior knowledge. In particular, we experiment with \lastweek, \lastday and \lasthour and focus on the \maxr and \maxu inference attacks. \adv's \textit{baseline} prediction is to \textit{replicate} the prior, as described in Section~\ref{sec:assignment-prior}. Figs~\ref{fig:task2-ass-priors-tfl}--\ref{fig:task2-ass-priors-sfc} plot the CDF of \adv's error in localizing commuters and cabs, over the inference week.

\descr{LAST\_WEEK.} \adv's average error localizing tube passengers with \lastweek is $0.24$. Both \maxr and \maxu inference strategies vaguely improve her performance and yield small privacy loss ($0.02)$. This indicates that users reporting lots of ROIs and ROIs themselves show regularity within weeks. Furthermore, \maxu attack is more consistent in improving \adv's localization success than \maxr, which increases \adv's error (compared to the prior) for $5\%$ of the users. For SFC cabs, we observe that \adv's avg. localization error is $0.73$, while both \maxr and \maxu do not reduce it further. Unlike TFL, we observe that the aggregates do not give any advantage to \adv in localizing taxis and there is no privacy loss.

\descr{LAST\_DAY.} With \lastday, \adv's mean error in predicting TFL passengers' locations during the inference week is $0.27$, thus, this prior is less revealing than \lastweek ($0.24$). This indicates that commuters show stronger weekly seasonality in their journeys. \maxr and \maxu achieve very small privacy loss ($0.01$ and $0.02$ resp.), thus, aggregate time-series enhance insignificantly \adv's inference goal. \maxu constantly reduces \adv's error over the prior, while \maxr increases it for a small percentage of users ($5\%$). For SFC, \adv's localization error with \lastday is $0.71$ indicating that cabs are a bit more likely to appear in the ROIs of last day rather than those of last week ($0.73$ error). Once again, the greedy inference strategies do not help \adv improve her predictions and there is negligible privacy loss.

\descr{LAST\_HOUR.} Finally, we plot \adv's error while localizing users with the \lasthour prior. For TFL, we observe that her error is now larger ($0.31$) compared to the two previous cases (\lastweek, \lastday) indicating that, in general, commuters do not show up in the ROIs of their last hour. The knowledge of the aggregate time-series enables \adv to improve her localization performance insignificantly and the greedy strategies \maxr and \maxu yield negligible amount of privacy loss ($0.01$ and $0.02$ resp.). When localizing SFC cabs with \lasthour, \adv's mean error is $0.64$. Thus, as this assignment prior helps \adv localize cabs better than \lastday ($0.71$) or \lastweek ($0.73$) and unlike tube commuters, taxis are more likely to appear in the locations they have recently reported. Both inference strategies lead to very small privacy loss.

\end{document}